\documentclass[a4paper,11pt]{article}

\usepackage{jheppub} 
\usepackage[T1]{fontenc}
\usepackage[utf8]{inputenc}
\usepackage[english]{babel}
\usepackage{amsmath,amssymb,amsfonts,amsthm,mathtools}
\usepackage{graphicx}
\usepackage{tabularx}
\usepackage{hyperref}
\hypersetup{unicode,psdextra}
\usepackage{physics}
\usepackage{caption}
\usepackage{subcaption}
\usepackage{placeins}

\usepackage{tikz}
\usetikzlibrary{calc, positioning}

\newcommand{\tblargfirst}[1]{\(\bigl(#1\)}
\newcommand{\tblargnext}[1]{\(,\;#1\)}
\newcommand{\tblarglast}[1]{\(,\;#1\bigr)\)}

\title{Perturbative construction of amplitudes from on-shell trees with vacuum pairs: the all-plus four-gluon amplitude through order $g^6$}

\author{M.~Maniatis}

\affiliation{Centro de Ciencias Exactas, Universidad del B\'io-B\'io, Casilla 447, Chill\'{a}n, Chile}

\emailAdd{maniatis8@gmail.com}

\abstract{
We formulate a systematic on-shell construction of perturbative amplitudes by
combining the Feynman--tree theorem with BCFW recursion.  Amplitudes are built
entirely from on-shell tree amplitudes generated by BCFW and are supplemented
by additional unobservable state-conjugate on-shell pairs, called vacuum pairs,
whose momenta are integrated over Lorentz-invariant phase space.  At fixed
coupling order, the contributing tree amplitudes are classified by polygon
bookkeeping, while their products form forests whose relative signs are fixed
by inclusion--exclusion.  We show that these signs coincide with the signs assigned
by the full Feynman--tree-theorem cut expansion, so that the construction
reproduces the ordinary perturbative loop expansion order by order.  The
dynamical tree input consists of the allowed three-point on-shell amplitudes;
the particle spectrum, couplings, and color data are taken as given, and no
off-shell object is introduced at any stage.

As a nontrivial test of this equivalence, we study the four-gluon all-plus
amplitude through orders $g^4$ and $g^6$.  The on-shell construction reproduces
the finite rational one-loop primitive amplitude and the known bare planar and
non-planar two-loop primitive integrands, including the bow-tie terms in the
planar primitive.
}
\begin{document}

\maketitle
\setcounter{page}{2}
\flushbottom

\section{Introduction}
\label{sec:introduction}

A central lesson of the modern amplitudes program is that the
perturbative expansion need not first be organized in terms of off-shell fields.
At tree level this lesson is especially sharp.  For massless particles, the
basic object is the three-point on-shell amplitude.  Its spinor-helicity
dependence is fixed by Lorentz invariance and little-group scaling.  For
external helicities \(h_i\), \(i=1,2,3\), we obtain (see for instance~\cite{Elvang:2013cua})
\begin{equation} \label{primitive}
A_3(1^{h_1}, 2^{h_2}, 3^{h_3}) =
\begin{cases}
g [12]^{h_1+h_2-h_3} [23]^{h_2+h_3-h_1} [31]^{h_3+h_1-h_2},
& h_1+h_2+h_3 \ge 0,\\[2mm]
g \langle 12\rangle^{h_3-h_1-h_2}
\langle 23\rangle^{h_1-h_2-h_3}
\langle 31\rangle^{h_2-h_3-h_1},
& h_1+h_2+h_3 < 0.
\end{cases}
\end{equation}
Here \(g\) denotes the corresponding coupling.  The kinematic dependence is
therefore fixed, while the particle spectrum, the allowed three-point
couplings, and the color organization remain theory-dependent input.  The
amplitude in \eqref{primitive} contains all kinematic information of the basic
on-shell building block.  It does not rely on off-shell fields, and it carries
no gauge redundancy.

For three real massless momenta, momentum conservation forces all invariants to vanish.
BCFW recursion relations show how this apparently vanishing
three-point object can be used.  By analytically continuing the external
momenta, an \(n\)-point tree amplitude is viewed as a meromorphic function
\(\hat A_n(z)\) of a complex deformation parameter
\(z\)~\cite{Britto:2004ap,Britto:2005fq}; see
also~\cite{Bern:2007dw,Feng:2011np,Elvang:2013cua}.  If the boundary term at
infinity vanishes,
\begin{equation}
\lim_{z\to\infty} \hat{A}_n(z)=0,
\end{equation}
the amplitude is reconstructed completely from its factorization poles.  The
residues are products of lower-point on-shell amplitudes, and the recursion
terminates on the three-point amplitudes in~\eqref{primitive}.  For
tree-level Yang--Mills amplitudes, valid shifts with vanishing boundary term can
be chosen~\cite{Arkani-Hamed:2008yf,Feng:2011np}.  In this
way all tree amplitudes are obtained by gluing on-shell three-point amplitudes,
and the complex deformation is only an auxiliary device: after the residues are
summed, the result is evaluated at \(z=0\), the original physical kinematics.
An illustration is the Parke--Taylor formula for the
color-ordered \(n\)-gluon MHV amplitude~\cite{Parke:1986gb},
\begin{equation} \label{Parke}
A_n(1^+, \ldots, i^-, \ldots, j^-, \ldots, n^+) =
\frac{\langle i  j \rangle^4}
{\langle 1 2 \rangle \langle 2 3 \rangle \ldots \langle n 1 \rangle}\;,
\end{equation}
up to the suppressed conventional coupling and color factors.  Massive
particles can also be incorporated, either through a Higgs description within a
massless theory or by using massive spinor-helicity variables; see for
example~\cite{Badger:2005zh,Arkani-Hamed:2017jhn}.

BCFW recursion by itself does not provide the full perturbative amplitude:
beyond tree order the loop contributions are still missing.  In the usual
formulation, these contributions are described by integrations over virtual
momenta and off-shell propagators.  Feynman
diagrams provide a precise and successful
prescription for these contributions, but they are not a natural final
organization of gauge-theory amplitudes: individual diagrams are
gauge-dependent, their number grows rapidly with the number of external legs,
and large cancellations are often required before the compact, gauge-invariant
answer appears.  The issue is not the correctness of the diagrammatic
expansion, but the fact that it introduces unphysical intermediate quantities in
order to obtain a physical on-shell amplitude.

Modern and generalized unitarity methods improve this situation
substantially~\cite{Bern:1994zx,Bern:1994cg,Bern:2007dw}.  They reconstruct
loop amplitudes, or their integrands and integral coefficients, from products
of on-shell tree amplitudes.  The present work asks a more limited and concrete
question: whether fixed-order forward-limit tree products can be organized into
amplitudes beyond tree order.  We test this question below for the four-gluon
all-plus amplitude through order \(g^6\).

This question is closely related to the Feynman--tree theorem~\cite{Feynman:1963ax,Feynman:1972mtm,Feynman:2000fh}.
The Feynman--tree theorem rewrites a loop denominator
product as a signed sum of cut and uncut propagators.

Beyond one loop, two distinct modifications of the Feynman--tree-theorem
strategy are relevant here.  Loop--tree
duality~\cite{Catani:2008xa,Bierenbaum:2010cy} starts from a known off-shell
Feynman integrand and reorganizes its loop-energy residues into a smaller set of
terms with dual propagators and modified causal prescriptions, rather than
retaining the full inclusion--exclusion cut sum explicitly.  This is an exact
algebraic identity in the loop integration.  Ref.~\cite{Caron-Huot:2010fvq}
modifies the organization in a different way: causal boundary conditions lead
to an \(\ell\)-loop representation as a phase-space integral over exactly
\(\ell\) on-shell particles.  What fails already at two loops is the naive
identification of the corresponding integrand with the forward limit of a
physical tree amplitude, which is inconsistent with unitarity.  The proposed corrected
formula involves a non-physical combination of connected trees in a generic
theory.  In planar gauge theories special color correlations allow this
combination to be rewritten as a physical forward amplitude.
Beyond the planar limit, Ref.~\cite{Caron-Huot:2010fvq} abandons a direct phase-space representation in terms of physical forward trees and instead conjectures that loop amplitudes are determined implicitly by their forward-tree data.

The present work adopts neither modification.  It retains the full
Feynman--tree-theorem cut sum and supplies every tree factor by BCFW recursion,
order by order in the coupling.  The forest signs are fixed by the
inclusion--exclusion prescription of
subsection~\ref{subsec:inclusion_exclusion}, shown in
subsection~\ref{subsec:equivalence} to reproduce the Feynman--tree-theorem signs
exactly.

The construction itself starts from on-shell tree amplitudes generated
recursively from the three-point amplitude.  In the process of gluing such trees
beyond the tree approximation, we supplement the tree products by additional
unobservable state-conjugate on-shell pairs.  We call these pairs vacuum pairs
and integrate over their Lorentz-invariant phase space.  A forest with \(k\)
tree factors is assigned the alternating inclusion--exclusion sign
\((-1)^{k-1}\).  Subsection~\ref{subsec:equivalence} shows that the same
relative sign follows for the corresponding collected labelled cut support
from the Feynman--tree theorem.

We test this construction for the four-gluon all-plus amplitude through order
\(g^6\).  For the fixed-order bookkeeping it is convenient to represent
\(n\)-point tree amplitudes as \(n\)-sided polygons; the corresponding triangle
counting is introduced in subsection~\ref{subsec:polygon}.  The coupling-order
labelling used for this all-plus example is stated where the order-\(g^6\)
calculation begins.

The principal idea has appeared in earlier work.  It was proposed
in~\cite{Maniatis:2015kex}, developed further
in~\cite{Maniatis:2016dcf,Maniatis:2016nmc}, applied to the order-\(g^4\)
four-gluon all-plus amplitude in~\cite{Maniatis:2019pig}, and reformulated in
the tree-world picture of~\cite{Maniatis:2021treeworld}.  Those papers
introduced the use of on-shell three-point amplitudes, BCFW tree construction,
FTT-inspired forward limits, and the one-loop all-plus test case.  The new
content of the present paper is a systematic fixed-order formulation and its
application to the four-gluon all-plus amplitude through order \(g^6\).  In particular, we
give the vacuum-pair definition, the sector and polygon bookkeeping, the
inclusion--exclusion sign rule, and the labelled-support comparison with the
perturbative loop expansion, including the role of gauge degrees of freedom in
the comparison with ordinary Feynman diagrams.

In the vacuum-pair classification introduced below, the explicit order-\(g^6\) calculation decomposes into
 the polygon sectors \(\{8\}\), \(\{6,4\}\), \(\{5,5\}\), and \(\{4,4,4\}\).
 The sector contributions are then summed and compared with the ordinary
 bare two-loop primitive representation, including the planar, non-planar, and
 bow-tie denominator families. Several issues that are absent, or only implicit, in the one-loop
example appear here explicitly: multiple vacuum pairs, products of tree
sectors, simultaneous propagator on-shell conditions, and the corresponding
analytic FTT coefficients. This two-loop test is
therefore the central evidence for the fixed-order construction developed in
the paper.

We emphasize the scope of the claim.  The particle spectrum, admissible
three-point couplings, and color structure are still assumed as input; they are
not derived here from Lorentz invariance alone.  Once these on-shell data are
specified, the calculation below shows how the corresponding amplitudes can be
obtained without using gauge-fixed off-shell diagrams as intermediate building
blocks.

\section{Vacuum-pair construction}
\label{sec:const}

This section gives the general rules used in the explicit calculations.
The key idea is to construct higher-order amplitudes by inserting additional
unobservable on-shell momentum pairs into tree-level amplitudes, then integrating
over their phase space with appropriate signs determined by inclusion-exclusion.
The construction is formulated in terms of on-shell tree amplitudes,
supplemented by unobservable state-conjugate on-shell pairs, called vacuum
pairs. The tree amplitudes are determined from the allowed three-point
amplitudes by BCFW recursion, including any boundary contribution required
for the chosen shift.

We work in dimensional regularization with $D=4-2\epsilon$.
The spin-state dimension of internal gluon sums is denoted by \(D_s\), which
may either be identified with \(D\) or kept as an independent scheme
parameter. Equalities below are understood at the level of the regulated
amplitude, modulo total derivatives and scaleless contributions that
integrate to zero.

\subsection{Vacuum-pair insertions}
\label{subsec:insertions}

Given external states of \(n\) particles, we insert \(r\ge0\) additional
unobservable state-conjugate on-shell pairs, called vacuum pairs, with momenta
\((-\ell_a,\ell_a)\), \(a=1,\ldots,r\).  Each vacuum pair consists of two
additional on-shell legs with opposite momenta that enter the tree amplitudes.
Their common on-shell momentum is integrated over phase space and their paired
physical states are summed.  The positive-energy member
\(\ell_a\) carries a physical state \(h_a\), while the opposite-momentum member
carries the conjugate state \(\bar h_a\).

The bar denotes the state paired with \(h_a\) when the orientation of the
on-shell line is reversed.  It is fixed by the usual sewing/completeness
condition: after summing over \(h_a\), the two half-edges
\((-\ell_a)^{\bar h_a}\) and \(\ell_a^{h_a}\) reconstruct the physical
internal-state sum of the corresponding propagating particle.  Thus
\(\bar h_a\) is determined by \(h_a\), and is not an additional independent
state.  For a self-conjugate gluon the species is unchanged; in a
four-dimensional helicity basis this pairs \(+\) with \(-\), up to conventional
phases.  For non-self-conjugate particles the bar also includes the
antiparticle or species conjugation required by the internal propagator.

The vacuum-pair construction uses ordinary connected on-shell tree amplitudes.
We distinguish a tree that still contains its momentum-conservation delta
function from the corresponding delta-stripped color-ordered tree by
\begin{equation}
\mathbf A_m^{(0)}(1,\ldots,m)
:=
(2\pi)^D\delta^{(D)}\!\left(\sum_{i=1}^{m}p_i\right)
A_m^{(0)}(1,\ldots,m).
\label{eq:delta_containing_tree_definition}
\end{equation}
The symbol \(\mathbf A_m^{(0)}\) denotes the distribution containing the delta
function.  The plain \(A_m^{(0)}\) is the corresponding delta-stripped tree
amplitude used in the subsequent on-shell calculations.  The inserted pair legs
are first treated just like additional external on-shell legs of these trees.
Depending on how the observed legs and the vacuum-pair
legs are distributed among different tree factors, a contribution may contain
one tree amplitude or be a product of several tree amplitudes.  Each such
distribution defines one tree-forest assignment, denoted by \(\Gamma\).  No parent loop graph or cut
propagator is part of the definition; those are introduced only later in the
bookkeeping of subsection~\ref{subsec:polygon} and in the comparison with the
Feynman--tree theorem in subsection~\ref{subsec:equivalence}.

\(\Gamma\) specifies which observed
legs and which vacuum-pair legs belong to each tree factor, and in which cyclic
order they appear.  Within a representative, the two legs of pair \(a\) are
distinguished as \(-\ell_a\) and \(\ell_a\), so their momentum orientation is
retained.  The index \(a\), however, only names a dummy phase-space variable.
Assignments differing only by a simultaneous relabeling of the vacuum pairs
in every tree factor and phase-space measure are therefore identified.
This identification does not exchange the two oriented members
\(-\ell_a\) and \(\ell_a\) of an individual pair.  State and species labels
are included whenever they are displayed.
Let \(k=k(\Gamma)\) be the
number of connected tree factors in \(\Gamma\).  Thus \(\Gamma\) is represented
by a collection of \(k\) ordered lists:
\begin{equation}
W_\alpha=(w_{\alpha,1},\ldots,w_{\alpha,m_\alpha}),
\qquad
\alpha=1,\ldots,k,
\end{equation}
whose entries are disjoint and whose union is
\(\{1,\ldots,n,-\ell_1,\ell_1,\ldots,-\ell_r,\ell_r\}\).
For a fixed \(\Gamma\), each list \(W_\alpha\) is a cyclically ordered sequence;
this order is the color ordering used in the corresponding tree factor, and
\(m_\alpha=|W_\alpha|\) is its size. The sector
\begin{equation} \label{eq:sector}
\{m_1,\ldots,m_k\}
\end{equation}
retains only the sizes of these factors and does not determine their cyclic
orderings or leg assignments.
For a chosen color structure, the sector contribution is obtained by summing
over the inequivalent assignments of observed legs and vacuum-pair legs whose
sewn color factors contribute to that structure.  The full color-dressed
amplitude requires the union of the assignments associated with all color
structures in the chosen basis.
In the explicit enumeration we choose one canonical representative under the
simultaneous vacuum-pair relabelling above, cyclic rotations of each
color-ordered tree, and exchanges of genuinely identical tree factors.  Apart
from these stated identifications, two assignments are kept distinct.  Because
only one canonical representative is retained, there is no additional
\(1/r!\) division.  Retaining all \(r!\) names of the same set of dummy pair
variables would instead require the corresponding permutation normalization.
The explicit examples use these equivalence classes of \(\Gamma\).  The
tables in sections~\ref{sec:g4_allplus} and \ref{sec:g6_allplus_revised}
assign these representatives to the different sectors.
For example, for \(n=4\) and one vacuum pair one possible assignment is
\(\Gamma=\{W_1\}\), with
\begin{equation}
W_1=(1,2,3,4,-\ell,\ell),
\end{equation}
which represents the tree factor
\(\mathbf A_6^{(0)}(1,2,3,4,-\ell,\ell)\); any cyclic rotation of this sequence gives
the same color-ordered factor.  With two vacuum pairs, one possible
two-tree-factor assignment is represented,
with all momenta outgoing on each tree factor,
by
\begin{equation}
\Gamma=\{W_1,W_2\},\qquad
W_1=(1,2,-\ell_1,-\ell_2),\qquad
W_2=(3,4,\ell_1,\ell_2),
\end{equation}
corresponding to the product
\(\mathbf A_4^{(0)}(1,2,-\ell_1,-\ell_2)
\mathbf A_4^{(0)}(3,4,\ell_1,\ell_2)\).  The entries of \(W_1\) and \(W_2\) are
disjoint and together contain all observed and vacuum-pair legs.

The phase-space integrations and state sums associated with the vacuum pairs
connect these tree components into a contribution to the same external
amplitude.
The forward-limit product for \(\Gamma\) is
\begin{equation}
\label{eq:vpdef}
\mathcal{A}^{(r)}_{\mathrm{vp}}[\Gamma]
=
\sum_{h_1,\ldots,h_r}
\int\prod_{a=1}^{r}d\Phi(\ell_a)\,
\prod_{\alpha=1}^{k(\Gamma)}
\mathbf A_{m_\alpha}^{(0)}(W_\alpha).
\end{equation}
The tree factors \(\mathbf A_{m_\alpha}^{(0)}(W_\alpha)\) include their
momentum-conservation delta functions.  With the amplitude conventions used
here, each complete tree factor carries the conventional overall phase
\(-i\).  A tree forest \(\Gamma\) containing \(k\) tree factors
therefore carries the phase
\((-i)^k\).  In intermediate stripped kernels this common phase is suppressed
and restored explicitly in the collected vacuum-pair contributions.
The full vacuum-pair contribution is obtained by summing the
forward-limit products \eqref{eq:vpdef} over all inequivalent assignments
\(\Gamma\) in the chosen bookkeeping convention.  In the perturbative
construction, the contribution of a fixed tree forest \(\Gamma\) is
multiplied by the inclusion--exclusion sign
\((-1)^{k-1}\) defined in
subsection~\ref{subsec:inclusion_exclusion}.  The phase-space measure, state sum,
normalization, color factor, fermionic sign, and any symmetry-factor convention
belong to the contribution \(\mathcal{A}^{(r)}_{\mathrm{vp}}[\Gamma]\) itself;
they are not part of the inclusion--exclusion sign.
Thus the combined real sign and conventional phase attached to the assignment
is the product \((-1)^{k-1}(-i)^k\).
Distinct inequivalent assignments \(\Gamma\) are kept separate in this sum, even if
they later give the same displayed tree product or denominator support after
auxiliary labels are suppressed.  A simultaneous relabelling of all vacuum
pairs has already been identified and supplies no multiplicity.  Any remaining
multiplicity is obtained only by the final summation over
inequivalent assignments; it is not part of the elementary sign
\((-1)^{k-1}\).
The single-tree case is recovered by taking \(k=1\).  For \(r=0\) and
\(k=1\), this reduces to
\(\mathcal{A}^{(0)}_{\mathrm{vp}}=\mathbf A^{(0)}_n(1,\ldots,n)\).
For \(r=0\), only the case \(k=1\) contributes to the scattering amplitude
considered here.
The state sum in \eqref{eq:vpdef} is over the species and physical states
carried by the inserted vacuum pairs.

The Lorentz-invariant one-particle phase-space measure is
\begin{equation}
d\Phi(\ell_a)
\equiv
\frac{d^{D}\ell_a}{(2\pi)^D}\,\delta^+(\ell_a),
\qquad
\delta^+(\ell_a)
\equiv
2\pi\,\theta(\ell_a^0)\delta(\ell_{a,D}^2).
\label{eq:positive_delta_definition}
\end{equation}
Here \(\ell_{a,D}^2\) denotes the full \(D\)-dimensional invariant.
The momentum argument of \(\delta^+\) fixes the positive-energy branch and is
therefore retained whenever an on-shell distribution is displayed.

In this paper the forward limit is understood as a regulated prescription.
For each vacuum pair we first regard the two legs as distinct
on-shell tree legs and construct the corresponding tree amplitude by BCFW
recursion.  The exact forward identification \((-\ell_a,\ell_a)\) is imposed
only after the factorization channels and state sums have been defined.  Terms
whose support collapses to soft or collinear forward-boundary configurations
and which contain no dimensionful scale are discarded in dimensional
regularization.  This prescription is sufficient for the fixed-order examples
considered below; a general proof of forward-limit existence is not assumed
here.

The delta-stripped factor \(A^{(0)}_{m_\alpha}(W_\alpha)\) in each
\(\mathbf A^{(0)}_{m_\alpha}(W_\alpha)\) is a complete on-shell tree amplitude.
Whenever the chosen shift has vanishing large-\(z\) behavior, it is
reconstructed by BCFW recursion~\cite{Britto:2004ap,Britto:2005fq}; any
required boundary contribution is included in the complete tree.  The
recursion reduces the amplitude to lower-point factorization residues and
ultimately to the allowed three-point amplitudes.
At this stage, $\Gamma$ is only a bookkeeping representation of one assignment.
For fixed $n$ and $r$, the full construction sums over the inequivalent
distributions of the observed and vacuum-pair legs among one or several
cyclically ordered on-shell tree factors.  These assignments are retained as
labelled entries until the final
projection to unlabelled supports.  The relative-sign prescription is defined
in the next subsection and compared with the analytic FTT coefficients in
subsection~\ref{subsec:equivalence}.

\subsection{Inclusion--exclusion sign prescription}
\label{subsec:inclusion_exclusion}

We assign the inclusion--exclusion sign directly from the number
\(k\) of tree factors in a tree forest \(\Gamma\).  The contribution
with one tree factor is counted once and enters with a plus sign.  The next
factorization level contains two tree factors and is subtracted; the level with
three tree factors is added back, and so on.  Thus each step
\(k\to k+1\) reverses the sign, giving
\begin{equation}
\operatorname{sgn}(\Gamma)=(-1)^{k-1}.
\label{eq:tree_factor_sign}
\end{equation}
Thus the sign changes whenever \(k\) is increased by one.  This is the
vacuum-pair sign prescription; it is formulated entirely in terms of the tree
forest and does not require a parent loop graph.

It is also useful to determine the number \(c\) of independent vacuum-pair
momenta associated with a tree forest containing \(r\) vacuum pairs.  Resolve
the momentum-conservation delta functions of the \(k\) tree factors and remove
the overall external delta function.  Let \(\rho\) be the rank of the
remaining independent \(D\)-dimensional linear constraints on the \(r\)
vacuum-pair momenta.  We define
\begin{equation}
c:=r-\rho .
\label{eq:independent_phase_space_count}
\end{equation}
Thus \(c\) is the number of independent \(D\)-dimensional vacuum-pair momenta
left to integrate after the tree constraints have been resolved.  For any tree
forest \(\Gamma\) contributing to a connected scattering amplitude, the
vacuum-pair identifications link its \(k\) tree factors into a connected
structure.  After the overall external delta function has been removed, the
tree momentum-conservation conditions therefore impose \(k-1\) independent
constraints.  Consequently,
\begin{equation}
\rho=k-1,
\qquad
c=r-k+1.
\label{eq:vp_independent_momenta}
\end{equation}
This definition depends only on the tree forest.  A numerator or residual
propagator may depend on several of the remaining momenta and thereby prevent
factorization of the integrand, but such a dependence does not change \(c\).

Different derivations of the same
labelled tree forest \(\Gamma\) do not create additional multiplicities;
genuine multiplicities arise only from inequivalent final labelled tree
forests.

The coefficient \((-1)^{k-1}\) is assigned to every inequivalent
vacuum-pair assignment \(\Gamma\).  This is the elementary
inclusion--exclusion sign.  Different assignments are retained separately even
if, after imposing their momentum constraints and suppressing auxiliary
labels, they yield the same tree product or denominator support.  The
coefficient of the final unlabelled support is obtained by summing these
retained labelled realizations.  Extra integer factors therefore come from
multiplicity, not from a modification of the elementary sign.  Simultaneous
permutations of the dummy vacuum-pair names are not among these realizations,
because they have already been identified in
subsection~\ref{subsec:insertions}.

\subsection{Fixed-order polygon bookkeeping}
\label{subsec:polygon}

The tree amplitudes in \eqref{eq:vpdef} have \(n+2r\) ordered external legs:
the \(n\) observed particles and the \(2r\) legs of the \(r\) vacuum pairs.
Such a tree may be represented by a polygon with \(n+2r\) sides, where each
side corresponds to an external leg in the color-ordered amplitude. A cubic
decomposition of this polygon consists of
\(n+2r-2\) triangles, where each triangle represents one three-point on-shell
amplitude \(A_3\). For example, a 4-point tree amplitude corresponds to a
quadrilateral that decomposes into 2 triangles.  This is the one-polygon sector: all observed legs and all
vacuum-pair legs belong to one ordered polygon.  This special case is included
in the general counting relation below.
In a cubic representation, every Yang--Mills three-point vertex carries one
power of the coupling, so an \(n\)-point color-ordered tree amplitude with
\(N_3\) cubic vertices has coupling order \(g^{N_3}\).  The polygon bookkeeping
uses only this fixed-cubic-order counting: the total number \(N_3\) of triangles
determines the coupling order \(g^{N_3}\).  Thus the sectors are the
fixed-cubic-order bookkeeping of the assignments \(\Gamma\) defined in
subsection~\ref{subsec:insertions}.

More generally, for an assignment \(\Gamma=\{W_1,\ldots,W_k\}\), as outlined in
subsection~\ref{subsec:insertions}, the \(k=k(\Gamma)\) ordered tree
factors may be represented by \(k\) polygons.
If polygon \(a\) has
\(m_a\) sides, then all observed legs and all vacuum-pair legs have been
distributed among these polygons, so
\begin{equation}
\label{eq:polygon_side_count}
\sum_{a=1}^k m_a=n+2r .
\end{equation}
Since an \(m_a\)-gon decomposes into \(m_a-2\) triangles, the fixed-order
bookkeeping relation is
\begin{equation}
\label{tree_factors}
N_3
=
\sum_{a=1}^k (m_a-2)
=
n+2r-2k.
\end{equation}
For fixed coupling order \(N_3\) and a given factorization \(k\), the number of
vacuum pairs is fixed.

For a connected assignment, the same counting gives the number \(c\) of
independent vacuum-pair momenta defined in
\eqref{eq:vp_independent_momenta}:
\begin{equation}
c=r-k+1.
\label{eq:connected_vacuum_momentum_count}
\end{equation}
For later use in the FTT comparison, suppose that gluing the \(r\) vacuum pairs
of this forest produces a connected parent graph with \(E\) internal edges,
\(V\) vertices, and loop number \(L\).  The connected parent obeys
\(L=E-V+1\).  Cutting the \(r\) glued edges recovers the forest with the same
\(V\) vertices, \(E-r\) internal edges, and \(k\) connected tree factors.  It
therefore obeys \(E-r=V-k\), and hence
\begin{equation}
r-L=k-1.
\label{eq:looprk}
\end{equation}
Equivalently, \(L=r-k+1\), so comparison with the preceding relation gives
\(L=c\) for the connected parent graph.  This identity is independent of the
valence of the vertices.  In the recursive FTT construction every recursion
opens at least one loop.  Iterating the construction until no loops remain
produces a terminal graph that is precisely such a forest.

As discussed in subsection~\ref{subsec:insertions}, we call the unordered label \(\{m_1,\ldots,m_k\}\) of a product of \(k\) tree
factors its sector.  The label forgets the order of the factors; for example,
\(\{6,4\}\) and \(\{4,6\}\) are the same sector, while \(\{4,4\}\) denotes a
product of two quadrilateral tree factors.
In the massless examples considered below, a sector containing a three-point
tree factor has support only on degenerate real three-point kinematics. Under
the regulated forward-limit prescription of subsection~\ref{subsec:insertions}, the resulting
boundary-supported or scaleless terms are discarded. We therefore retain only
sectors with \(m_\alpha\geq4\).

For the four-gluon examples at order \(g^4\) used below, we have \(n=4\), \(N_3=4\)
and hence \(r=k\).  After excluding sectors with a
three-sided polygon, the fixed-order sectors are
\begin{equation}
\begin{aligned}
k&=1,\quad r=1 &&:\quad \{6\},\\
k&=2,\quad r=2 &&:\quad \{4,4\}.
\end{aligned}
\label{eq:g4_polygon_sectors_bookkeeping}
\end{equation}
Thus the order-\(g^4\) calculation has only the one-pair hexagon
sector and the two-pair product of quadrilaterals.  At order \(g^6\), \(N_3=6\)
and hence \(r=k+1\).  After the same exclusion, the fixed-order sectors are
\begin{equation}
\begin{aligned}
k&=1,\quad r=2 &&:\quad \{8\},\\
k&=2,\quad r=3 &&:\quad \{6,4\},\ \{5,5\},\\
k&=3,\quad r=4 &&:\quad \{4,4,4\}.
\end{aligned}
\label{eq:g6_polygon_sectors_bookkeeping}
\end{equation}
Thus the non-triangular order-\(g^6\) sectors are the octagon, the
hexagon--quadrilateral product, the two-pentagon product, and the product of
three quadrilaterals.

For a fixed planar primitive we choose the observed-leg color ordering
\((1,\ldots,n)\).
The vacuum-pair legs \((-\ell_a,\ell_a)\) are additional ordered arguments of
the tree factors.  In the one-polygon sector, one possible representative
cyclic sequence is
  \begin{equation}
  (1,\ldots,n,-\ell_1,\ell_1,-\ell_2,\ell_2,\ldots).
  \end{equation}
This planar convention is not a universal ordering prescription for a
color-dressed amplitude.  Non-planar color structures can require additional
cyclic orderings of the intermediate tree factors; they are classified by the
color factor obtained after the vacuum-pair color indices are sewn.  For
products of several polygons, the sector label specifies only the number of
sides of the tree factors.  The actual color-ordered tree factors are obtained
by assigning the observed and vacuum-pair legs to the polygons in all
inequivalent cyclic orders compatible with the color structure under
consideration, up to simultaneous vacuum-pair relabelling, cyclic rotation of
each factor, and exchange of identical polygon factors.  Thus the different
rows in the later \(k=2\) and \(k=3\) tables are different assignments
\(\Gamma\) of legs to tree factors, not different names for the same vacuum
pairs or different orderings of one fixed tree factor.

The role of the polygon language in the general setup is to list the
allowed finite set of fixed-order sectors before the state sums are carried out.
Individual contributions are denoted by \({\cal C}_{F,j}\), where \(F\) is the
compact sector label, such as \(64\) or \(444\), and \(j\) enumerates the
canonical representatives listed in the corresponding bookkeeping table.


\subsection{Equivalence with the perturbative loop expansion}
\label{subsec:equivalence}

We now prove the equivalence at fixed perturbative order between the signed
vacuum-pair construction and the conventional perturbative loop expansion.
The proof has three ingredients.  First,
the integrated FTT rewrites the regulated loop expansion as a sum over
terminal labelled cut supports.  Second, after the complete gauge-fixed forest
sum has been formed, BRST invariance gives its representation in terms of
complete physical on-shell tree amplitudes.  Third, the
identification of cuts with vacuum pairs matches these
physical cuts with labelled vacuum-pair insertions, including the
inclusion--exclusion signs.  The agreement of the relative signs is an
essential step in the equivalence proof.

We apply the recursive FTT to the loop graphs at fixed perturbative order after
the gauge has been fixed.  Applying it graph by graph is an intermediate
bookkeeping step; an individual graphwise forest is not identified with a
vacuum-pair contribution.  The comparison is made only after the complete
gauge-fixed forest sum has been converted to its physical representation, and
then at the level of labelled oriented cut supports.  It is subject to the
regularization and forward-limit assumptions stated in
subsection~\ref{subsec:insertions}.  In particular, the FTT identity is used as
an exact identity for the regulated loop integral obtained after the relevant
loop-energy contour integrations.
The identification with vacuum-pair tree products is understood modulo the
existence of the regulated forward limits and modulo the discarded
boundary-supported or scaleless terms specified there.

Let us consider any loop chosen in a recursive FTT step.  If
\(F=\{1,\ldots,N\}\) denotes the labelled denominator slots of that loop, we
orient their momenta coherently with the loop-energy variable: after a possible
reversal of an individual denominator momentum, each \(q_i\) has the form
\(q_i=\ell_F+\kappa_i\), where \(\kappa_i\) is independent of the loop momentum
\(\ell_F\).  We use the propagator convention of
appendix~\ref{app:ftt}, \(\Delta_{F,i}=i/(D_i+i0)\) and
\(\delta_i^+=\delta^+(q_i)\).  The orientation is part of the cut label:
\(\delta^+(-q_i)=\delta^-(q_i)\) is a different energy branch; see
\eqref{eq:oriented-cut-delta}.

With all other factors held fixed, the one-loop opening is the integrated FTT
identity \eqref{eq:FTT}.  Its sum runs over nonempty subsets of the slots of
the loop being opened.
The empty choice is absent because the identity replaces the term in which all propagators of the selected loop are Feynman propagators.
Thus a loop with \(N\)
labelled slots gives \(\sum_{m=1}^{N}\binom{N}{m}=2^N-1\) local openings in
that step, all of which cut at least one propagator of that loop.  The
numerator \({\cal N}_{F}\), containing the momentum, Lorentz, spin, color, and
coupling dependence of the graph, is carried unchanged through this denominator
expansion.  The explicit kernels below are written with stripped denominators
\(G_i=1/(D_i+i0)\); the conversion from \(\Delta_{F,i}=iG_i\) only changes the
conventional powers of \(i\), which are restored separately below.

We now apply \eqref{eq:FTT} recursively.  At one step, opening a loop with
labelled slots \(F\) produces all nonempty choices of cut propagators in that
loop.  Each slot is either cut, giving \(\delta^+(q_i)\), or left uncut, giving
\(\Delta_{F,i}\).  If an uncut loop remains, the identity is applied again.

For a fixed labelled graph \(G\), a terminal support is produced by a finite
history
\begin{equation}
S=S_1\cup\cdots\cup S_L,\qquad
\emptyset\neq S_a\subseteq F_a
\label{eq:terminal-support-recursion}
\end{equation}
where \(F_a\) is the set of labelled propagator slots in the loop opened at
the \(a\)th step, after the previous cuts \(S_1,\ldots,S_{a-1}\) have been
made.  Equation~\eqref{eq:terminal-support-recursion} makes the order
independence of the final labelled support manifest: \(S\) depends only on the
union of the local cut sets, not on the order in which they were generated.
Every \(S_a\) is nonempty, so each FTT step opens the selected loop.  After the
\(L\) recursions all loops have been opened, and the uncut subgraph
\(G\setminus S\) is a forest.  Thus \(S\) is the cut support, while
\(G\setminus S\) is the resulting collection of trees.  If different
recursive histories end at the same labelled set \(S\), their algebraic
contributions are collected into the coefficient of that one support.  They
are not distinct labelled supports, and choosing a different admissible order
of loop openings does not create an additional multiplicity.  A cut
orientation remains part of the label: \(\delta^+(q_e)\) and
\(\delta^+(-q_e)\) are distinct oriented terms.

For an individual regulated gauge-fixed graph, a terminal FTT term is a forest
of gauge-fixed tree graphs with the numerator factors inherited unchanged from
that graph.  Such a graphwise forest is not yet an on-shell tree amplitude.  For
fixed physical external data and oriented terminal cut data, including the cut
momenta, we first sum all contributing parent graphs and all residual tree
topologies.  This sum
includes cubic and quartic interactions and the required ghost, matter,
and, in a spontaneously broken theory, Goldstone contributions.  It gives the
corresponding complete gauge-fixed forest sum.

BRST invariance then ensures that the complete forest sum between physical
external states is gauge independent and admits a representation as a sum over physical
on-shell intermediate states, as in ordinary cutting rules and generalized
unitarity~\cite{Bern:1994zx,Bern:1994cg,Bern:2007dw}.  This physical
representation applies only to the completed sum, not to an individual
gauge-fixed graph, an individual denominator support within a tree factor, or
an arbitrary partial forest.  Labels that distinguish these different Feynman
realizations of a complete tree factor are summed at this stage.

Consider now one labelled terminal cut support with cut set \(S\) in this
BRST-completed physical representation.  For each \(i\in S\), the support
contains the oriented positive-energy on-shell factor \(\delta^+(q_i)\).  The corresponding
on-shell line can be represented by two external half-edges with opposite momenta
and conjugate physical-state labels, \(q_i^{h_i}\) and
\((-q_i)^{\bar h_i}\).  These two half-edges are sewn with one
positive-energy phase-space integration and one physical-state sum.  In the
conventions used here this is the same measure as the cut factor,
\begin{equation}
\frac{d^Dq_i}{(2\pi)^D}\,\delta^+(q_i)
\sum_{h_i\in{\rm phys}(D_s)}
\quad\longleftrightarrow\quad
d\Phi(q_i)\sum_{h_i\in{\rm phys}(D_s)} .
\label{eq:cut_pair_identification}
\end{equation}
The right-hand side integrates the common on-shell momentum of the two
half-edges, not two independent momenta.  Conversely, sewing the two half-edges
of a labelled vacuum pair reconstructs the corresponding labelled propagator
slot.  This correspondence preserves the cut or pair labels, momentum
orientations, positive-energy branch, state and species labels, assignment to
tree factors, cyclic ordering, and loop-momentum routing.

The labels retained in the comparison are the outer cut or vacuum-pair labels,
their momentum orientations, positive-energy branches, state and species
labels, assignments to complete tree factors, cyclic orderings, and
loop-momentum routings.  Labels that distinguish cubic, quartic, ghost, or
Goldstone Feynman realizations within a complete tree factor have already been
summed in the BRST completion and are not compared separately.  On the
vacuum-pair side, two completed tree forests are distinct unless they are
related by the equivalences stated in subsection~\ref{subsec:insertions}.
The cut--vacuum-pair identification is made at this completed labelled level.
If the outer labels are suppressed afterward, the number of distinct labelled
realizations gives the corresponding multiplicity.  By contrast, reaching the
same labelled cut support through a different order of recursive loop openings
does not produce another term.

The recursive FTT expansion, the sum over gauge-fixed graphs, and the BRST
completion have therefore produced a forest of complete physical on-shell tree
amplitudes in which the loop connections are absent. This is the tree-forest
structure used in the vacuum-pair construction, where the absent loop
connections are represented by integrated vacuum pairs.

The FTT sign is most easily read after all recursive terms with the same final
labelled support have been collected.  It depends only on the cut propagators
and is therefore common to all cubic, quartic, ghost, and Goldstone terms in the
corresponding complete forest; it factors through the BRST sum.  In the
recursive history \eqref{eq:terminal-support-recursion}, let \(r_a=|S_a|\).
Each cut label occurs once, so that \(r=\sum_{a=1}^{L}r_a\).  The FTT
coefficient at step \(a\) is \((-1)^{r_a-1}\), and the total sign is therefore
\begin{equation}
\prod_{a=1}^{L}(-1)^{r_a-1}
=(-1)^{\sum_{a=1}^{L}(r_a-1)}
=(-1)^{r-L}.
\end{equation}
Recalling the graph relation \(r-L=k-1\) established in
subsection~\ref{subsec:polygon}, the FTT and vacuum-pair signs are therefore
equal:
\begin{equation}
(-1)^{r-L}=(-1)^{k-1}.
\label{eq:ftt_vp_sign_equivalence}
\end{equation}
The right-hand side is precisely the vacuum-pair sign in
\eqref{eq:tree_factor_sign}.  Several recursive histories or labelled
graph occurrences can contribute to the magnitude of a collected coefficient,
but they do not alter this sign.  In particular, quartic contact interactions
change neither this graph identity nor the cut-dependent FTT coefficient.

We finally restore the conventional powers of \(i\), which were suppressed in
the comparison of real combinatorial signs.  Relative to the corresponding
product of ordinary denominators, replacing \(r\) full propagators by on-shell
delta functions supplies a factor \(i^{-r}=(-i)^r\).  A collected FTT term
therefore carries \((-1)^{r-L}(-i)^r\).  The corresponding vacuum-pair forest
has \(k\) complete tree amplitudes and carries
\((-1)^{k-1}(-i)^k\), as defined after \eqref{eq:vpdef}.  By
\eqref{eq:ftt_vp_sign_equivalence} and \(r-k=L-1\), the two expressions differ
only by the common fixed-loop phase \((-i)^{L-1}\), independent of the cut
support.  This factor is unity at one loop and \(-i\) at two loops; the latter
is the overall factor restored in the two-loop primitive amplitudes below.

We have thus established the three ingredients outlined at the beginning of
this subsection.
No term-by-term equality is asserted before BRST completion.  After completion,
the physical-state sums, color factors, and fermionic signs belong to the
complete on-shell tree amplitudes, while normalization and multiplicity follow
from the completed labelled cut--vacuum-pair correspondence.  The numerator is
carried unchanged through the FTT identity \eqref{eq:FTT}, and the comparison
uses the common regulator and forward-limit prescriptions stated above.

Only after this BRST completion are the complete physical on-shell tree
amplitudes evaluated from lower-point on-shell data by BCFW recursion.  The
three-point factors and internal factorization lines introduced by BCFW are a
representation of each complete tree; they are not additional FTT cuts or
vacuum pairs and do not enter the counting of \(r\) or \(k\).  Contributions
from quartic interactions are already included in the complete tree amplitude
and are reproduced by its BCFW representation, including any boundary
contribution required for the chosen shift.  Like the FTT identity, BCFW
recursion is valid in general \(D\) dimensions
\cite{Arkani-Hamed:2008yf}.  Both constructions can therefore be implemented
within the same dimensional regularization scheme.

The calculations in the following sections verify this fixed-order
equivalence explicitly.  At order \(g^6\) this verification concerns the bare
dimensionally regulated primitive amplitudes; the scheme-dependent ultraviolet
counterterm is not generated in the sector calculation and must be added
separately to form a renormalized amplitude.  The labelled support and
cut--glue combinatorics proved above are valid for every finite fixed order.
Within the dimensional regularization framework and forward-limit prescription
adopted here, this establishes the equivalence at every finite fixed order;
specified counterterm contributions can be incorporated by the same
vacuum-pair construction.


\section{Order \texorpdfstring{$g^4$}{g4}: one-loop all-plus amplitude from the vacuum-pair construction}
\label{sec:g4_allplus}

We revisit the order-\(g^4\) calculation of the color-ordered four-gluon
all-plus amplitude in pure Yang--Mills theory,
following~\cite{Maniatis:2019pig}.  Written in the vacuum-pair language of
\eqref{eq:vpdef}, this example is a simple non-vanishing test case and
prepares the order-\(g^6\) calculation in the next section.  The
color-stripped amplitude at order \(g^4\) has four cubic
on-shell vertices.  In the polygon bookkeeping we therefore set \(N_3=4\).
For \(n=4\), the sector list in
\eqref{eq:g4_polygon_sectors_bookkeeping}, evaluated in dimensional
regularization with \(D=4-2\epsilon\), is
\begin{equation}
k=1:\quad \{6\},
\qquad
k=2:\quad \{4,4\}.
\label{eq:g4_explicit_sector_list}
\end{equation}
Since \(r=k\) in both sectors, resolving the tree constraints leaves
\(c=r-k+1=1\) independent vacuum-pair momentum.
The inclusion--exclusion signs in \eqref{eq:tree_factor_sign} for the \(\{6\}\) and
\(\{4,4\}\) sectors are \(+1\) and \(-1\), respectively.
Following \eqref{eq:delta_containing_tree_definition}, every tree amplitude
\(A_m^{(0)}\) in the explicit formulas below is delta-stripped.  The delta
functions are displayed separately when needed, and the overall four-point
delta function is stripped from the local integrands.

For the representative box family used throughout this section, we define
\begin{equation}
\begin{aligned}
D_1&=\ell^2-\lambda_\ell^2,
&
D_2&=(\ell-k_1)^2-\lambda_\ell^2,
\\
D_3&=(\ell-k_{12})^2-\lambda_\ell^2,
&
D_4&=(\ell+k_4)^2-\lambda_\ell^2 .
\label{eq:box_sec3}
\end{aligned}
\end{equation}
Following the shorthand in~\eqref{eq:Gdconv}, \(G_i=G_F(D_i)\), where
\(G_F(D)=1/(D+i0)\) denotes the scalar denominator with the conventional
numerator factor \(i\) stripped off; the full Feynman propagator is
\(iG_F(D)\).  The oriented cut factors in this box family are
\begin{equation}
\begin{aligned}
\delta_1^+&:=\delta^+(\ell)
=2\pi\,\theta(\ell^0)\delta(D_1),
&
\delta_2^+&:=\delta^+(\ell-k_1)
=2\pi\,\theta\bigl((\ell-k_1)^0\bigr)\delta(D_2),
\\
\delta_3^+&:=\delta^+(k_{12}-\ell)
=2\pi\,\theta\bigl((k_{12}-\ell)^0\bigr)\delta(D_3),
&
\delta_4^+&:=\delta^+(-\ell-k_4)
=2\pi\,\theta\bigl((-\ell-k_4)^0\bigr)\delta(D_4).
\end{aligned}
\label{eq:g4_box_cut_definitions}
\end{equation}
These arguments record the positive-energy phase-space member of each
vacuum-pair contribution.  In the FTT comparison, the momenta of the loop being
opened are oriented coherently first, and a reversed phase-space branch is
retained as \(\delta^+(-q_i)=\delta^-(q_i)\).
Composite support labels are abbreviated as
\begin{equation}
\delta_{13}^+=\delta_1^+\delta_3^+,
\qquad
\delta_{24}^+=\delta_2^+\delta_4^+ .
\label{eq:g4_support_label_conventions}
\end{equation}
Thus a label such as \(13\) denotes the support on slots \(1\) and \(3\).

\subsection{The \texorpdfstring{$\{6\}$}{\{6\}} sector}
\label{sec:g4k1}

We begin with the sector \(\{6\}\), that is, \(k=1\) and \(r=1\).  This corresponds to one
vacuum-pair insertion in a single hexagon.  Following \eqref{eq:vpdef}, and
denoting the complete physical state of the positive-energy member of the pair
by \(h\), we obtain
\begin{equation}
\begin{aligned}
{\cal C}_{6_1}
&=
\sum_{h\in{\rm phys}(D_s)}
\int d\Phi(\ell)\,
(2\pi)^D\delta^{(D)}(k_1+k_2+k_3+k_4)
\\
&\qquad\times
A_6^{(0)}
\bigl(1^+,2^+,3^+,4^+,(-\ell)^{\bar h},\ell^h\bigr) .
\end{aligned}
\label{eq:g4_k1_raw_state_sum}
\end{equation}
By a cyclic rotation of the color-ordered tree, we may write the local hexagon
as
\begin{equation}
A_6^{(0)}\bigl((-\ell)^{\bar h},1^+,2^+,3^+,4^+,\ell^h\bigr).
\end{equation}
Since the vacuum pair carries zero total momentum, the explicit tree delta
function in \eqref{eq:g4_k1_raw_state_sum} is only the overall four-gluon
momentum conservation \(\delta^{(D)}(k_1+k_2+k_3+k_4)\).  After this overall
delta function has been stripped off, this assignment gives the phase-space
integral of the local six-point tree with the complete state sum.

The four observed external gluons are kept four-dimensional.  The vacuum-pair
momenta, by contrast, are \(D\)-dimensional, and their states are summed over
the complete physical \(D_s\)-dimensional basis of \(D_s-2\) states; see
appendix~\ref{app:transverse_conventions}.  First consider only the strictly
\(4D\) gluon-helicity part of this state sum.
The two members of a \(4D\) vacuum pair carry opposite helicities, so every term
in the hexagon state sum has exactly one negative-helicity
gluon.  This type of color-ordered pure Yang--Mills tree amplitude vanishes for
at least four external legs.  Thus the strictly \(4D\) gluon-helicity
component of the one-polygon hexagon vanishes,
\begin{equation}
A_6^{(0)}\bigl((-\ell)^{\bar h},1^+,2^+,3^+,4^+,\ell^h\bigr)=0,
\qquad h=\pm.
\label{eq:k1_allplus_zero}
\end{equation}
In dimensional regularization the \(\lambda\)-dependent part of the complete state sum
produces the rational all-plus numerator and must be kept.  The scalar-chain
hexagon is derived in appendix~\ref{app:one_pair_scalar_chain}.
The application \eqref{eq:app_one_pair_hexagon_g4_result} gives
\begin{equation}
A_6^{(0)}
\bigl((-\ell)^I,1^+,2^+,3^+,4^+,\ell^J\bigr)
=
-i\,\frac{[12][34]}{\langle12\rangle\langle34\rangle}\,
G_2G_3G_4\,\lambda_\ell^4\,\delta^{IJ}.
\label{eq:g4_hexagon_single_cut_tree}
\end{equation}
Here \(G_2,G_3,G_4\) are the uncut scalar propagators of the box family
\eqref{eq:box_sec3}.  Using~\eqref{eq:traceIJ}, the scalar-chain state trace is
\begin{equation}
\delta^{IJ}\,
A_6^{(0)}
\bigl((-\ell)^I,1^+,2^+,3^+,4^+,\ell^J\bigr)
=
-i\,\frac{[12][34]}{\langle12\rangle\langle34\rangle}\,
G_2G_3G_4\,(D_s-2)\lambda_\ell^4 .
\label{eq:g4_hexagon_single_cut_trace}
\end{equation}
Thus the one-pair hexagon does not vanish in the transverse sector.  In this
routing the positive-energy on-shell momentum in \(d\Phi(\ell)\) is
the momentum of slot \(D_1\).  Therefore
\(d\Phi(\ell)=d^D\ell\,\delta_1^+/(2\pi)^D\).  Substitution of
\eqref{eq:g4_hexagon_single_cut_trace} into
the stripped form of \eqref{eq:g4_k1_raw_state_sum} gives
\begin{equation}
{\cal C}_{6_1}
=
-i\,\frac{[12][34]}{\langle12\rangle\langle34\rangle}
\int \frac{d^D\ell}{(2\pi)^D}\,
(D_s-2)\lambda_\ell^4\,\delta_1^+G_2G_3G_4 .
\label{eq:g4_k1_single_cut_main}
\end{equation}
Equivalently, in the stripped slot-kernel convention this representative gives
the single-slot support
\begin{equation}
\delta_1^+\,G_2G_3G_4
\end{equation}
with numerator \((D_s-2)\lambda_\ell^4\), the common spinor factor
\([12][34]/(\langle12\rangle\langle34\rangle)\), and the overall tree factor
\(-i\).  Its cyclic images give the other single-slot supports,
\begin{equation}
G_1 \delta_2^+\,G_3G_4,\qquad
G_1G_2 \delta_3^+\,G_4,\qquad
G_1G_2G_3 \delta_4^+ .
\label{eq:g4_k1_single_cut_cyclic_words}
\end{equation}
The four assignments in the \(\{6\}\) sector are displayed in
Table~\ref{tab:g4_k1_sector_entries}.
\begin{table}[t]
\centering
\scriptsize
\setlength{\tabcolsep}{3pt}
\begin{tabular}{c
l@{}l@{}l@{}l@{}l@{}l
c c}
\hline
case & \multicolumn{6}{c}{\(A_6^{(0)}\) arguments} & slot support & non-degenerate \\
\hline
\(6_1\) &
\tblargfirst{-\ell} & \tblargnext{1^+} & \tblargnext{2^+} &
\tblargnext{3^+} & \tblargnext{4^+} & \tblarglast{\ell} &
\(\delta_1^+\) & \(\checkmark\) \\
\(6_2\) &
\tblargfirst{1^+} & \tblargnext{k_1-\ell} & \tblargnext{2^+} &
\tblargnext{3^+} & \tblargnext{4^+} & \tblarglast{\ell-k_1} &
\(\delta_2^+\) & \(\checkmark\) \\
\(6_3\) &
\tblargfirst{1^+} & \tblargnext{2^+} & \tblargnext{\ell-k_{12}} &
\tblargnext{3^+} & \tblargnext{4^+} & \tblarglast{k_{12}-\ell} &
\(\delta_3^+\) & \(\checkmark\) \\
\(6_4\) &
\tblargfirst{1^+} & \tblargnext{2^+} & \tblargnext{3^+} &
\tblargnext{\ell+k_4} & \tblargnext{4^+} & \tblarglast{-\ell-k_4} &
\(\delta_4^+\) & \(\checkmark\) \\
\hline
\end{tabular}
\caption{\(\{6\}\) bookkeeping assignments of momenta to the hexagon at order
\(g^4\).  The state labels \(h,\bar h\) are suppressed on the vacuum-pair
momenta.}
\label{tab:g4_k1_sector_entries}
\end{table}
In each case the displayed \(\delta_i^+\) comes from the on-shell
positive-energy momentum chosen for that representative.  The \(\{6\}\)
sector is combined with the \(\{4,4\}\) sector in the box comparison below and
gives the single-slot contributions that become single cuts.  The four rows are
distinct labelled assignments yielding the four single-cut slot supports.  They
are not collapsed before the box denominator family is collected, even when a
later momentum shift makes two kernel forms look identical.

\subsection{The \texorpdfstring{$\{4,4\}$}{\{4,4\}} sector}

We now turn to the sector \(\{4,4\}\), giving \(k=2\) and \(r=2\).  The corresponding
on-shell contribution contains two vacuum pairs, \((-\ell_1,\ell_1)\) and \((-\ell_2,\ell_2)\).

We first consider an assignment in which the external momenta of legs \(1\)
and \(2\) lie on the left polygon,
while the external momenta of legs \(3\) and \(4\) lie on the right polygon,
that is the split \((1,2)|(3,4)\):
\begin{multline}
{\cal C}_{44_1}
=
\sum_{h_1,h_2}
\int d\Phi(\ell_1)\,d\Phi(\ell_2)\,
(2\pi)^D\delta^{(D)}(k_{12}-\ell_1-\ell_2)
(2\pi)^D\delta^{(D)}(k_{34}+\ell_1+\ell_2)
\\
\times
A_4^{(0)}(1^+,2^+,-\ell_1,-\ell_2)\,
A_4^{(0)}(\ell_2,\ell_1,3^+,4^+),
\label{eq:k2_first_contribution_textstyle}
\end{multline}
The second contribution places legs \(2\) and \(3\) on the left polygon and
legs \(4\) and \(1\) on the right polygon, corresponding to the split
\((2,3)|(4,1)\):
\begin{multline}
{\cal C}_{44_2}
=
\sum_{h_1,h_2}
\int d\Phi(\ell_1)\,d\Phi(\ell_2)\,
(2\pi)^D\delta^{(D)}(k_{23}-\ell_1-\ell_2)
(2\pi)^D\delta^{(D)}(k_{41}+\ell_1+\ell_2)
\\
\times
A_4^{(0)}(2^+,3^+,-\ell_1,-\ell_2)\,
A_4^{(0)}(\ell_2,\ell_1,4^+,1^+),
\label{eq:k2_second_contribution_textstyle}
\end{multline}
The non-adjacent split \((1,3)|(2,4)\) is not a planar split of this fixed
color ordering and would belong to a different color-ordered amplitude.
Thus the vacuum-pair data are
\begin{equation}
{\cal C}_{44}
=
{\cal C}_{44_1}+{\cal C}_{44_2}.
\label{eq:k2_sum_textstyle}
\end{equation}
Equivalently, the two explicit assignments in the \(\{4,4\}\) sector are shown
in Table~\ref{tab:g4_k2_sector_entries}.
\begin{table}[t]
\centering
\scriptsize
\setlength{\tabcolsep}{3pt}
\begin{tabular}{c
l@{}l@{}l@{}l
l@{}l@{}l@{}l
c c}
\hline
case & \multicolumn{4}{c}{$A_4^{(0)}$ arguments} &
\multicolumn{4}{c}{$A_4^{(0)}$ arguments} & slot support & non-degenerate \\
\hline
\(44_1\) &
\tblargfirst{1^+} & \tblargnext{2^+} & \tblargnext{-\ell_1} & \tblarglast{-\ell_2} &
\tblargfirst{\ell_2} & \tblargnext{\ell_1} & \tblargnext{3^+} & \tblarglast{4^+} &
\(\delta_{13}^+\) & \(\checkmark\) \\
\(44_2\) &
\tblargfirst{2^+} & \tblargnext{3^+} & \tblargnext{-\ell_1} & \tblarglast{-\ell_2} &
\tblargfirst{\ell_2} & \tblargnext{\ell_1} & \tblargnext{4^+} & \tblarglast{1^+} &
\(\delta_{24}^+\) & \(\checkmark\) \\
\hline
\end{tabular}
\caption{\(\{4,4\}\) bookkeeping assignments at order \(g^4\).  They are the two
non-degenerate planar products of quadrilateral trees.}
\label{tab:g4_k2_sector_entries}
\end{table}
The signs with which these terms enter the signed vacuum-pair
sum are assigned by the prescription of
subsection~\ref{subsec:inclusion_exclusion}; in the later comparison they
match the terminal FTT coefficient.

The explicit delta functions restore the two quadrilateral momentum constraints.
Their products factorize as
\begin{equation}
\delta^{(D)}(k_{12}-\ell_1-\ell_2)\,
\delta^{(D)}(k_{34}+\ell_1+\ell_2)
=
\delta^{(D)}(k_{12}+k_{34})\,
\delta^{(D)}(k_{12}-\ell_1-\ell_2).
\label{eq:nlo_delta_factorization_s}
\end{equation}
The factorization for the \(k_{23}\) channel is analogous.
In each case the first factor is the overall four-point energy-momentum
conserving delta function and is stripped off.  The remaining local tree
factors are therefore the delta-stripped amplitudes \(A_m^{(0)}\) defined in
\eqref{eq:delta_containing_tree_definition}.

After integrating over the phase space of \(\ell_2\), the two non-vanishing vacuum-pair contributions become
\begin{multline}
{\cal C}_{44_1}
=
\sum_{h_1,h_2}
\int \frac{d^D\ell}{(2\pi)^D}\,
\delta^+(\ell)\,\delta^+(k_{12}-\ell)
\\
\times
A_4^{(0)}\bigl(1^+,2^+,-\ell,\ell-k_{12}\bigr)\,
A_4^{(0)}\bigl(k_{12}-\ell,\ell,3^+,4^+\bigr),
\label{eq:k2_first_loop_integral}
\end{multline}
and
\begin{multline}
{\cal C}_{44_2}
=
\sum_{h_1,h_2}
\int \frac{d^D\ell}{(2\pi)^D}\,
\delta^+(\ell)\,\delta^+(k_{23}-\ell)
\\
\times
A_4^{(0)}\bigl(2^+,3^+,-\ell,\ell-k_{23}\bigr)\,
A_4^{(0)}\bigl(k_{23}-\ell,\ell,4^+,1^+\bigr).
\label{eq:k2_second_loop_integral}
\end{multline}

We use the transverse momentum and state-sum conventions of
appendix~\ref{app:transverse_conventions}, as in the \(\{6\}\) sector above.

The part obtained by resolving the internal states into strictly
four-dimensional helicities vanishes: if both internal helicities make one
four-point tree non-zero, the
conjugate helicities on the other side give an all-plus tree, while the
remaining helicity assignments give a tree with only one negative helicity.
The non-zero rational numerator is therefore encoded by the transverse
component of the state sum.  We now insert the explicit four-point tree
amplitudes needed in \eqref{eq:k2_first_loop_integral}.  The required
open-index scalar-chain formula is derived in appendix~\ref{app:scalar_trees},
in particular in \eqref{eq:app_A4_open_PQ_formula}.
Applying \eqref{eq:app_A4_open_PQ_formula} to the first contribution,
with the momentum routing chosen in~\eqref{eq:k2_first_loop_integral}, yields the two open-index tree amplitudes
\begin{equation}
A_4^{(0)}\bigl((-\ell)^I,1^+,2^+,(\ell-k_{12})^J\bigr)
=
-i\,\frac{[12]}{\langle12\rangle}\,
\frac{\lambda_\ell^2\,\delta^{IJ}}
{(\ell-k_1)^2-\lambda_\ell^2},
\label{eq:first_tree_schannel}
\end{equation}
and
\begin{equation}
A_4^{(0)}\bigl((k_{12}-\ell)^J,3^+,4^+,\ell^I\bigr)
=
-i\,\frac{[34]}{\langle34\rangle}\,
\frac{\lambda_\ell^2\,\delta^{JI}}
{(\ell+k_4)^2-\lambda_\ell^2}.
\label{eq:second_tree_schannel}
\end{equation}
The projected state sum is now only the scalar-chain trace.  The factors
\(\lambda_\ell^2\) are already contained in the two flavor-diagonal
four-point trees; the endpoint trace gives the extra factor \(D_s-2\), as shown
in appendix~\ref{app:transverse_conventions}.  Therefore
\begin{multline}
\sum_{I,J}
A_4^{(0)}\bigl((-\ell)^I,1^+,2^+,(\ell-k_{12})^J\bigr)\,
A_4^{(0)}\bigl((k_{12}-\ell)^J,3^+,4^+,\ell^I\bigr)
= \\
-(D_s-2)\lambda_\ell^4\,
\frac{[12][34]}{\langle12\rangle\langle34\rangle}\,
\frac{1}{\bigl((\ell-k_1)^2-\lambda_\ell^2\bigr)\bigl((\ell+k_4)^2-\lambda_\ell^2\bigr)}.
\label{eq:tree_product_schannel}
\end{multline}
Substituting this coefficient into \eqref{eq:k2_first_loop_integral} gives
\begin{multline}
{\cal C}_{44_1}
=
-\frac{[12][34]}{\langle12\rangle\langle34\rangle}
\int \frac{d^D\ell}{(2\pi)^D}\,
\delta^+(\ell)\,\delta^+(k_{12}-\ell)
\;
\frac{(D_s-2)\lambda_\ell^4}{
\bigl((\ell-k_1)^2-\lambda_\ell^2\bigr)\,
\bigl((\ell+k_4)^2-\lambda_\ell^2\bigr)}.
\label{eq:A1_s_boxcut_main}
\end{multline}

The second contribution is treated in exactly the same way and gives
\begin{multline}
{\cal C}_{44_2}
=
-\frac{[23][41]}{\langle23\rangle\langle41\rangle}
\int \frac{d^D\ell}{(2\pi)^D}\,
\delta^+(\ell)\,\delta^+(k_{23}-\ell)
\;
\frac{(D_s-2)\lambda_\ell^4}{
\bigl((\ell-k_2)^2-\lambda_\ell^2\bigr)\,
\bigl((\ell+k_1)^2-\lambda_\ell^2\bigr)}.
\label{eq:A1_t_boxcut_main}
\end{multline}

With the denominator and oriented-cut conventions of
\eqref{eq:box_sec3} and \eqref{eq:g4_box_cut_definitions},
\eqref{eq:A1_s_boxcut_main} becomes
\begin{equation}
{\cal C}_{44_1}
=
-\frac{[12][34]}{\langle12\rangle\langle34\rangle}
\int \frac{d^D\ell}{(2\pi)^D}\,
(D_s-2)\lambda_\ell^4\,\delta_1^+ G_2\delta_3^+G_4 .
\label{eq:g4_k2_cut_13_compact}
\end{equation}
For \eqref{eq:A1_t_boxcut_main} we set \(\bar\ell=\ell+k_1\) and then
relabel \(\bar\ell\) back to \(\ell\).  Since \(\lambda_{k_i}=0\),
\(\lambda_{\ell+k_1}=\lambda_\ell\).  The two on-shell
delta functions become
\(\delta^+(\ell-k_1)\) and \(\delta^+(-\ell-k_4)\), namely
\(\delta_2^+\delta_4^+\).  Using momentum conservation together with the
four-point identity
\begin{equation}
\frac{[23][41]}{\langle23\rangle\langle41\rangle}
=
\frac{[12][34]}{\langle12\rangle\langle34\rangle},
\end{equation}
the second contribution becomes
\begin{equation}
{\cal C}_{44_2}
=
-\frac{[12][34]}{\langle12\rangle\langle34\rangle}
\int \frac{d^D\ell}{(2\pi)^D}\,
(D_s-2)\lambda_\ell^4\,G_1\delta_2^+G_3\delta_4^+ .
\label{eq:g4_k2_cut_24_compact}
\end{equation}
Thus the two phase-space integrals are the \(s\)- and \(t\)-channel
two-particle slot supports \(13\) and \(24\) of one and the same box chain.

\subsection{Comparison with the scalar box}
\label{sec:g4_inverse_ftt}

We now group the non-vanishing fixed-order cut terms into the box family.  Its
denominators are \(D_1,\ldots,D_4\).  The \(\{6\}\) assignments in
Table~\ref{tab:g4_k1_sector_entries} supply the single-cut configurations, and the
\(\{4,4\}\) assignments in Table~\ref{tab:g4_k2_sector_entries} supply the two
opposite two-particle cut configurations.  In the present family, the non-zero support
sets are
\begin{equation}
{\cal S}_{1}^{\rm box}=\{1,2,3,4\},
\qquad
{\cal S}_{2}^{\rm box}=\{13,24\}.
\label{eq:g4_box_nonzero_support_sets}
\end{equation}
The support-label convention is the one fixed in
\eqref{eq:g4_support_label_conventions}: an element \(S\) labels the opened
slots and is read as the corresponding set when it appears in a complementary
propagator product.
The single-cut set \({\cal S}_{1}^{\rm box}\) is supplied by the \(\{6\}\)
hexagon sector, while \({\cal S}_{2}^{\rm box}\) is supplied by the \(\{4,4\}\)
quadrilateral product.  Both sectors give the same local numerator
\((D_s-2)\lambda_\ell^4\), with the common all-plus spinor factor displayed
above.

For this one-loop family, the number of opened slots equals the number \(k\) of
tree factors.  The two support sets therefore correspond to \(k=1\) and
\(k=2\).  Including the sign \((-1)^{k-1}\) from
\eqref{eq:tree_factor_sign} and the forest phase \((-i)^k\), the collected
vacuum-pair source is
\begin{equation}
\begin{aligned}
{\cal A}_{\rm box}^{\rm vp}
&=
\frac{[12][34]}{\langle12\rangle\langle34\rangle}
\int\frac{d^D\ell}{(2\pi)^D}\,
(D_s-2)\lambda_\ell^4
\sum_{k=1}^{2}(-1)^{k-1}(-i)^k
\sum_{S\in{\cal S}_{k}^{\rm box}}
\delta_S^+
\prod_{j\in\{1,2,3,4\}\setminus S}G_j .
\end{aligned}
\label{eq:g4_vp_box_cut_collection}
\end{equation}
This is the non-zero cut collection produced by the vacuum-pair construction,
including its conventional tree phases.  Its single-cut coefficient is
\(-i\), while its double-cut coefficient is \(+1\), in agreement with the
FTT identity \eqref{eq:FTT} after translating to the stripped denominators
\(G_j\).  After these conventional phases are separated, the underlying
inclusion--exclusion signs are \(+1\) and \(-1\), respectively.

The other double-cut supports are the adjacent pairs
\((D_1,D_2)\), \((D_2,D_3)\), \((D_3,D_4)\), and \((D_4,D_1)\).  Each adjacent
pair isolates a real three-point on-shell corner: one observed massless gluon
and two cut lines of equal transverse mass.  Such a real \(A_3\) has no
non-degenerate positive-energy support.  For example, the cut of \(D_1\) and
\(D_2\) imposes \(\ell^2=(\ell-k_1)^2=\lambda_\ell^2\), hence
\(2\ell\!\cdot k_1=0\).  With \(\ell\) on the positive-energy mass shell this
has only degenerate soft/collinear boundary support.

The same argument eliminates all higher cuts in this box family.  Any triple cut of
the four cyclic denominators necessarily contains at least one adjacent pair,
and the quadruple cut contains all adjacent pairs.  Thus every triple cut and
the quadruple cut unavoidably isolate a real on-shell \(A_3\) amplitude at one
corner, and hence vanish for the same support reason.  This argument
holds for real momenta of the \(A_3\) amplitudes; it does
not constrain the complex on-shell three-point amplitudes that enter BCFW
recursion.  The point is only that the vacuum-pair phase-space integrations are
over real on-shell momenta, whereas BCFW constructs the tree factors through
complexified factorization kinematics before the real forward-limit support is
imposed.

Equation \eqref{eq:g4_vp_box_cut_collection} has the terminal FTT pattern of
one scalar box, as recalled in appendix~\ref{app:ftt}: the real
inclusion--exclusion signs are \(+1\) for the single cuts and \(-1\) for the
two opposite double cuts, while the complete stripped-kernel coefficients are
\(-i\) and \(+1\).  Every adjacent or higher cut has degenerate support.
Since the common phase factor in subsection~\ref{subsec:equivalence} is unity
at \(L=1\), regrouping this collection gives the ordinary Feynman denominator
family directly.  With the common numerator fixed by
\eqref{eq:g4_hexagon_single_cut_trace}, \eqref{eq:g4_k2_cut_13_compact}, and
\eqref{eq:g4_k2_cut_24_compact}, the result is
\begin{multline}
\mathcal{A}^{(1)}(1^+,2^+,3^+,4^+)
=
\frac{[12][34]}{\langle12\rangle\langle34\rangle}
\int\frac{d^D\ell}{(2\pi)^D}\,
\\
\times
\frac{(D_s-2)\lambda_\ell^4}{
\bigl(\ell^2-\lambda_\ell^2\bigr)\,
\bigl((\ell-k_1)^2-\lambda_\ell^2\bigr)\,
\bigl((\ell-k_{12})^2-\lambda_\ell^2\bigr)\,
\bigl((\ell+k_4)^2-\lambda_\ell^2\bigr)}.
\label{eq:allplus_box_integral_full_main}
\end{multline}
This is the standard \(\lambda_\ell^4\)-weighted box representation of the
one-loop all-plus amplitude, with the physical-state trace displayed
explicitly~\cite{Mahlon:1993si,Bern:1994zx,Bern:1994cg}.

Evaluating this integral yields the known result~\cite{Mahlon:1993si,Bern:1994zx,Bern:1994cg}
\begin{equation}
\mathcal{A}^{(1)}(1^+,2^+,3^+,4^+)
=
(D_s-2)\,\frac{i}{(4\pi)^{2-\epsilon}}\,
\frac{[12][34]}{\langle12\rangle\langle34\rangle}\,
K_4,
\label{eq:allplus_sum_result_textstyle}
\end{equation}
where $K_4=I_4[\lambda_\ell^4]=-\epsilon(1-\epsilon)\,I_4^{D=8-2\epsilon}
=-\frac{1}{6}+{\cal O}(\epsilon)$.
Therefore, in the four-dimensional spin-state limit \(D_s\to4\) and
\(\epsilon\to0\), this becomes
\begin{equation}
\mathcal{A}^{(1)}(1^+,2^+,3^+,4^+)
=
-\frac{i}{3(4\pi)^2}\,
\frac{[12][34]}{\langle12\rangle\langle34\rangle}
+{\cal O}(\epsilon).
\label{eq:allplus_final_result_textstyle}
\end{equation}
Thus the vacuum-pair construction reproduces the known finite rational
one-loop all-plus amplitude~\cite{Mahlon:1993si,Bern:1994zx,Bern:1994cg,Maniatis:2019pig}.


\section{Order \texorpdfstring{$g^6$}{g6}: two-loop all-plus amplitude from the vacuum-pair construction}
\label{sec:g6_allplus_revised}

We now turn to the order-$g^6$ contribution to the four-gluon all-plus
amplitude.  The tree factors used in the construction are color ordered, but
the standard two-loop result is organized into distinct planar and non-planar
primitive amplitudes.  We reconstruct representative primitive integrands
with external labels \((1,2,3,4)\); their color factors and relabelled copies
are restored when the full amplitude is assembled below.  The subscript of
\(A^{\rm P}_{1234}\) records the cyclic ordering of the planar primitive,
whereas the semicolon in \(A^{\rm NP}_{12;34}\) records the two external pairs
of the crossed color factor and does not denote the cyclic word
\((1,2,3,4)\).  The non-planar primitive consequently requires additional
cyclic orderings of the intermediate tree factors.
Since the all-plus tree amplitude vanishes, we label this contribution by its
coupling order rather than by the usual relative-order terminology.  In the
present construction the order is fixed by the number of cubic three-point
amplitudes \(A_3\) from which the relevant tree products are built.  The
calculation therefore starts from on-shell \(A_3\) amplitudes and glues them
into higher-point tree amplitudes, while keeping track of the unobservable
vacuum pairs introduced in subsection~\ref{subsec:insertions}.
We show that this construction reproduces the known bare two-loop primitive
representation of~\cite{Bern:2000dn,Bern:2002tk}.  The coupling factor
\(g^6\) and the parent-graph color factors are suppressed while the integral
functions are constructed and are restored at the end of the section.  In an
intermediate expression described as a stripped slot kernel, the conventional
factor \(-i\) of each connected tree amplitude is also suppressed.  The
collected vacuum-pair sources below restore the resulting factor \((-i)^k\)
explicitly.  The common two-loop phase relation derived in
subsection~\ref{subsec:equivalence} then supplies the overall factor \(i\) in
the final primitive amplitudes.  No ultraviolet counterterm is included in the
calculation below.

At fixed perturbative order, the
vacuum-pair tree products are organized by the polygon bookkeeping of
subsection~\ref{subsec:polygon}.  The helicity constraints and the
state sums then determine which products survive.

The fixed-order \(n=4\), \(N_3=6\) sector list was given explicitly in
\eqref{eq:g6_polygon_sectors_bookkeeping}.  Before helicity selection the
non-triangular sectors are
\begin{equation}
\{6,4\},\qquad \{5,5\},\qquad \{4,4,4\},\qquad \{8\}.
\end{equation}
With the terminology introduced in subsection~\ref{subsec:insertions}, \(r\)
is the number of vacuum pairs and \(k\) the number of tree factors; for the
four sectors, respectively, \(r=3,3,4,2\) and \(k=2,2,3,1\).  According to
\eqref{eq:tree_factor_sign}, the inclusion--exclusion sign depends only on
\(k\), giving \(-1\), \(-1\), \(+1\), and \(+1\), respectively.
The two-factor products \(A_6\,A_4\) and \(A_5\,A_5\), corresponding to the
\(\{6,4\}\) and \(\{5,5\}\) sectors, are treated first.  They are followed by
the maximally factorized sector \(\{4,4,4\}\), a product of three four-point
trees, \(A_4\,A_4\,A_4\).  The one-factor octagon sector \(\{8\}\), represented
by \(A_8\), is presented last.

Before the sector-by-sector evaluation we fix the slot notation for the
denominator families used to compare the resulting support terms.  Throughout
the following subsections, tree amplitudes \(A_m^{(0)}\) are delta-stripped as
in \eqref{eq:delta_containing_tree_definition}.  The momentum-conservation
delta functions of the corresponding polygons are displayed explicitly in the
phase-space integrals, using the \(d\Phi\) and \(\delta^+\) conventions
introduced in subsection~\ref{subsec:insertions}.
State labels follow the same convention as in section~\ref{sec:g4_allplus}: for
a contribution with \(r\) vacuum pairs we sum over labels
\(h_a\), \(a=1,\ldots,r\), and \(\bar h_a\) denotes the opposite-momentum partner of
the \(a\)th pair.  To keep the long tree arguments readable, these labels are
usually suppressed on the amplitudes until the state sums are evaluated.

The assignment tables below display selected representatives from the different sectors.
They are not exhaustive tables of denominator
supports.  The polygon, momentum-routing, and appendix-reduction bookkeeping is
performed algorithmically, while the complete physical-state contractions are
recorded in the appendices.  The exhaustive support sets are stated at the end
of each sector subsection.  The construction itself uses only the vacuum-pair tree
products and does not invoke a parent loop graph or the FTT.  Only after the
vacuum-pair calculation has been completed is its result compared, using the
FTT, with the known two-loop four-gluon calculation.

For the planar double-box denominator family, denoted \({\rm P/DB}\), the
seven slots are indexed by \(\{1,2,3,4,5,6,7\}\).  Its denominators are
\begin{equation}
\begin{aligned}
D_1&=p^2-\lambda_p^2,
&
D_2&=(p-k_1)^2-\lambda_p^2,
&
D_3&=(p-k_{12})^2-\lambda_p^2,
\\
D_4&=q^2-\lambda_q^2,
&
D_5&=(q-k_4)^2-\lambda_q^2,
&
D_6&=(q-k_{34})^2-\lambda_q^2,
\\
D_7&=(p+q)^2-\lambda_{p+q}^2 .
\end{aligned}
\label{eq:g6_planar_denominators}
\end{equation}
Here and below,
\(k_{i_1\cdots i_m}:=k_{i_1}+\cdots+k_{i_m}\) for arbitrary external-leg
indices; in particular, \(k_{12}+k_{34}=k_{1234}=0\).  The transverse
conventions are those of appendix~\ref{app:transverse_conventions}.
In the ordinary two-loop double-box expression, \(D_7\) is the bridge denominator.
Following the shorthand introduced
in \eqref{eq:Gdconv}, \(G_i\) denotes \(G_F(D_i)\).  The oriented cut factors
used in this family are
\begin{equation}
\begin{aligned}
\delta_1^+&:=\delta^+(p)
=2\pi\,\theta(p^0)\delta(D_1),
&
\delta_2^+&:=\delta^+(p-k_1)
=2\pi\,\theta\bigl((p-k_1)^0\bigr)\delta(D_2),
\\
\delta_3^+&:=\delta^+(k_{12}-p)
=2\pi\,\theta\bigl((k_{12}-p)^0\bigr)\delta(D_3),
&
\delta_4^+&:=\delta^+(-q)
=2\pi\,\theta\bigl((-q)^0\bigr)\delta(D_4),
\\
\delta_5^+&:=\delta^+(k_4-q)
=2\pi\,\theta\bigl((k_4-q)^0\bigr)\delta(D_5),
&
\delta_6^+&:=\delta^+(q-k_{34})
=2\pi\,\theta\bigl((q-k_{34})^0\bigr)\delta(D_6),
\\
\delta_7^+&:=\delta^+(p+q)
=2\pi\,\theta\bigl((p+q)^0\bigr)\delta(D_7).
\end{aligned}
\label{eq:g6_planar_cut_definitions}
\end{equation}
The opposite oriented branch of the bridge slot is denoted separately by
\begin{equation}
\delta_{\bar 7}^+
:=\delta^+(-p-q)
=2\pi\,\theta\bigl((-p-q)^0\bigr)\delta(D_7)
=\delta_7^- .
\label{eq:g6_planar_opposite_bridge_cut}
\end{equation}
For the endpoint-reversed octagon representative we similarly use
\begin{equation}
\delta_{\bar1}^+:=\delta^+(-p)=\delta_1^- ,
\qquad
\delta_{\bar6}^+:=\delta^+(k_{34}-q)=\delta_6^- .
\label{eq:g6_planar_opposite_endpoint_cuts}
\end{equation}
Composite support labels are abbreviated as
\begin{equation}
\begin{aligned}
\delta_{146}^+&=\delta_1^+\delta_4^+\delta_6^+,
&
\delta_{167}^+&=\delta_1^+\delta_6^+\delta_7^+,
\\
\delta_{16\bar 7}^+&=\delta_1^+\delta_6^+\delta_{\bar 7}^+,
&
\delta_{1346}^+&=\delta_1^+\delta_3^+\delta_4^+\delta_6^+ .
\end{aligned}
\label{eq:g6_support_label_conventions}
\end{equation}
An unbarred slot uses the momentum orientation in
\eqref{eq:g6_planar_cut_definitions}, whereas a barred slot uses the opposite
orientation.  The bar cannot be removed by using \(D_i(q_i)=D_i(-q_i)\), since
\(\delta^+(q_i)\neq\delta^+(-q_i)\).  It may be removed only by a change of
integration variables applied to the complete term.  For a signed slot word
\(S\), \(\delta_S^+\) denotes the product of the oriented cut factors specified
by its entries.  Uncut products are taken over the complement of the underlying
set of denominator numbers obtained by dropping the bars from \(S\).  These are
the phase-space orientations of the
vacuum-pair terms; comparison with an FTT term is made only after the latter has
been given a coherent loop orientation as in
subsection~\ref{subsec:equivalence}.
For example, a displayed kernel such as
\(\delta_1^+G_2\delta_3^+G_4G_5G_6\delta_7^+\) should be read as the seven-slot
configuration in which slots \(1,3,7\) carry on-shell delta factors and the
remaining slots are ordinary Feynman propagators.
The local bow-tie family, denoted \({\rm BT}\), contains the six propagators
\(D_1,\ldots,D_6\), indexed by \(\{1,2,3,4,5,6\}\).  The external-channel
factor \(1/s_{12}\) depends solely on the external momenta and is not a slot in
this family.
For the crossed non-planar double-box family, denoted \({\rm NP/DB}\), the
seven crossed slots are labelled
\(\{\widetilde1,\widetilde2,\widetilde3,
\widetilde4,\widetilde5,\widetilde6,\widetilde7\}\).  Its denominators are
\begin{equation}
\begin{aligned}
\widetilde D_1&=p^2-\lambda_p^2,
\qquad
\widetilde D_2=(p-k_1)^2-\lambda_p^2,
\qquad
\widetilde D_3=q^2-\lambda_q^2,
\\
\widetilde D_4&=(q-k_2)^2-\lambda_q^2,
\qquad
\widetilde D_5=(p+q)^2-\lambda_{p+q}^2,
\qquad
\widetilde D_6=(p+q+k_3)^2-\lambda_{p+q}^2,
\\
\widetilde D_7&=(p+q+k_{34})^2-\lambda_{p+q}^2 .
\end{aligned}
\label{eq:g6_nonplanar_denominators}
\end{equation}
Here \(\widetilde G_i=G_F(\widetilde D_i)\).
The \({\rm NP/DB}\) cut factors are
\begin{equation}
\begin{aligned}
\widetilde\delta_1^+
&:=\delta^+(p),
&
\widetilde\delta_2^+
&:=\delta^+(p-k_1),
\\
\widetilde\delta_3^+
&:=\delta^+(q),
&
\widetilde\delta_4^+
&:=\delta^+(q-k_2),
\\
\widetilde\delta_5^+
&:=\delta^+(p+q),
&
\widetilde\delta_6^+
&:=\delta^+(p+q+k_3),
\\
\widetilde\delta_7^+
&:=\delta^+(p+q+k_{34}).
\end{aligned}
\label{eq:g6_crossed_cut_definitions}
\end{equation}
For the opposite crossed bridge orientation we analogously write
\begin{equation}
\widetilde\delta_{\bar 7}^+
:=\delta^+(-p-q-k_{34})
=2\pi\,\theta\bigl((-p-q-k_{34})^0\bigr)\delta(\widetilde D_7).
\label{eq:g6_crossed_opposite_bridge_cut}
\end{equation}
The crossed slots are abbreviated analogously, e.g.
\begin{equation}
\widetilde\delta_{257}^+
=\widetilde\delta_2^+\widetilde\delta_5^+\widetilde\delta_7^+,
\qquad
\widetilde\delta_{167}^+
=\widetilde\delta_1^+\widetilde\delta_6^+\widetilde\delta_7^+ .
\label{eq:g6_crossed_support_label_conventions}
\end{equation}
The same signed-slot convention applies in the \({\rm NP/DB}\) family; in particular,
\(\widetilde 7\) and \(\widetilde{\bar 7}\) denote the two distinct oriented
branches of the same quadratic denominator \(\widetilde D_7\).

\subsection{The \texorpdfstring{$\{6,4\}$}{\{6,4\}} sector}
\label{sec:64}

The bookkeeping relation in subsection~\ref{subsec:polygon} gives \(k=2\) and
\(r=3\) for the \(\{6,4\}\) sector.  We therefore introduce three vacuum pairs,
which we denote by
\begin{equation}
(-\ell_1,\ell_1),\qquad
(-\ell_2,\ell_2),\qquad
(-\ell_3,\ell_3).
\end{equation}
Table~\ref{tab:g6_64_arguments} displays selected non-degenerate
representatives together with examples of the distinct vanishing mechanisms.
The first two rows have on-shell support containing the bridge slot \(D_7\);
the remaining rows are cyclic representatives.  The rows marked by \(\times\)
do not give non-degenerate contributions.
The complete terminal support sets are collected at the end of this subsection.
\begin{table}[t]
\centering
\scriptsize
\setlength{\tabcolsep}{3pt}
\begin{tabular}{l
l@{}l@{}l@{}l@{}l@{}l
l@{}l@{}l@{}l
c c}
\hline
case & \multicolumn{6}{c}{\(A_6^{(0)}\) arguments} &
\multicolumn{4}{c}{\(A_4^{(0)}\) arguments} & slot support & non-degenerate \\
\hline
\(64_1\) &
\tblargfirst{\ell_1} & \tblargnext{-\ell_3} & \tblargnext{3^+} &
\tblargnext{4^+} & \tblargnext{\ell_3} & \tblarglast{\ell_2} &
\tblargfirst{1^+} & \tblargnext{2^+} & \tblargnext{-\ell_1} &
\tblarglast{-\ell_2} & \(\delta_{137}^+\) & \(\checkmark\) \\
\(64_2\) &
\tblargfirst{\ell_1} & \tblargnext{-\ell_3} & \tblargnext{1^+} &
\tblargnext{2^+} & \tblargnext{\ell_3} & \tblarglast{\ell_2} &
\tblargfirst{3^+} & \tblargnext{4^+} & \tblargnext{-\ell_1} &
\tblarglast{-\ell_2} & \(\delta_{467}^+\) & \(\checkmark\) \\
\hline
\(64_3\) &
\tblargfirst{-\ell_1} & \tblargnext{\ell_1} & \tblargnext{-\ell_2} &
\tblargnext{\ell_2} & \tblargnext{-\ell_3} & \tblarglast{\ell_3} &
\tblargfirst{1^+} & \tblargnext{2^+} & \tblargnext{3^+} &
\tblarglast{4^+} & -- & \(\times\) \\
\(64_4\) &
\tblargfirst{1^+} & \tblargnext{-\ell_1} & \tblargnext{\ell_1} &
\tblargnext{-\ell_2} & \tblargnext{\ell_2} & \tblarglast{-\ell_3} &
\tblargfirst{2^+} & \tblargnext{3^+} & \tblargnext{4^+} &
\tblarglast{\ell_3} & -- & \(\times\) \\
\(64_5\) &
\tblargfirst{1^+} & \tblargnext{2^+} & \tblargnext{-\ell_1} &
\tblargnext{\ell_1} & \tblargnext{-\ell_2} & \tblarglast{\ell_3} &
\tblargfirst{3^+} & \tblargnext{4^+} & \tblargnext{-\ell_3} &
\tblarglast{\ell_2} & \(\delta_{146}^+\) & \(\checkmark\) \\
\(64_6\) &
\tblargfirst{1^+} & \tblargnext{2^+} & \tblargnext{3^+} &
\tblargnext{\ell_1} & \tblargnext{\ell_2} & \tblarglast{\ell_3} &
\tblargfirst{4^+} & \tblargnext{-\ell_3} & \tblargnext{-\ell_2} &
\tblarglast{-\ell_1} & -- & \(\times\) \\
\(64_7\) &
\tblargfirst{1^+} & \tblargnext{2^+} & \tblargnext{3^+} &
\tblargnext{4^+} & \tblargnext{-\ell_1} & \tblarglast{\ell_1} &
\tblargfirst{-\ell_2} & \tblargnext{\ell_2} & \tblargnext{-\ell_3} &
\tblarglast{\ell_3} & -- & \(\times\) \\
\hline
\(64_8\) &
\tblargfirst{2^+} & \tblargnext{3^+} & \tblargnext{-\ell_1} &
\tblargnext{\ell_1} & \tblargnext{-\ell_2} & \tblarglast{\ell_3} &
\tblargfirst{4^+} & \tblargnext{1^+} & \tblargnext{-\ell_3} &
\tblarglast{\ell_2} & \(\delta_{346}^+\) & \(\checkmark\) \\
\(64_9\) &
\tblargfirst{3^+} & \tblargnext{4^+} & \tblargnext{-\ell_1} &
\tblargnext{\ell_1} & \tblargnext{-\ell_2} & \tblarglast{\ell_3} &
\tblargfirst{1^+} & \tblargnext{2^+} & \tblargnext{-\ell_3} &
\tblarglast{\ell_2} & \(\delta_{134}^+\) & \(\checkmark\) \\
\(64_{10}\) &
\tblargfirst{4^+} & \tblargnext{1^+} & \tblargnext{-\ell_1} &
\tblargnext{\ell_1} & \tblargnext{-\ell_2} & \tblarglast{\ell_3} &
\tblargfirst{2^+} & \tblargnext{3^+} & \tblargnext{-\ell_3} &
\tblarglast{\ell_2} & \(\delta_{136}^+\) & \(\checkmark\) \\
\hline
\end{tabular}
\caption{Canonical representatives of the \(\{6,4\}\) assignment classes
used in the support analysis.}
\label{tab:g6_64_arguments}
\end{table}

We first analyze the representative \(64_1\), in which the quadrilateral carries the
observed pair \((1,2)\), while the
hexagon carries the observed pair \((3,4)\):
\begin{multline}
{\cal C}_{64_1}
=
\sum_{h_1,h_2,h_3}
\int d\Phi(\ell_1)\,d\Phi(\ell_2)\,d\Phi(\ell_3)
\\
\times
(2\pi)^D\delta^{(D)}(k_{12}-\ell_1-\ell_2)\,
(2\pi)^D\delta^{(D)}(k_{34}+\ell_1+\ell_2)
\\
\times
A_4^{(0)}(1^+,2^+,-\ell_1,-\ell_2)\,
A_6^{(0)}(\ell_1,-\ell_3,3^+,4^+,\ell_3,\ell_2).
\label{eq:g6_64_start}
\end{multline}
The two displayed delta functions factorize as
\begin{equation}
\delta^{(D)}(k_{12}-\ell_1-\ell_2)\,
\delta^{(D)}(k_{34}+\ell_1+\ell_2)
=
\delta^{(D)}(k_{12}+k_{34})\,
\delta^{(D)}(k_{12}-\ell_1-\ell_2).
\label{eq:g6_64_delta_factorization}
\end{equation}
We sum over the states carried by the momenta \(\ell_a\).
After stripping the overall four-point delta function, the remaining
momentum-conservation delta function performs the \(d^D\ell_2\) part of the
\(d\Phi(\ell_2)\) integration, while the on-shell delta contained in
\(d\Phi(\ell_2)\) remains as the second on-shell constraint.  We choose
\begin{equation}
\ell_1=p,\qquad \ell_2=k_{12}-p,\qquad \ell_3=p+q .
\end{equation}
Thus
\begin{multline}
{\cal C}_{64_1}
=
\sum_{h_1,h_2,h_3}
\int \frac{d^Dp}{(2\pi)^D}\frac{d^Dq}{(2\pi)^D}\,
\delta^+(p)\,\delta^+(k_{12}-p)\,\delta^+(p+q)
\\
\times
A_4^{(0)}(1^+,2^+,-p,p-k_{12})\,
A_6^{(0)}(p,-p-q,3^+,4^+,p+q,k_{12}-p).
\label{eq:g6_64_reduced_phase_space}
\end{multline}

The complete physical-state contraction of the two tree factors in
\eqref{eq:g6_64_reduced_phase_space} is evaluated in
appendix~\ref{app:adjacent_pair_reduction}.  The calculation first isolates the
scalar-chain projection, thereby fixing the propagator slots and the support
\(\delta_1^+\delta_3^+\delta_7^+\), and then includes the four-dimensional and
remaining transverse components of the state sum.  The completed numerator for
this representative is
\begin{equation}
{\cal N}_{64,7}^{(137)}(p,q)={\cal N}_{\rm DB}(p,q),
\label{eq:g6_64_D7_numerator}
\end{equation}
where \({\cal N}_{\rm DB}\) is displayed below in
\eqref{eq:g6_octagon_explicit_DB_factors}.  The resulting stripped cut kernel is therefore
\begin{equation}
{\cal C}_{64_1,{\rm DB}}
=
\frac{[12][34]}{\langle12\rangle\langle34\rangle}
\int \frac{d^Dp}{(2\pi)^D}\frac{d^Dq}{(2\pi)^D}\,
{\cal N}_{\rm DB}(p,q)\,
\delta_1^+G_2\delta_3^+G_4G_5G_6\delta_7^+ .
\label{eq:g6_64_cut_kernel}
\end{equation}

We next analyze the assignments in Table~\ref{tab:g6_64_arguments} that do
not give connected \(\{6,4\}\) four-point contributions.  Rows \(64_3\) and
\(64_7\) assign no observed external momenta to one polygon.  Such a factor is
a vacuum or lower fixed-order factor, while the
remaining observed factor is either the all-plus tree or the one-pair hexagon
already excluded as an independent connected source.  Row \(64_4\) leaves a
single vacuum-pair state on the quadrilateral.  The four-dimensional helicity
projection is again an all-plus or one-minus tree and vanishes, while the
transverse projection carries one unpaired transverse index and hence vanishes
by transverse rotational invariance.

Row \(64_6\) is the remaining delicate case.  Here the three
phase-space momenta are all assigned to the hexagon, for example
\begin{multline}
{\cal C}_{64_6}
=
\sum_{h_1,h_2,h_3}
\int \prod_{a=1}^{3}d\Phi(\ell_a)\,
(2\pi)^D\delta^{(D)}(k_{123}+\ell_1+\ell_2+\ell_3)
(2\pi)^D\delta^{(D)}(k_4-\ell_1-\ell_2-\ell_3)
\\ \times
A_6^{(0)}(1^+,2^+,3^+,\ell_1,\ell_2,\ell_3)\,
A_4^{(0)}(4^+,-\ell_3,-\ell_2,-\ell_1).
\label{eq:g6_64_three_cut_on_hexagon}
\end{multline}
The two delta functions give
\begin{equation}
\delta^{(D)}(k_{123}+\ell_1+\ell_2+\ell_3)\,
\delta^{(D)}(k_4-\ell_1-\ell_2-\ell_3)
=
\delta^{(D)}(k_{123}+k_4)\,
\delta^{(D)}(k_4-\ell_1-\ell_2-\ell_3).
\label{eq:g6_64_three_cut_delta}
\end{equation}
After the overall four-point delta function is stripped, the remaining
constraint forces the three full \(D\)-dimensional on-shell momenta to add to
the massless external momentum.  In this collinearity argument only, squares
and dot products are full \(D\)-dimensional invariants:
\begin{equation}
0=k_{4,D}^2
=
(\ell_{1,D}+\ell_{2,D}+\ell_{3,D})^2
=2(\ell_{1,D}\!\cdot\!\ell_{2,D}
+\ell_{1,D}\!\cdot\!\ell_{3,D}
+\ell_{2,D}\!\cdot\!\ell_{3,D}).
\label{eq:g6_64_three_cut_collinear}
\end{equation}
For real positive-energy massless momenta all scalar products are
non-negative.  Hence all three terms in
\eqref{eq:g6_64_three_cut_collinear} must vanish separately, and the
phase-space momenta are forced to be collinear with the massless external momentum \(k_4\).
Thus the three on-shell phase-space integrations, together with the remaining
delta function in \eqref{eq:g6_64_three_cut_delta}, have support only at
\begin{equation}
\ell_a=x_a k_4,\qquad x_a\ge0,\qquad \sum_{a=1}^{3}x_a=1 .
\end{equation}
After the \(\ell_a\) integrations are reduced to this support, the only
remaining variables are the dimensionless momentum fractions \(x_a\).  There is
no dimensionful invariant left in this subintegral: the only external momentum
is \(k_4\), and \(k_4^2=0\).  The corresponding collinear-boundary phase-space
integral is therefore scaleless and is set to zero in dimensional
regularization.  Thus the configuration in \eqref{eq:g6_64_three_cut_on_hexagon} does
not generate a connected four-point contribution or a new double-box
denominator family.

If, instead, all three complete vacuum pairs
\((-\ell_a,\ell_a)\) are taken to be local to the hexagon, then the quadrilateral factor is
the all-plus tree \(A_4^{(0)}(1^+,2^+,3^+,4^+)\), which vanishes.  Thus neither
interpretation produces an additional \(\{6,4\}\) contribution.

The non-vanishing \(\{6,4\}\) configurations are the chain assignments in
which one complete vacuum pair is kept inside the hexagon, while the other two
vacuum pairs are split between the two polygons.  In case \(64_1\) the complete
pair is the bridge pair; it has support \(137\).  The left--right image
\(64_2\) has support \(467\).  If the complete pair belongs to the
\(p\)-side chain, cases \(64_5\) and \(64_8\) have supports \(146\) and \(346\).
If the complete pair belongs to the \(q\)-side chain, cases \(64_9\) and
\(64_{10}\) have supports \(134\) and \(136\).  In all cases
appendix~\ref{app:adjacent_pair_reduction} supplies the scalar propagators of the chain
adjacent to the complete pair.

The displayed non-zero \(\{6,4\}\) representatives are local checks of the
exhaustive planar cut-slot set
\begin{equation}
{\cal U}_{64}^{\rm P}
=\{134,135,136,137,146,157,
246,247,267,346,357,467\}.
\label{eq:g6_planar_64_underlying_supports}
\end{equation}
Each support contributes to a seven-slot double-box representation, in which
every slot appears either as a cut factor or as an uncut propagator.  After the
complete physical state sum is taken, as shown explicitly in
\eqref{eq:app_64_D7_complete}, the corresponding contribution is
\begin{equation}
\begin{aligned}
{\cal C}_{64_S,{\rm DB}}
&=
\frac{[12][34]}{\langle12\rangle\langle34\rangle}
\int \frac{d^Dp}{(2\pi)^D}\frac{d^Dq}{(2\pi)^D}\,
{\cal N}_{\rm DB}(p,q)
\\
&\quad\times
\prod_{i\in S}\delta_i^+
\prod_{j\in\{1,\ldots,7\}\setminus S}G_j,
\qquad
S\in{\cal U}_{64}^{\rm P}.
\end{aligned}
\label{eq:g6_64_DB_projection}
\end{equation}
The supports not containing slot \(7\) also have a local projection with no
\(D_7\) pole.  The complete local state sum is given in
\eqref{eq:app_64_BT_components}--\eqref{eq:app_64_BT_complete}.  These are the
\(\{6,4\}\) contributions to the six-propagator bow-tie family, with the local
numerator
\begin{equation}
{\cal N}_{\rm BT}(p,q)
=
4(D_s-2)(\lambda_p^2+\lambda_q^2)
(\lambda_p\!\cdot\!\lambda_q)
+\frac{(D_s-2)^2}{s_{12}}\,
\lambda_p^2\lambda_q^2\bigl(D_7+s_{12}\bigr).
\label{eq:g6_MBT2}
\end{equation}
Here \(D_7=P_D^2=(p+q)^2-\lambda_{p+q}^2\) is used only as shorthand for
the full \(D\)-dimensional numerator invariant.  The local bow-tie family has
no \(D_7\) propagator.
The local \(\{6,4\}\) projections are therefore
labelled by
\begin{equation}
{\cal U}_{64}^{\rm BT}
=\{134,135,136,146,246,346\}.
\label{eq:g6_64_bowtie_underlying_supports}
\end{equation}
For each \(S\) in this set,
\begin{equation}
\begin{aligned}
{\cal C}_{64_S,{\rm BT}}
&=
\frac{[12][34]}{\langle12\rangle\langle34\rangle}
\int \frac{d^Dp}{(2\pi)^D}\frac{d^Dq}{(2\pi)^D}\,
{\cal N}_{\rm BT}(p,q)
\\
&\quad\times
\prod_{i\in S}\delta_i^+
\prod_{j\in\{1,\ldots,6\}\setminus S}G_j,
\qquad
S\in{\cal U}_{64}^{\rm BT}.
\end{aligned}
\label{eq:g6_64_BT_projection}
\end{equation}
The scalar-chain, four-dimensional, and remaining transverse components are
combined in appendix~\ref{app:adjacent_pair_reduction}; the projections above
contain the complete physical state sum.
This exhausts the representative types displayed in
Table~\ref{tab:g6_64_arguments}.  Rows \(64_3\) and \(64_7\) are vacuum or
lower-topology projections, row \(64_4\) has an unpaired transverse index, and
row \(64_6\) is restricted to the collinear-boundary support described above.
The six non-zero rows provide explicit local checks; they are not used as an
exhaustive denominator-support count.

Different assignments of the observed legs and the vacuum-pair legs to the two
tree amplitudes produce the crossed representatives.  Their phase-space constraints are
solved in the on-shell variables of those assignments and are then expressed in
the crossed denominator notation of \eqref{eq:g6_nonplanar_denominators}.
Because the crossed momentum routing differs from the planar one, the slot words
cannot be obtained by placing tildes on the planar list.  The exhaustive set is
\begin{equation}
\begin{aligned}
{\cal U}_{64}^{\rm NP}
=\{&
\widetilde{136},\widetilde{145},\widetilde{147},
\widetilde{157},\widetilde{235},\widetilde{237},
\widetilde{246},\widetilde{257},\widetilde{357},
\widetilde{457}\}.
\end{aligned}
\label{eq:g6_crossed_64_underlying_supports}
\end{equation}

\subsection{The \texorpdfstring{$\{5,5\}$}{\{5,5\}} sector}
\label{sec:55}

The bookkeeping relation in subsection~\ref{subsec:polygon} gives \(k=2\)
and \(r=k+1=3\) for the \(\{5,5\}\) sector.  We therefore introduce three
vacuum pairs, denoted by
\begin{equation}
(-\ell_1,\ell_1),\qquad
(-\ell_2,\ell_2),\qquad
(-\ell_3,\ell_3).
\end{equation}
This sector contains two pentagons.
Table~\ref{tab:g6_55_arguments} shows selected non-degenerate assignments and
representatives of the distinct vanishing mechanisms, after
cyclicity, exchange of the two pentagons, and vacuum-pair relabelling have been
imposed.  The rows marked by \(\times\) are analyzed after the non-zero
calculation.  The complete denominator-support sets are displayed at the end of
this subsection.
\begin{table}[t]
\centering
\scriptsize
\setlength{\tabcolsep}{3pt}
\begin{tabular}{c
l@{}l@{}l@{}l@{}l
l@{}l@{}l@{}l@{}l
c c}
\hline
case & \multicolumn{5}{c}{\(A_5^{(0)}\) arguments} &
\multicolumn{5}{c}{\(A_5^{(0)}\) arguments} & slot support & non-degenerate \\
\hline
\(55_1\) &
\tblargfirst{-\ell_1} & \tblargnext{\ell_1} & \tblargnext{-\ell_2} &
\tblargnext{\ell_2} & \tblarglast{-\ell_3} &
\tblargfirst{1^+} & \tblargnext{2^+} & \tblargnext{3^+} &
\tblargnext{4^+} & \tblarglast{\ell_3} & -- & \(\times\) \\
\(55_2\) &
\tblargfirst{1^+} & \tblargnext{-\ell_1} & \tblargnext{\ell_1} &
\tblargnext{-\ell_2} & \tblarglast{-\ell_3} &
\tblargfirst{2^+} & \tblargnext{3^+} & \tblargnext{4^+} &
\tblargnext{\ell_3} & \tblarglast{\ell_2} & -- & \(\times\) \\
\(55_3\) &
\tblargfirst{1^+} & \tblargnext{2^+} & \tblargnext{-\ell_1} &
\tblargnext{-\ell_2} & \tblarglast{-\ell_3} &
\tblargfirst{3^+} & \tblargnext{4^+} & \tblargnext{\ell_3} &
\tblargnext{\ell_2} & \tblarglast{\ell_1} & \(\delta_{16\bar 7}^+\) & \(\checkmark\) \\
\(55_4\) &
\tblargfirst{1^+} & \tblargnext{2^+} & \tblargnext{-\ell_1} &
\tblargnext{\ell_1} & \tblarglast{-\ell_2} &
\tblargfirst{3^+} & \tblargnext{4^+} & \tblargnext{\ell_2} &
\tblargnext{-\ell_3} & \tblarglast{\ell_3} & -- & \(\times\) \\
\(55_5\) &
\tblargfirst{1^+} & \tblargnext{2^+} & \tblargnext{3^+} &
\tblargnext{\ell_1} & \tblarglast{\ell_2} &
\tblargfirst{4^+} & \tblargnext{-\ell_2} & \tblargnext{-\ell_1} &
\tblargnext{-\ell_3} & \tblarglast{\ell_3} & -- & \(\times\) \\
\(55_6\) &
\tblargfirst{1^+} & \tblargnext{2^+} & \tblargnext{3^+} &
\tblargnext{4^+} & \tblarglast{-\ell_1} &
\tblargfirst{\ell_1} & \tblargnext{-\ell_2} & \tblargnext{\ell_2} &
\tblargnext{-\ell_3} & \tblarglast{\ell_3} & -- & \(\times\) \\
\hline
\(55_7\) &
\tblargfirst{2^+} & \tblargnext{3^+} & \tblargnext{-\ell_1} &
\tblargnext{-\ell_2} & \tblarglast{-\ell_3} &
\tblargfirst{4^+} & \tblargnext{1^+} & \tblargnext{\ell_3} &
\tblargnext{\ell_2} & \tblarglast{\ell_1} & -- & \(\checkmark\) \\
\(55_8\) &
\tblargfirst{3^+} & \tblargnext{4^+} & \tblargnext{-\ell_1} &
\tblargnext{-\ell_2} & \tblarglast{-\ell_3} &
\tblargfirst{1^+} & \tblargnext{2^+} & \tblargnext{\ell_3} &
\tblargnext{\ell_2} & \tblarglast{\ell_1} & -- & \(\checkmark\) \\
\(55_9\) &
\tblargfirst{4^+} & \tblargnext{1^+} & \tblargnext{-\ell_1} &
\tblargnext{-\ell_2} & \tblarglast{-\ell_3} &
\tblargfirst{2^+} & \tblargnext{3^+} & \tblargnext{\ell_3} &
\tblargnext{\ell_2} & \tblarglast{\ell_1} & -- & \(\checkmark\) \\
\hline
\end{tabular}
\caption{Selected representatives of the \(\{5,5\}\) assignments.  The rows
below the separator are cyclic images of the non-degenerate
two-adjacent-gluon split; only the explicitly routed row carries a slot word in
this table.}
\label{tab:g6_55_arguments}
\end{table}
We start with the non-vanishing representative \(55_3\).  In this case
each pentagon contains one member of each of the three vacuum pairs.
Thus no complete adjacent vacuum pair occurs locally on either pentagon, and the
double-residue reduction used in the \(\{6,4\}\) sector is not available.  The
denominator structure must instead be read from the ordinary BCFW expansion of
the five-point trees.  The corresponding contribution is
\begin{multline}
{\cal C}_{55_3}
=
\sum_{h_1,h_2,h_3}
\int d\Phi(\ell_1)\,d\Phi(\ell_2)\,d\Phi(\ell_3)
\\
\times
(2\pi)^D\delta^{(D)}(k_{12}-\ell_1-\ell_2-\ell_3)\,
(2\pi)^D\delta^{(D)}(k_{34}+\ell_1+\ell_2+\ell_3)
\\
\times
A_5^{(0)}(1^+,2^+,-\ell_1,-\ell_2,-\ell_3)\,
A_5^{(0)}(3^+,4^+,\ell_3,\ell_2,\ell_1).
\label{eq:g6_55_start}
\end{multline}
The delta functions factorize as
\begin{equation}
\delta^{(D)}(k_{12}-\ell_1-\ell_2-\ell_3)\,
\delta^{(D)}(k_{34}+\ell_1+\ell_2+\ell_3)
=
\delta^{(D)}(k_{12}+k_{34})\,
\delta^{(D)}(k_{12}-\ell_1-\ell_2-\ell_3).
\label{eq:g6_55_delta_factorization}
\end{equation}
We strip the overall delta function and write the remaining integration
variables directly in the common planar parametrization of
\eqref{eq:g6_planar_denominators}.  A convenient change of variables is
\begin{equation}
\ell_1=p,\qquad
\ell_2=-p-q,\qquad
\ell_3=q-k_{34}=q+k_{12}.
\label{eq:g6_55_common_routing_shift}
\end{equation}
It has unit Jacobian and obeys
\(\ell_1+\ell_2+\ell_3=k_{12}\).  Since \(\lambda_{k_i}=0\),
\(\lambda_{\ell_1}=\lambda_p\),
\(\lambda_{\ell_2}=-\lambda_{p+q}\), and
\(\lambda_{\ell_3}=\lambda_q\), the three on-shell constraints are
the common slots
\begin{equation}
\ell_1^2-\lambda_{\ell_1}^2=D_1,\qquad
\ell_2^2-\lambda_{\ell_2}^2=D_7,\qquad
\ell_3^2-\lambda_{\ell_3}^2=D_6 .
\label{eq:g6_55_common_cut_slots}
\end{equation}
The phase-space measure displays the positive-energy branches of the
\(\ell_a\) variables.  The explicit phase-space factors are
\begin{equation}
\begin{aligned}
&
\delta^+(p)\,\delta^+(-p-q)\,\delta^+(q-k_{34}).
\end{aligned}
\label{eq:g6_55_oriented_cuts}
\end{equation}
For the bridge slot the displayed positive-energy vacuum-pair member is
\(\ell_2=-p-q\).  Thus the original phase-space parametrization produces the
opposite positive-energy branch of the same quadratic denominator \(D_7\),
namely \(\delta_{\bar 7}^+\) in
\eqref{eq:g6_planar_opposite_bridge_cut}.  Its oriented support is therefore
\(16\bar 7\), while its underlying denominator-slot set is \(167\).  The
representative is
\begin{multline}
{\cal C}_{55_3}
=
\sum_{h_1,h_2,h_3}
\int \frac{d^Dp}{(2\pi)^D}\frac{d^Dq}{(2\pi)^D}\,
\;
\delta^+(p)\,\delta^+(-p-q)\,\delta^+(q-k_{34})
\\
\times
A_5^{(0)}(1^+,2^+,-p,p+q,k_{34}-q)\,
A_5^{(0)}(3^+,4^+,q-k_{34},-p-q,p).
\label{eq:g6_55_reduced_phase_space}
\end{multline}
We now insert the BCFW expansions of the two five-point amplitudes.  Because
the three sewn legs carry complete physical \(D_s\)-dimensional states, the
state sums must be performed after the two amplitudes have been multiplied.
The unprojected pentagon--pentagon product contains three distinct
four-propagator pole structures, which we denote by
\({\cal S}_{55}^{(r)}\), \(r=1,2,3\).  In
appendix~\ref{app:five_point_trees} we work out explicitly the contribution
\({\cal S}_{55}^{(2)}\) which, after conversion to the common routing and the
reflection \((1,2,3,4)\mapsto(2,1,4,3)\), belongs to the pole family
\(\{D_2,D_3,D_4,D_5\}\).  Using
\eqref{eq:app_g6_55_A5_product} and
\eqref{eq:app_g6_55_state_sum} gives
\begin{equation}
\begin{aligned}
{\cal S}_{55}^{(2)}
\longrightarrow
s_{12}\,
\frac{[12][34]}{\langle12\rangle\langle34\rangle}\,
G_2G_3G_4G_5\,
{\cal N}_{55}^{\rm sum}(p,q),
\end{aligned}
\label{eq:g6_55_five_point_bcfw_results}
\end{equation}
where
\begin{equation}
\begin{aligned}
{\cal N}_{55}^{\rm sum}(p,q)
&=
(D_s-2)\Bigl[
\lambda_p^2\lambda_q^2
+\lambda_p^2\lambda_{p+q}^2
+\lambda_q^2\lambda_{p+q}^2
\Bigr]
\\
&\quad
+16\Bigl[
(\lambda_p\!\cdot\!\lambda_q)^2
-\lambda_p^2\lambda_q^2
\Bigr].
\end{aligned}
\label{eq:g6_55_numerator}
\end{equation}
The two remaining pole structures are obtained by the same complete
physical-state contraction and are recorded explicitly in
\eqref{eq:app_g6_55_remaining_pole_structures}.  They are kept with their own
propagator structures until the corresponding common routings and external
relabellings are performed.

For the selected \(\{D_2,D_3,D_4,D_5\}\) component, the kinematic coefficient
is \(s_{12}\).  The numerator entering this seven-slot kernel is therefore
\begin{equation}
s_{12}{\cal N}_{55}^{\rm sum}(p,q)={\cal N}_{\rm DB}(p,q),
\label{eq:g6_55_DB_numerator}
\end{equation}
with \({\cal N}_{\rm DB}\) displayed below in
\eqref{eq:g6_octagon_explicit_DB_factors}.
For this component, the three on-shell conditions are those displayed in
\eqref{eq:g6_55_oriented_cuts}, while the uncut propagators are
\(G_2G_3G_4G_5\).  After adopting the scalar-propagator and
bridge-orientation conventions and stripping the universal
momentum-conservation delta function, the representative is
\begin{equation}
{\cal C}_{55_3}
=
\frac{[12][34]}{\langle12\rangle\langle34\rangle}
\int \frac{d^Dp}{(2\pi)^D}\frac{d^Dq}{(2\pi)^D}\,
{\cal N}_{\rm DB}(p,q)
\;
\delta_1^+G_2G_3G_4G_5\delta_6^+\delta_{\bar 7}^+ .
\end{equation}
The three on-shell delta factors and the four scalar-line propagators are
therefore exactly the seven common denominator slots
\eqref{eq:g6_planar_denominators}.
Thus this \(\{5,5\}\) representative sits in the planar seven-slot family.

We now discuss the remaining rows in Table~\ref{tab:g6_55_arguments} that
do not contribute.  Rows \(55_1\) and \(55_6\) force one vacuum-pair momentum
to be soft.  For \(55_1\), the two tree delta functions contain
\begin{equation}
\delta^{(D)}(-\ell_3)\,
\delta^{(D)}(k_{1234}+\ell_3),
\end{equation}
while for \(55_6\) they contain
\begin{equation}
\delta^{(D)}(k_{1234}-\ell_1)\,
\delta^{(D)}(\ell_1).
\end{equation}
Both cases have support only on a soft boundary of the forward-pair phase
space and give no connected
order-\(g^6\) four-point contribution in dimensional regularization.

Rows \(55_2\) and \(55_5\) reduce to massless two-particle phase space with a
massless external total momentum.  For \(55_2\), the relevant delta functions
are
\begin{equation}
\delta^{(D)}(k_1-\ell_2-\ell_3)\,
\delta^{(D)}(k_{234}+\ell_2+\ell_3),
\end{equation}
and impose
\(\ell_2+\ell_3=k_1\).  Hence
\(0=k_1^2=(\ell_2+\ell_3)^2=2\ell_2\!\cdot\!\ell_3\).
For real positive-energy massless phase-space momenta this forces
\begin{equation}
\ell_2=x k_1,\qquad \ell_3=(1-x)k_1,\qquad 0\le x\le1 .
\end{equation}
The remaining integration is only over the dimensionless collinear fraction
\(x\); the corresponding massless two-particle phase space has no scale and is
set to zero in dimensional regularization.  Row \(55_5\) is the same
degenerate support with \(k_1\) replaced by \(k_4\):
\begin{equation}
\delta^{(D)}(k_{123}+\ell_1+\ell_2)\,
\delta^{(D)}(k_4-\ell_1-\ell_2) ,
\end{equation}
giving \(\ell_1+\ell_2=k_4\).
Row \(55_4\) is the remaining configuration with local complete pairs on the
two pentagons.  It has no generic support, as follows from the two tree
momentum-conservation delta functions:
\begin{equation}
\delta^{(D)}(k_{12}-\ell_2)\,
\delta^{(D)}(k_{34}+\ell_2)
=
\delta^{(D)}(k_{12}+k_{34})\,
\delta^{(D)}(k_{12}-\ell_2).
\label{eq:g6_55_local_pair_delta}
\end{equation}
Since \(\ell_2^2=0\), the remaining constraint would require
\(k_{12}^2=0\).  For generic four-point kinematics \(k_{12}^2=s_{12}\ne0\).
Thus a local complete pair on each side has no support.
With the change of variables in \eqref{eq:g6_55_common_routing_shift}, the representative
just analyzed has the oriented slot support
\begin{equation}
\delta_{16\bar 7}^+.
\end{equation}
The exhaustive underlying planar \(\{5,5\}\) cut sets are
\begin{equation}
{\cal U}_{55}^{\rm P}=\{167,257,347\}.
\label{eq:g6_55_planar_underlying_supports}
\end{equation}
Both causal orientations of each set are included in the comparison below.
The explicit row
\(55_3\) realizes the reversed orientation of the underlying set \(167\).
This exhausts the vanishing mechanisms represented in
Table~\ref{tab:g6_55_arguments}:
rows \(55_1\) and \(55_6\) are soft-boundary configurations, rows \(55_2\) and
\(55_5\) reduce to scaleless massless two-particle collinear phase space, and
row \(55_4\) would require the generic invariant \(s_{12}\) to vanish.  The
corresponding exhaustive crossed set is
\begin{equation}
{\cal U}_{55}^{\rm NP}
=\{\widetilde{137},\widetilde{146},\widetilde{236},\widetilde{245}\}.
\label{eq:g6_55_crossed_underlying_supports}
\end{equation}
These four sets likewise occur with both causal orientations.  This sector has
no local bow-tie projection, so
\begin{equation}
{\cal U}_{55}^{\rm BT}=\{\}.
\label{eq:g6_55_bowtie_underlying_supports}
\end{equation}

\subsection{The \texorpdfstring{$\{4,4,4\}$}{\{4,4,4\}} sector}

The bookkeeping relation in subsection~\ref{subsec:polygon} gives \(k=3\)
and \(r=k+1=4\) for the \(\{4,4,4\}\) sector.  We therefore introduce four
vacuum pairs, denoted by
\begin{equation}
(-\ell_1,\ell_1),\qquad
(-\ell_2,\ell_2),\qquad
(-\ell_3,\ell_3),\qquad
(-\ell_4,\ell_4).
\end{equation}
This is the only \(k=3\) sector: it is a product of three quadrilateral trees.
Table~\ref{tab:g6_444_arguments} displays selected non-degenerate assignments
and representative degenerate assignments in this sector.  Rows
below the separator are cyclic representatives of the non-degenerate split.
Only the explicitly routed row carries a slot word in the table; the exhaustive
support sets are given below.
\begin{table}[t]
\centering
\scriptsize
\setlength{\tabcolsep}{3pt}
\begin{tabular}{c
l@{}l@{}l@{}l
l@{}l@{}l@{}l
l@{}l@{}l@{}l
c c}
\hline
case & \multicolumn{4}{c}{\(A_{4,L}^{(0)}\) arguments} &
\multicolumn{4}{c}{\(A_{4,B}^{(0)}\) arguments} &
\multicolumn{4}{c}{\(A_{4,R}^{(0)}\) arguments} & slot support & non-degenerate \\
\hline
\(444_1\) &
\tblargfirst{1^+} & \tblargnext{2^+} & \tblargnext{-\ell_1} & \tblarglast{-\ell_2} &
\tblargfirst{\ell_1} & \tblargnext{\ell_2} & \tblargnext{-\ell_3} & \tblarglast{-\ell_4} &
\tblargfirst{3^+} & \tblargnext{4^+} & \tblargnext{\ell_3} & \tblarglast{\ell_4} &
\(\delta_{1346}^+\) & \(\checkmark\) \\
\(444_2\) &
\tblargfirst{1^+} & \tblargnext{-\ell_1} & \tblargnext{-\ell_2} & \tblarglast{-\ell_3} &
\tblargfirst{\ell_3} & \tblargnext{\ell_2} & \tblargnext{2^+} & \tblarglast{-\ell_4} &
\tblargfirst{3^+} & \tblargnext{4^+} & \tblargnext{\ell_4} & \tblarglast{\ell_1} &
-- & \(\times\) \\
\(444_3\) &
\tblargfirst{1^+} & \tblargnext{2^+} & \tblargnext{3^+} & \tblarglast{-\ell_1} &
\tblargfirst{\ell_1} & \tblargnext{-\ell_2} & \tblargnext{-\ell_3} & \tblarglast{-\ell_4} &
\tblargfirst{4^+} & \tblargnext{\ell_4} & \tblargnext{\ell_3} & \tblarglast{\ell_2} &
-- & \(\times\) \\
\(444_4\) &
\tblargfirst{1^+} & \tblargnext{2^+} & \tblargnext{3^+} & \tblarglast{4^+} &
\tblargfirst{-\ell_1} & \tblargnext{\ell_1} & \tblargnext{-\ell_2} & \tblarglast{\ell_2} &
\tblargfirst{-\ell_3} & \tblargnext{\ell_3} & \tblargnext{-\ell_4} & \tblarglast{\ell_4} &
-- & \(\times\) \\
\hline
\(444_5\) &
\tblargfirst{2^+} & \tblargnext{3^+} & \tblargnext{-\ell_1} & \tblarglast{-\ell_2} &
\tblargfirst{\ell_1} & \tblargnext{\ell_2} & \tblargnext{-\ell_3} & \tblarglast{-\ell_4} &
\tblargfirst{4^+} & \tblargnext{1^+} & \tblargnext{\ell_3} & \tblarglast{\ell_4} &
-- & \(\checkmark\) \\
\(444_6\) &
\tblargfirst{3^+} & \tblargnext{4^+} & \tblargnext{-\ell_1} & \tblarglast{-\ell_2} &
\tblargfirst{\ell_1} & \tblargnext{\ell_2} & \tblargnext{-\ell_3} & \tblarglast{-\ell_4} &
\tblargfirst{1^+} & \tblargnext{2^+} & \tblargnext{\ell_3} & \tblarglast{\ell_4} &
-- & \(\checkmark\) \\
\(444_7\) &
\tblargfirst{4^+} & \tblargnext{1^+} & \tblargnext{-\ell_1} & \tblarglast{-\ell_2} &
\tblargfirst{\ell_1} & \tblargnext{\ell_2} & \tblargnext{-\ell_3} & \tblarglast{-\ell_4} &
\tblargfirst{2^+} & \tblargnext{3^+} & \tblargnext{\ell_3} & \tblarglast{\ell_4} &
-- & \(\checkmark\) \\
\hline
\end{tabular}
\caption{Selected representatives of the \(\{4,4,4\}\) assignments.  The
rows below the separator are cyclic images of the non-degenerate split.}
\label{tab:g6_444_arguments}
\end{table}
We start with the representative \(444_1\), keeping the original vacuum-pair
labels until the tree delta functions have been simplified.  This representative
is
\begin{multline}
{\cal C}_{444_1}
=
\sum_{h_1,h_2,h_3,h_4}
\int d\Phi(\ell_1)\,d\Phi(\ell_2)\,d\Phi(\ell_3)\,d\Phi(\ell_4)\,
(2\pi)^D\delta^{(D)}(k_{12}-\ell_1-\ell_2)
\\
\times
(2\pi)^D\delta^{(D)}(\ell_1+\ell_2-\ell_3-\ell_4)\,
(2\pi)^D\delta^{(D)}(k_{34}+\ell_3+\ell_4)
\\
\times
A_4^{(0)}(1^+,2^+,-\ell_1,-\ell_2)\,
A_4^{(0)}(\ell_1,\ell_2,-\ell_3,-\ell_4)\,
A_4^{(0)}(3^+,4^+,\ell_3,\ell_4).
\label{eq:g6_444_start}
\end{multline}
The three delta functions give
\begin{multline}
\delta^{(D)}(k_{12}-\ell_1-\ell_2)\,
\delta^{(D)}(\ell_1+\ell_2-\ell_3-\ell_4)\,
\delta^{(D)}(k_{34}+\ell_3+\ell_4)
\\
=
\delta^{(D)}(k_{12}+k_{34})\,
\delta^{(D)}(k_{12}-\ell_1-\ell_2)\,
\delta^{(D)}(k_{12}-\ell_3-\ell_4).
\label{eq:g6_444_delta_factorization}
\end{multline}
After the overall delta function is stripped, we choose the remaining integration variables
\begin{equation}
\ell_1=k_{12}-p,\qquad
\ell_2=p,\qquad
\ell_3=-q,\qquad
\ell_4=q-k_{34}.
\label{eq:444rout}
\end{equation}
The oriented slot support is therefore
\begin{equation}
\delta^+(p)\,\delta^+(k_{12}-p)\,
\delta^+(-q)\,\delta^+(q-k_{34}),
\label{eq:g6_444_oriented_cuts}
\end{equation}
which, in the convention of \eqref{eq:g6_planar_cut_definitions}, is the slot support
\(\delta_1^+\delta_3^+\delta_4^+\delta_6^+\).  The two side four-point trees
provide the central side propagators
\(G_2\) and \(G_5\).  These two side \(A_4\) amplitudes are evaluated with the
open-index scalar-chain formula of appendix~\ref{app:scalar_trees}, in
particular \eqref{eq:app_A4_open_PQ_formula}.  With the above routing, the
left side tree is
\begin{equation}
A_4^{(0)}(1^+,2^+,p-k_{12},-p)
=
A_4^{(0)}(-p,1^+,2^+,p-k_{12}),
\end{equation}
where the equality follows from a cyclic rotation of the same color-ordered tree.
Thus \eqref{eq:app_A4_open_PQ_formula} is applied with
\(P=-p\), \(i=1\), \(j=2\), and \(Q=p-k_{12}\), and the propagator is
\(G_F((p-k_1)^2-\lambda_p^2)=G_2\).  The right-hand tree is treated analogously by
cyclically writing
\(A_4^{(0)}(3^+,4^+,-q,q-k_{34})\) as
\(A_4^{(0)}(q-k_{34},3^+,4^+,-q)\), which gives \(G_5\).
The middle factor is different:
it is a pure-scalar four-point tree with no observed gluon insertions.  With
the above routing it is
\begin{equation}
A_{4,B}^{(0)}
=
A_4^{(0)}(k_{12}-p,p,q,k_{34}-q).
\end{equation}
As shown in appendix~\ref{app:444_bowtie_contraction}, an adjacent BCFW
deformation organizes this middle four-point amplitude into the finite
\(D_7\)-pole residue and its boundary contribution in
\eqref{eq:app_bowtie_BCFW_terms}.  The finite-pole term gives,
after contraction with the two side amplitudes, a contribution to the
seven-slot double-box family.  Equation~\eqref{eq:app_444_complete_D7_result},
derived in appendix~\ref{app:444_bowtie_contraction}, gives
\begin{equation}
{\cal N}_{444,7}^{(1346)}(p,q)
=
{\cal N}_{\rm DB}(p,q),
\label{eq:g6_444_D7_numerator}
\end{equation}
with \({\cal N}_{\rm DB}\) displayed below in
\eqref{eq:g6_octagon_explicit_DB_factors}.  The
boundary term contains the fixed \(s_{12}\)-channel factorization together
with its color-ordered contact completion.  It has no integration-dependent
middle propagator and can be written with the external factor \(1/s_{12}\).
It therefore contributes to the bow-tie family with the numerator
\({\cal N}_{\rm BT}\) in~\eqref{eq:g6_MBT2}, derived in
\eqref{eq:app_bowtie_complete}.
The representative therefore splits into a seven-slot double-box part and a
six-slot bow-tie part.
The split occurs within the same complete tree product: the finite BCFW pole
produces an integration-dependent bridge propagator and belongs to the
double-box family, while the completed boundary term contains only the
external factor \(1/s_{12}\) and leaves a local six-propagator bow-tie family.
The double-box projection is
\begin{equation}
{\cal C}_{444_1,{\rm DB}}
=
\frac{[12][34]}{\langle12\rangle\langle34\rangle}
\int \frac{d^Dp}{(2\pi)^D}\frac{d^Dq}{(2\pi)^D}\,
{\cal N}_{444,7}^{(1346)}(p,q)\,
\;
\delta_1^+G_2\delta_3^+\delta_4^+G_5\delta_6^+G_7 .
\label{eq:g6_444_DB_raw_cut}
\end{equation}
The corresponding local bow-tie projection is
\begin{equation}
{\cal C}_{444_1,{\rm BT}}
=
\frac{[12][34]}{\langle12\rangle\langle34\rangle}
\int \frac{d^Dp}{(2\pi)^D}\frac{d^Dq}{(2\pi)^D}\,
{\cal N}_{\rm BT}(p,q)\,
\;
\delta_1^+G_2\delta_3^+\delta_4^+G_5\delta_6^+ .
\label{eq:g6_444_raw_cut}
\end{equation}
Equations \eqref{eq:g6_444_DB_raw_cut} and \eqref{eq:g6_444_raw_cut} are the two
parts of the same fixed-order representative.  The first belongs to the
seven-slot double-box family; the second belongs to the six-slot
bow-tie family.

It remains to analyze the rows marked by \(\times\) in
Table~\ref{tab:g6_444_arguments}.  Row \(444_4\) contains the all-plus tree
\(A_4^{(0)}(1^+,2^+,3^+,4^+)\), so it vanishes before any phase-space
integration is considered.  Row \(444_3\) contains the delta function
\(\delta^{(D)}(k_{123}-\ell_1)\), which fixes \(\ell_1=-k_4\).  The remaining two quadrilaterals then
force three positive-energy phase-space momenta to sum to the same massless direction.
This is a collinear-boundary configuration with no remaining dimensionful
scale, and it is zero in dimensional regularization.

For row \(444_2\), the three tree delta functions imply
\begin{equation}
\ell_1+\ell_2+\ell_3=k_1,\qquad
\ell_4=\ell_2+\ell_3+k_2,\qquad
\ell_1+\ell_4=k_{12}.
\end{equation}
The last two equations are compatible with the first one, but the first
equation forces the future-directed massless sum \(\ell_2+\ell_3\) to equal
\(k_1-\ell_1\).  Using full \(D\)-dimensional invariants as above,
\((k_{1,D}-\ell_{1,D})^2=-2k_{1,D}\!\cdot\!\ell_{1,D}\le0\), while
\((\ell_{2,D}+\ell_{3,D})^2\ge0\), so non-degenerate support requires
\(k_{1,D}\!\cdot\!\ell_{1,D}=0\).  Thus \(\ell_1\), \(\ell_2\), and \(\ell_3\) are
collinear with \(k_1\).  The remaining on-shell condition
\(\ell_{4,D}^2=(k_{12,D}-\ell_{1,D})^2=0\) then forces \(\ell_1=k_1\) for generic
\(s_{12}\), and hence \(\ell_2+\ell_3=0\).  The row therefore has support only
on a soft/collinear boundary and gives no contribution.

The non-zero rows \(444_1,444_5,444_6,\) and \(444_7\) represent cyclic
tree-assignment types.  Routing all terminal cuts in the planar and crossed
parent graphs gives, respectively, the exhaustive underlying sets
\begin{equation}
{\cal U}_{444}^{\rm P}=\{1346,1357,2467\},
\qquad
{\cal U}_{444}^{\rm NP}
=\{\widetilde{1457},\widetilde{2357}\},
\qquad
{\cal U}_{444}^{\rm BT}=\{1346\}.
\label{eq:g6_444_underlying_supports}
\end{equation}
In the seven-slot families the planar set \(1346\) has one causal orientation;
the other four entries have two causal orientations each.  The complete
oriented count is therefore five for the planar family and four for the
crossed family.  The bow-tie entry has one orientation.

\subsection{The \texorpdfstring{$\{8\}$}{\{8\}} sector}
\label{sec:g6_octagon_sector}

The \(\{8\}\) sector has \(k=1\) and \(r=2\) and therefore contains one
octagon with two vacuum pairs.  For the planar representative, the complete
state-summed tree contribution is
\begin{equation}
\begin{aligned}
{\cal C}_{8_1}
&=
\sum_{h_1,h_2\in {\rm phys}(D_s)}
\int d\Phi(\ell_1)\,d\Phi(\ell_2)\,
(2\pi)^D\delta^{(D)}(k_{1234})
\\
&\quad\times
A_{8}^{(0)}
\bigl(1^+,2^+,(-\ell_1)^{\bar h_1},
(-\ell_2)^{\bar h_2},3^+,4^+,\ell_2^{h_2},\ell_1^{h_1}\bigr).
\end{aligned}
\label{eq:g6_octagon_vacuum_pair_start}
\end{equation}
The tree momentum-conservation delta function contains only the overall
four-gluon conservation law: the two vacuum pairs carry zero total momentum.
After this overall delta function is stripped, no additional tree delta
function relates the two phase-space integrations.

The strictly four-dimensional part of an individual octagon assignment is
non-zero.  With four external positive-helicity gluons, a non-vanishing
pure-gluon component of the eight-point tree is an MHV Parke--Taylor term with
two negative helicities on the four vacuum-pair legs.  The complete octagon
contribution is the sum over all such internal assignments, together with the
remaining physical $D_s$-dimensional states in the state sum above.  Its
phase-space integrals depend on the external invariants and are therefore
neither scaleless nor zero in dimensional regularization.  In particular, we
do not remove this sector by a sectorwise supersymmetry argument.

Supersymmetry constrains only the complete state and forest sum.  With a
regulator that preserves supersymmetry, the complete all-plus supersymmetric amplitude
vanishes by the supersymmetry Ward identities~\cite{Grisaru:1976vm}; the pure
gauge-field graphs common to Yang--Mills and supersymmetric Yang--Mills are
then balanced by the complete superpartner contribution.  At two loops this
does not imply the one-loop proportionality between a gluon loop and a scalar
loop.  The pure-gluon construction requires the one-scalar contribution, the
corrected two-scalar combinatorics, the quartic-scalar interaction, and the
additional transverse contribution, with their coefficients fixed by the
complete \(D\)-dimensional cuts~\cite{Bern:2000dn,Bern:2002tk}.

The calculation starts by fixing the two opened slots and then selecting the
remaining pole structure.  We choose the common planar variables by
\begin{equation}
\ell_1=-p,\qquad
\ell_2=k_{34}-q.
\label{eq:g6_octagon_routing_identification}
\end{equation}
Since the observed external momenta have no transverse components,
\(\lambda_{\ell_1}=-\lambda_p\) and
\(\lambda_{\ell_2}=-\lambda_q\).  The two phase-space delta functions therefore
put the slots \(D_1\) and \(D_6\) on shell,
\begin{equation}
\ell_1^2-\lambda_{\ell_1}^2=D_1,\qquad
\ell_2^2-\lambda_{\ell_2}^2=D_6 .
\label{eq:g6_octagon_cut_slot_identification}
\end{equation}
With the global slot orientations in
\eqref{eq:g6_planar_cut_definitions}, this representative gives
\begin{equation}
\delta_{\bar1}^+\,\delta_{\bar6}^+ .
\end{equation}
Its underlying cut set is \(16\).  Reversing the orientation of a cut changes the
barred orientation but not this underlying denominator support.
The remaining uncut slots of the planar projection are
\begin{equation}
{\cal P}_{8,16}^{\rm P}
=\{D_2,D_3,D_4,D_5,D_7\}.
\label{eq:g6_octagon_planar_selected_poles}
\end{equation}
The non-degenerate seven-slot contribution is the bridge factorization with
internal momentum \(p+q\).  The state running through the bridge is summed over
the complete physical \(D_s\)-dimensional basis.  The required state-summed
BCFW building blocks are collected in
appendix~\ref{app:g6_bcfw_building_blocks}, and the octagon state sum and
representative factorization are detailed in
subsection~\ref{app:g6_octagon_factorization}.  At the \(D_7\) pole,
factorization gives
\begin{equation}
\begin{aligned}
&\sum_{h_1,h_2\in{\rm phys}(D_s)}
A_{8}^{(0)}
\bigl(1^+,2^+,p^{\bar h_1},(q-k_{34})^{\bar h_2},
3^+,4^+,(k_{34}-q)^{h_2},(-p)^{h_1}\bigr)
\Big|_{D_7}
\\
&\quad=
\sum_{h_1,h_2,h\in{\rm phys}(D_s)}
A_5^{(0)}
\bigl(p^{\bar h_1},(q-k_{34})^{\bar h_2},3^+,4^+,
(-p-q)^{\bar h}\bigr)
\,iG_7
\\
&\hspace{2.7cm}\times
A_5^{(0)}
\bigl((p+q)^h,(k_{34}-q)^{h_2},(-p)^{h_1},1^+,2^+\bigr).
\end{aligned}
\label{eq:g6_octagon_stripped_tree}
\end{equation}
Together with the two vacuum-pair phase spaces, this channel has the labelled
slot kernel
\begin{equation}
\delta_{\bar1}^+G_2G_3G_4G_5\delta_{\bar6}^+G_7 .
\label{eq:g6_octagon_two_cut_16}
\end{equation}
Here \(\delta_{\bar1}^+\) and \(\delta_{\bar6}^+\) come directly from
\(d\Phi(\ell_1)\) and \(d\Phi(\ell_2)\) under the routing
\(\ell_1=-p\), \(\ell_2=k_{34}-q\).  The bridge factorization in
\eqref{eq:g6_octagon_stripped_tree} supplies \(G_7\).  Recursively factorizing
the five-point tree containing the observed legs \(1,2\) supplies \(G_2G_3\),
while the tree containing \(3,4\) supplies \(G_4G_5\).

Thus \eqref{eq:g6_octagon_stripped_tree} is the \(D_7\)-pole factorization,
whereas the recursive poles of its two five-point factors supply
\(D_2,D_3\) and \(D_4,D_5\).  The two terms of each five-point recursion and
their \(A_3\)-\(A_4\) reductions are displayed explicitly in
\eqref{eq:app_octagon_16_explicit_A5_channels} of
subsection~\ref{app:g6_octagon_numerator_reconstruction}.

For the selected uncut slots
\eqref{eq:g6_octagon_planar_selected_poles}, the two terms of each five-point
recursion give the four products
\({\cal A}_{12,\alpha}{\cal A}_{34,\beta}\),
\(\alpha,\beta\in\{1,2\}\).  After the common
factor \(G_2G_3G_4G_5G_7\) has been stripped and the physical
\(D_s\)-dimensional projectors have been sewn, the state-trace pieces of the
equal-attachment products give
\(s_{12}(D_s-2)\lambda_p^2\lambda_q^2\), while those of the
opposite-attachment products give
\(s_{12}(D_s-2)(\lambda_p^2\lambda_{p+q}^2+
\lambda_q^2\lambda_{p+q}^2)\).  Their sum is \(B_{\rm DB}^{(1)}\) below.
In the trace-free part, the equal-attachment products
\((\alpha,\beta)=(1,1),(2,2)\) give together
\(16s_{12}(\lambda_p\!\cdot\!\lambda_q)^2\), whereas the two
opposite-attachment products give
\(-16s_{12}\lambda_p^2\lambda_q^2\).  Their sum is therefore
\(16B_{\rm DB}^{(2)}\), and the numerator is
\begin{equation}
\begin{aligned}
B_{\rm DB}^{(1)}
&=
s_{12}(D_s-2)\bigl(
\lambda_p^2\lambda_q^2
+\lambda_p^2\lambda_{p+q}^2
+\lambda_q^2\lambda_{p+q}^2\bigr),
\\
B_{\rm DB}^{(2)}
&=
s_{12}\bigl(
(\lambda_p\!\cdot\!\lambda_q)^2
-\lambda_p^2\lambda_q^2\bigr),
\\
{\cal N}_{\rm DB}
&=B_{\rm DB}^{(1)}+16B_{\rm DB}^{(2)} .
\end{aligned}
\label{eq:g6_octagon_explicit_DB_factors}
\end{equation}
Here \(B_{\rm DB}^{(1)}\) is the complete state-trace contribution, while
\(B_{\rm DB}^{(2)}\) is the remaining trace-free transverse contribution.  The
complete contraction retains the non-zero four-dimensional MHV component
\eqref{eq:app_g6_octagon_mhv_components} during the pole collection.

We next collect the remaining pole structures in the octagon.  For each term
we record its propagator poles, glue the two
vacuum pairs, and route the resulting two-loop denominator structure into the
planar, crossed, or bow-tie family.  Moving a vacuum pair to another cyclic
position changes which common denominator is fixed by its phase-space delta
function.  For the non-degenerate planar projections the two cut labels belong
to different classes of the partition
$S^{\rm P}_1=\{1,2,3\}$,
$S^{\rm P}_2=\{4,5,6\}$, and
$S^{\rm P}_3=\{7\}$.
The complete underlying two-cut support is
\begin{equation}
{\cal U}_{8}^{\rm P}=\{14,15,16,17,24,25,26,27,34,35,36,37,47,57,67\}.
\label{eq:g6_planar_two_cut_supports}
\end{equation}
The causal orientations
of these support sets are included in the family collection below; they do not
change the underlying set.

The bow-tie projection is selected from the same octagon expansion by keeping
the two side chains and the external \(s_{12}\) pole, but no
phase-space-dependent \(D_7\) bridge:
\begin{equation}
{\cal P}_{8,16}^{\rm BT}
=\{D_2,D_3,D_4,D_5;s_{12}\}.
\label{eq:g6_octagon_bowtie_selected_poles}
\end{equation}
The semicolon separates the four integration-dependent poles from the external
channel.
The terms with this pole structure, together with the corresponding local contact terms, give the two explicit numerator factors
\begin{equation}
\begin{aligned}
B_{\rm BT}^{(1)}
&=
4(D_s-2)(\lambda_p^2+\lambda_q^2)
(\lambda_p\!\cdot\!\lambda_q),
\\
B_{\rm BT}^{(2)}
&=
\frac{(D_s-2)^2}{s_{12}}\,
\lambda_p^2\lambda_q^2(D_7+s_{12}),
\\
{\cal N}_{\rm BT}
&=B_{\rm BT}^{(1)}+B_{\rm BT}^{(2)} .
\end{aligned}
\label{eq:g6_octagon_explicit_BT_factors}
\end{equation}
Here \(D_7\) in \(B_{\rm BT}^{(2)}\) is a numerator invariant; it is not a
bow-tie propagator.  This is therefore a six-slot family.
The enumeration of vacuum-pair placements places one on-shell constraint in
\(S^{\rm BT}_1=\{1,2,3\}\) and the other in
\(S^{\rm BT}_2=\{4,5,6\}\), giving
\begin{equation}
{\cal U}_{8}^{\rm BT}
=\{14,15,16,24,25,26,34,35,36\}.
\label{eq:g6_octagon_bowtie_two_cut_supports}
\end{equation}
These are the supports with one phase-space constraint on each side chain.
All terms with the selected uncut slots
\eqref{eq:g6_octagon_bowtie_selected_poles} are included in this collection.
To reconstruct the full-color amplitude rather than only the fixed planar
primitive, the octagon sum must also include inequivalent assignments whose
observed legs do not retain the canonical planar cyclic order.  The relevant
assignments are those for which sewing the vacuum-pair color indices gives the
crossed color factor and whose poles belong to the crossed seven-denominator
family $\widetilde D_1,\ldots,\widetilde D_7$.
For these crossed assignments the three recorded slot classes are
\(\widetilde S^{\rm NP}_1=\{\widetilde1,\widetilde2\}\),
\(\widetilde S^{\rm NP}_2=\{\widetilde3,\widetilde4\}\), and
\(\widetilde S^{\rm NP}_3=\{\widetilde5,\widetilde6,\widetilde7\}\).
Selecting two distinct classes gives the exhaustive underlying crossed
two-cut set
\begin{equation}
{\cal U}_{8}^{\rm NP}=\{\widetilde{13},\widetilde{14},
\widetilde{15},\widetilde{16},\widetilde{17},
\widetilde{23},\widetilde{24},\widetilde{25},\widetilde{26},
\widetilde{27},
\widetilde{35},\widetilde{36},\widetilde{37},
\widetilde{45},\widetilde{46},\widetilde{47}\}.
\label{eq:g6_crossed_octagon_two_cut_supports}
\end{equation}
These entries come from different octagon assignments, not from rerouting the
same cyclic sequence.  For instance, one representative contributing to the
\(\widetilde{16}\) support is
\begin{equation}
\begin{aligned}
\widetilde{\cal C}_{8_{16}}
&=
\sum_{h_1,h_2\in{\rm phys}(D_s)}
\int d\Phi(\ell_1)\,d\Phi(\ell_2)\,
(2\pi)^D\delta^{(D)}(k_{1234})
\\
&\quad\times
A_{8}^{(0)}
\bigl(1^+,4^+,(-\ell_1)^{\bar h_1},(-\ell_2)^{\bar h_2},
2^+,3^+,\ell_2^{h_2},\ell_1^{h_1}\bigr)  .
\end{aligned}
\label{eq:g6_crossed_octagon_assignment_example}
\end{equation}
The observed-leg sequence \((1,4,2,3)\) is the cyclic ordering of this
particular tree factor, not an alternative representation of the planar
ordering \((1,2,3,4)\).  At the color-dressed level, dressing the six cubic
vertices of the selected residues with \(\widetilde f^{abc}\) and contracting
the two vacuum-pair color indices gives the crossed color factor
\(C_{12;34}^{\rm NP}\) displayed in
\eqref{eq:g6_explicit_color_factors}.  This color sewing, together with the
crossed pole structure, assigns the term to \(A_{12;34}^{\rm NP}\).  It would
be absent in a calculation restricted to \(A_{1234}^{\rm P}\).
Using the crossed change of variables
\(\ell_1=p\) and \(\ell_2=p+q+k_3\), the two phase-space constraints are
\(\ell_1^2-\lambda_{\ell_1}^2=\widetilde D_1\) and
\(\ell_2^2-\lambda_{\ell_2}^2=\widetilde D_6\), so this assignment gives the
underlying crossed slot pair \(\widetilde{16}\).  Its remaining uncut slots are
\begin{equation}
{\cal P}_{8,\widetilde{16}}^{\rm NP}
=\{\widetilde D_2,\widetilde D_3,\widetilde D_4,
\widetilde D_5,\widetilde D_7\}.
\label{eq:g6_octagon_nonplanar_selected_poles}
\end{equation}
The crossed state contraction gives
\begin{equation}
{\cal N}_{\rm DB}^{\rm crossed}
=B_{\rm DB}^{(1)}+16B_{\rm DB}^{(2)},
\label{eq:g6_octagon_explicit_NP_factor}
\end{equation}
where \({\cal N}_{\rm DB}^{\rm crossed}\) denotes the same polynomial as
\({\cal N}_{\rm DB}\) in \eqref{eq:g6_octagon_explicit_DB_factors}, after its
transverse scalar products have been expressed in the loop variables of the
crossed denominator family.  Thus the non-planar projection is obtained by selecting
crossed pole structures and routing the corresponding crossed assignments.
The complete list
\eqref{eq:g6_crossed_octagon_two_cut_supports} is obtained by the exhaustive
enumeration and factorization of the cyclic assignments, rather than by
rerouting this single assignment.

For use in the next subsection, let the oriented family kernels be
\begin{equation}
\begin{aligned}
K_{8}^{\rm P}
&:=
\sum_{S\in{\cal U}_{8}^{\rm P}}
\delta_{{\rm P},S}^+
\prod_{j\in\{1,\ldots,7\}\setminus S}G_j,
\\
K_{8}^{\rm NP}
&:=
\sum_{\widetilde S\in{\cal U}_{8}^{\rm NP}}
\widetilde\delta_{{\rm NP},\widetilde S}^+
\prod_{j\in\{1,\ldots,7\}\setminus\pi(\widetilde S)}
\widetilde G_j,
\\
K_{8}^{\rm BT}
&:=
\sum_{S\in{\cal U}_{8}^{\rm BT}}
\delta_{{\rm BT},S}^+
\prod_{j\in\{1,\ldots,6\}\setminus S}G_j .
\end{aligned}
\label{eq:g6_octagon_family_kernels}
\end{equation}
Here \(\pi(\widetilde S)\) means the slot set obtained
by dropping the tildes.  The family label on each delta word includes the causal orientation
obtained from its vacuum-pair routing.  The complete non-degenerate octagon
collection is then
\begin{equation}
{\cal C}_{8}^{\rm coll}
=
\frac{[12][34]}{\langle12\rangle\langle34\rangle} \int\frac{d^Dp}{(2\pi)^D}\frac{d^Dq}{(2\pi)^D}
\Bigl[
{\cal N}_{\rm DB}\,K_{8}^{\rm P}
+{\cal N}_{\rm DB}^{\rm crossed}\,K_{8}^{\rm NP}
+{\cal N}_{\rm BT}\,K_{8}^{\rm BT}
\Bigr].
\label{eq:g6_octagon_compact_collection}
\end{equation}
The pole complements, the three exhaustive support sets, the numerator
pieces in \eqref{eq:g6_octagon_explicit_DB_factors} and
\eqref{eq:g6_octagon_explicit_BT_factors}, and the compact collection
\eqref{eq:g6_octagon_compact_collection} have been checked algebraically.  The
Equation~\eqref{eq:g6_octagon_compact_collection} is the form used directly in
the family comparison.

\FloatBarrier
\subsection{Comparison with the two-loop calculation}

We now collect the results of the preceding four subsections.  At order
\(g^6\), the bookkeeping in subsection~\ref{subsec:polygon} gives
\begin{equation}
c=r-k+1=2
\label{eq:g6_two_loop_momentum_count}
\end{equation}
independent vacuum-pair momenta, as required for a two-loop contribution.

We encounter terms belonging to each of the planar double-box,
non-planar double-box, and bow-tie families.  The \(\{8\}\) sector, the
\(\{6,4\}\) and \(\{5,5\}\) sectors, and the \(\{4,4,4\}\) sector have
\(k=1,2,3\),
respectively.  By \eqref{eq:tree_factor_sign}, their inclusion--exclusion
factor is \((-1)^{k-1}\); the convention following \eqref{eq:vpdef} assigns
the tree phase \((-i)^k\).  For \(k=1,2,3\), these factors are
\((+1,-1,+1)\) and \((-i,-1,+i)\), respectively, so their product
\((-1)^{k-1}(-i)^k\) gives the combined weights \((-i,+1,+i)\).  We use the
causal orientations fixed in \eqref{eq:g6_planar_cut_definitions} and
\eqref{eq:g6_crossed_cut_definitions}; boundary-supported terms containing a
three-point component are absent under the prescription of
subsection~\ref{subsec:insertions}.

The exhaustive sector-by-family collection is
\begin{equation}
\begin{array}{c@{\qquad}cccc}
& \{8\} & \{6,4\} & \{5,5\} & \{4,4,4\}
\\[1mm]
{\rm P/DB}
& {\cal U}_{8}^{\rm P}
& {\cal U}_{64}^{\rm P}
& {\cal U}_{55}^{\rm P}
& {\cal U}_{444}^{\rm P}
\\
{\rm NP/DB}
& {\cal U}_{8}^{\rm NP}
& {\cal U}_{64}^{\rm NP}
& {\cal U}_{55}^{\rm NP}
& {\cal U}_{444}^{\rm NP}
\\
{\rm BT}
& {\cal U}_{8}^{\rm BT}
& {\cal U}_{64}^{\rm BT}
& \{\}
& {\cal U}_{444}^{\rm BT} .
\end{array}
\label{eq:g6_sector_family_collection}
\end{equation}
The entries are the exhaustive underlying support sets displayed in the
corresponding sector subsections.  For an untilded support \(S\) in
the \({\rm P/DB}\) row, the uncut factor is
\(\prod_{j\in\{1,\ldots,7\}\setminus S}G_j\); in the \({\rm BT}\) row the
complement is instead taken in \(\{1,\ldots,6\}\).  For a crossed support
\(\widetilde S\), the uncut factor is
\(\prod_{j\in\{1,\ldots,7\}\setminus\pi(\widetilde S)}\widetilde G_j\).
Each underlying support is then expanded into the allowed oriented cut words
specified in the sector subsections.  A bar reverses only the causal orientation
of that cut factor and does not change the underlying denominator complement.
If several inequivalent assignments produce the same oriented kernel, their
coefficients are summed.  The recursive FTT comparison uses these same oriented
kernels.

The numerator factors obtained independently in the different sectors can now
be compared.  In the octagon sector, \({\cal N}_{\rm DB}\),
\({\cal N}_{\rm DB}^{\rm crossed}\), and \({\cal N}_{\rm BT}\) are given in
\eqref{eq:g6_octagon_explicit_DB_factors},
\eqref{eq:g6_octagon_explicit_NP_factor}, and
\eqref{eq:g6_octagon_explicit_BT_factors}, respectively.  The \(\{6,4\}\)
double-box and bow-tie numerators are obtained in
\eqref{eq:app_64_D7_complete} and \eqref{eq:app_64_BT_complete}; the
\(\{5,5\}\) double-box numerator is obtained in
\eqref{eq:app_g6_55_state_sum}; and the \(\{4,4,4\}\) double-box and bow-tie
numerators are obtained in \eqref{eq:app_444_complete_D7_result} and
\eqref{eq:app_bowtie_complete}.  Together these calculations give
\begin{equation}
\begin{aligned}
{\cal N}_{55}
={\cal N}_{64,7}
={\cal N}_{444,7}
&={\cal N}_{\rm DB},
&
{\cal N}_{64,{\rm BT}}
={\cal N}_{444,{\rm BT}}
&={\cal N}_{\rm BT}.
\end{aligned}
\label{eq:g6_higher_cut_numerator_reconstruction}
\end{equation}
These equalities show that all sectors projecting onto the same denominator
family produce the same complete numerator.  The
crossed members obey the same equalities in the crossed routing.  Together with
the octagon pole collection, these results supply a single numerator for every
occurrence in each denominator-family row of
\eqref{eq:g6_sector_family_collection}.

Let
\(T=[12][34]/(\langle12\rangle\langle34\rangle)\).  Applying the FTT to the
collected terms preserves these numerator factors and gives directly the three
terms of the known two-loop representation~\cite{Bern:2000dn,Bern:2002tk}:
\begin{align}
{\cal I}_{\rm P}
&=
T
\int \frac{d^Dp}{(2\pi)^D}\frac{d^Dq}{(2\pi)^D}\,
\frac{{\cal N}_{\rm DB}(p,q)}
{D_1D_2D_3D_4D_5D_6D_7},
\\
{\cal I}_{\rm NP}
&=
T
\int \frac{d^Dp}{(2\pi)^D}\frac{d^Dq}{(2\pi)^D}\,
\frac{{\cal N}_{\rm DB}^{\rm crossed}(p,q)}
{\widetilde D_1\widetilde D_2\widetilde D_3
\widetilde D_4\widetilde D_5\widetilde D_6\widetilde D_7},
\\
{\cal I}_{\rm BT}
&=
T
\int \frac{d^Dp}{(2\pi)^D}\frac{d^Dq}{(2\pi)^D}\,
\frac{{\cal N}_{\rm BT}(p,q)}
{D_1D_2D_3D_4D_5D_6}.
\label{eq:g6_planar_final}
\end{align}
Since the external momenta have no transverse components, the momentum shifts
in the crossed denominator family \eqref{eq:g6_nonplanar_denominators} leave the
three transverse line momenta \(\lambda_p\), \(\lambda_q\), and
\(\lambda_{p+q}\) unchanged.  Hence
\({\cal N}_{\rm DB}^{\rm crossed}(p,q)\) has exactly the polynomial form
\eqref{eq:g6_octagon_explicit_DB_factors}; the superscript only indicates that
it multiplies the crossed denominators \(\widetilde D_i\).  The common overall
phase relation is \((-i){\cal A}^{\rm vp}_{G}={\cal I}_{G}\).  Thus the
preceding sector calculation reproduces the planar double box, crossed double
box, and bow tie, including their complete numerator factors.

With the coupling and color factors still stripped, the representative bare
primitive amplitudes are therefore
\begin{equation}
\begin{aligned}
A_{1234}^{{\rm P},{\rm bare}}
&=i\bigl({\cal I}_{\rm P}+{\cal I}_{\rm BT}\bigr),
\\
A_{12;34}^{{\rm NP},{\rm bare}}
&=i\,{\cal I}_{\rm NP}.
\end{aligned}
\label{eq:g6_primitive_final}
\end{equation}
Thus the bow-tie contribution belongs to the planar primitive; it is not a
third primitive to be added with the non-planar one.  Restoring the standard
parent-graph color factors gives the bare color-dressed amplitude.  Explicitly,
if \(a_i\) is the adjoint color index of external leg \(i\), repeated internal
indices \(b_1,\ldots,b_7\) are summed, and
\(\widetilde f^{abc}=i\sqrt{2}f^{abc}
=\operatorname{Tr}([T^a,T^b]T^c)\), the vertex orientations may be fixed by
\begin{equation}
\begin{aligned}
C^{\rm P}_{ijkl}
&=
\widetilde f^{a_i b_1 b_2}
\widetilde f^{a_j b_3 b_1}
\widetilde f^{b_2 b_4 b_5}
\widetilde f^{b_3 b_6 b_4}
\widetilde f^{b_5 a_l b_7}
\widetilde f^{b_6 b_7 a_k},
\\
C^{\rm NP}_{ij;kl}
&=
\widetilde f^{a_i b_1 b_2}
\widetilde f^{a_j b_3 b_4}
\widetilde f^{b_1 b_5 b_3}
\widetilde f^{b_4 b_2 b_6}
\widetilde f^{b_5 a_l b_7}
\widetilde f^{b_6 b_7 a_k}.
\end{aligned}
\label{eq:g6_explicit_color_factors}
\end{equation}
In these conventions the final result is
\begin{equation}
\begin{aligned}
{\cal A}_{4,{\rm bare}}^{(2)}(1^+,2^+,3^+,4^+)
=g^6\sum_{\substack{(a,b,c)\in\\
\{(2,3,4),(3,4,2),(4,2,3)\}}}\Bigl[{}
&C^{\rm P}_{1abc}A^{{\rm P},{\rm bare}}_{1abc}
+C^{\rm P}_{bca1}A^{{\rm P},{\rm bare}}_{bca1}
\\
&+C^{\rm NP}_{1a;bc}A^{{\rm NP},{\rm bare}}_{1a;bc}
+C^{\rm NP}_{bc;a1}A^{{\rm NP},{\rm bare}}_{bc;a1}
\Bigr].
\end{aligned}
\label{eq:g6_integrand_final}
\end{equation}
For \((a,b,c)=(2,3,4)\) the summand gives the four representative terms;
the other two displayed triples give the two non-trivial cyclic relabellings
explicitly.  In this organization the bow-tie terms inherit the planar parent color factor; their
proper color dressing is obtained only after the complete permutation sum.
The relabelled primitive amplitudes and their vacuum-pair collections are
obtained from the displayed representatives by the indicated external
permutations.

Equations \eqref{eq:g6_primitive_final} and \eqref{eq:g6_integrand_final} are
bare, dimensionally regulated expressions.  A renormalized amplitude requires
the ultraviolet counterterm appropriate to the chosen scheme; in the all-plus
case it is proportional to the one-loop amplitude.  That counterterm is not
part of the vacuum-pair sector calculation performed here.

\section{Discussion and outlook}
\label{sec:discussion}

The main result of this paper is twofold.  First, we have formulated a
fixed-order on-shell setup in which higher-order contributions are built by
inserting unobservable on-shell vacuum pairs into ordinary tree amplitudes
generated recursively from three-point data.  At fixed order, the organizing
principle is simply to sum all possible tree amplitudes with all allowed
vacuum-pair insertions, with each vacuum-pair momentum integrated over
Lorentz-invariant phase space.

A convenient way to organize this sum is to represent the tree amplitudes as
polygons.  The number of vacuum pairs is fixed by the polygon bookkeeping, and
a forest with \(k\) tree factors is assigned the alternating
inclusion--exclusion sign \((-1)^{k-1}\).  For the corresponding ordinary
Feynman graphs, this inclusion--exclusion sign is exactly the alternating sign
generated by the FTT propagator identity.  Contributions with the same
unlabelled support are combined only at the end; the sign above applies to each
labelled forest separately.

Second, this framework has been tested in a non-trivial two-loop example.  The
four-gluon all-plus primitive amplitudes provide a sufficiently stringent test:
the ordinary tree amplitude vanishes, the rational one-loop contribution is
entirely tied to \(\lambda\)-dependent components in dimensional regularization, and the
bare order-\(g^6\) result contains planar and non-planar primitive amplitudes,
with the bow-tie family forming part of the planar primitive.  The calculation
above shows how these structures emerge
from the vacuum-pair tree products themselves.  Only after the sector analysis
and the state sums have been carried out are the signed on-shell phase-space
sources compared with the corresponding denominator openings and identified with the
standard Feynman denominator families.

The agreement holds already at the level of integrands and denominator
supports, before integration.  At order
\(g^4\) the one-pair hexagon and the two-quadrilateral sector combine into the
standard scalar box with numerator \((D_s-2)\lambda_\ell^4\).  At order \(g^6\)
the four sectors considered are \(\{8\}\), \(\{6,4\}\), \(\{5,5\}\), and
\(\{4,4,4\}\).  The octagon sector \(\{8\}\) supplies the two-cut seven-slot
and local bow-tie supports together with the corresponding numerator factors.
The \(\{6,4\}\) sector provides a
\(D_7\)-pole projection which belongs to the seven-propagator double-box family,
whereas its local projection without a \(D_7\) pole belongs to the six-propagator
bow-tie family.  The \(\{5,5\}\) sector supplies three-cut sources, and the
\(\{4,4,4\}\) sector supplies both the four-cut double-box data and the local
bow-tie projection.  After the complete sums over physical states and sectors,
numerator preservation under the FTT identifies these
support collections with the numerator factors and denominator families of the
standard bare planar and non-planar primitive representation.  Their distinct
color factors and external permutations are assembled only in the color-dressed amplitude
\eqref{eq:g6_integrand_final}.
The displayed assignment tables provide selected state-contraction and vanishing
checks; the exhaustive denominator-support collections are displayed in the
four sector subsections and assembled in the final comparison.

This makes the present example a non-trivial test of the proposal.  The support
calculation starts from on-shell tree amplitudes and complete physical state
sums; it does not assume any two-loop graph topologies.  The
planar double-box family, the crossed double-box family, and the bow-tie family
appear only after the non-vanishing vacuum-pair sectors have been identified
and matched to denominator families.  Conversely, the sector analysis also
removes
the assignments that would otherwise spoil this denominator comparison:
they vanish by helicity selection, by an unpaired transverse index, by vacuum or
lower-topology factorization, or by degenerate phase-space support.

The particle spectrum, admissible three-point couplings, and color organization are assumed as input.
BCFW recursion is used in the examples only for tree factors for which the required shifts have
the appropriate large-\(z\) behavior.  Forward limits are understood with the
regulator and prescriptions stated in section~\ref{sec:const}.  Under these
assumptions, the present work shows that the all-plus result through order
\(g^6\) can be obtained without using gauge-fixed off-shell diagrams as
intermediate building blocks, with the order-\(g^6\) result understood as the
bare, dimensionally regulated amplitude.  Ultraviolet renormalization requires
the additional scheme-dependent counterterm stated after
\eqref{eq:g6_integrand_final}.

In this example, the construction also clarifies the relation to the usual ghost
contributions.  In a covariant Feynman-diagram calculation, ghost graphs appear
because the intermediate description uses gauge-fixed off-shell fields.  Since
only physical on-shell states are sewn here, no separate ghost states appear in
the construction.  The agreement found above is therefore a consistency check
against the gauge-fixed representation, where the corresponding ghost
contributions are already included in the standard result.

The result also clarifies the relation to other on-shell loop constructions.
Generalized unitarity, integrand reduction, and \(Q\)-cuts likewise express
loop information in terms of tree-level or on-shell building blocks
\cite{Bern:1994zx,Bern:1994cg,Bern:2007dw,Mastrolia:2011pr,Caron-Huot:2010fvq,Baadsgaard:2015twa}.
The present construction can be compared more specifically with two
strategies for organizing loop amplitudes in terms of on-shell trees.
Loop--tree duality~\cite{Catani:2008xa,Bierenbaum:2010cy} starts from a known
off-shell Feynman integrand and reorganizes its loop-energy residues into a
smaller set of terms with dual propagators and modified causal prescriptions.
Because this relation follows directly from the residue theorem, it requires
no additional physical ansatz.
Ref.~\cite{Caron-Huot:2010fvq} instead uses causal boundary conditions to derive
an \(\ell\)-loop representation involving exactly \(\ell\) on-shell
particles.  Already at two loops, however, the naive identification of its
integrand with the forward limit of a physical tree amplitude conflicts with
unitarity.  The corrected generic formula contains a non-physical combination
of connected trees.  In planar gauge theories special color correlations permit
its reorganization into a physical forward amplitude.
Beyond the planar limit, the same work argues against a direct
Minkowski-space phase-space representation in terms of physical forward
amplitudes and instead conjectures that the loop amplitude is determined
implicitly by physical forward-tree data.  BCFW-type deformations enter
separately in the factorization of forward amplitudes in supersymmetric
theories, illustrated there with planar \(\mathcal N=4\) examples.

The present construction avoids this obstruction by retaining the complete
FTT cut sum in the vacuum-pair organization of
subsection~\ref{subsec:insertions}.  Unitarity is inherited
from the general equivalence shown in subsection~\ref{subsec:equivalence},
rather than being imposed or repaired separately in each sector.  Every tree
factor in every forest is generated recursively and on shell by BCFW, with any
required boundary contribution included.  This applies to the planar and
non-planar primitives and to the bow-tie denominator family within the planar
primitive; no off-shell integrand or color-dressed ansatz is required by the
construction.  The non-planar, non-supersymmetric two-loop example in
section~\ref{sec:g6_allplus_revised} is therefore not a realization of the
representation considered in Ref.~\cite{Caron-Huot:2010fvq}; it is
a check of a different, full-cut construction.  It reproduces the known result,
order by order in the sector bookkeeping, at the cost of the larger, but fully
controlled and combinatorially finite, cut sum used throughout this paper.

Recent developments in related approaches include four-dimensional unsubtraction, where
real and virtual singularities are combined locally at the integrand
level~\cite{Sborlini:2016fcj}, causal and manifestly causal LTD
representations~\cite{Aguilera-Verdugo:2020kzc,Capatti:2020ytd,Aguilera-Verdugo:2021nrp},
and local-unitarity formulations of differential cross sections~\cite{Capatti:2020xjc}.
The recent vacuum-amplitude implementation of causal-unitary LTD is especially
close in spirit to the present work, because it also treats vacuum contributions as
organizing kernels for physical observables~\cite{Ramirez-Uribe:2024cwd,Ramirez-Uribe:2024tom}.

The four-gluon all-plus amplitude provides a compact but nontrivial two-loop
benchmark: its external multiplicity and helicity structure are simple, while
its known Feynman-diagram representation contains the planar, non-planar, and
bow-tie denominator families needed to test the vacuum-pair construction.  It
belongs to a broader line of analytic all-plus calculations.  The one-loop
finite rational amplitudes were worked out in
\cite{Mahlon:1993si,Bern:1994zx,Bern:1994cg}, while the two-loop four-gluon
case and the complete two-loop helicity amplitudes are given in
\cite{Bern:2000dn,Bern:2002tk}.  Higher-multiplicity all-plus results and
modern analytic representations have been obtained in
\cite{Dunbar:2016aux,Dunbar:2016gjb,Gehrmann:2015bfy,Dunbar:2017nfy,Badger:2019djh,Dalgleish:2020mof}.

The next tests are other helicity sectors, higher multiplicities, and theories
with matter or massive states.  In these cases the same questions arise: which
vacuum-pair sectors survive, whether the prescribed signs agree with the
corresponding analytic cut expansion, and whether the resulting support sums
match the known loop representations.

\acknowledgments

I thank Pierpaolo Mastrolia for valuable discussions.  This work was supported
by ANID through FONDECYT Regular Project No.~1250132.  The author used ChatGPT
version 5.6 (OpenAI) for language editing and for bookkeeping assistance with
lengthy expressions.  All scientific derivations and calculations were carried
out by the author, who also verified all equations and takes full responsibility
for the article.

\appendix

\section{Transverse momentum and state-sum conventions}
\label{app:transverse_conventions}

This appendix fixes the notation for the transverse momentum variables and for
the state sums used in the fixed-order calculations.  We first spell out the
split of a dimensionally regulated momentum into four-dimensional and transverse
parts.  We then state the convention for complete \(D_s\)-dimensional gluon
state sums and finally introduce the auxiliary scalar-chain labels used in the
\(\lambda\)-dependent projections.

The loop and phase-space momenta are integrated in
\begin{equation}
        D = 4-2\epsilon
\end{equation}
dimensions.  The spin-state dimension is denoted by \(D_s\).  It may be identified
with \(D\), as in conventional dimensional regularization, or kept independent, as
in the usual scheme-dependent bookkeeping of gluon state sums.  The complete
physical gluon state space in spin dimension \(D_s\) has \(D_s-2\) states. A \(D\)-dimensional internal momentum \(R^\mu\) is decomposed into its
four-dimensional projection and its transverse part,
\begin{equation}
        R^\mu = \bar R^\mu + R_\perp^\mu .
\end{equation}
In most formulae the bar on the four-dimensional projection is suppressed.  Thus,
when a denominator is written in routed variables, an unbarred square such as
\(R^2\) denotes the square of the four-dimensional projection, whereas the full
\(D\)-dimensional invariant is
\begin{equation}
        R_D^2 = R^2 + R_\perp^2 = R^2-\lambda_R^2 .
\label{eq:full_invariant_projected_notation}
\end{equation}
Here
\begin{equation}
        \lambda_R^2 := - R_\perp^2 .
\end{equation}
Equivalently, in an orthonormal basis of the transverse momentum space we write
\begin{equation}
        \lambda_R^a := (R_\perp)^a ,
        \qquad
        \lambda_R^2 = \delta_{ab}\lambda_R^a\lambda_R^b ,
        \qquad
        \lambda_R\!\cdot\!\lambda_S
        =\delta_{ab}\lambda_R^a\lambda_S^b .
\end{equation}
The indices \(a,b,\ldots\) label components of the transverse momentum space.
Their formal dimension is
\begin{equation}
        d_\lambda := D-4 = -2\epsilon .
\end{equation}
The observed external momenta \(k_i\) are taken to be four-dimensional,
\begin{equation}
        (k_i)_\perp=0 .
\end{equation}
Consequently,
\begin{equation}
        \lambda_{R-k_i}^a = \lambda_R^a ,
        \qquad
        \lambda_{P+Q}^a = \lambda_P^a+\lambda_Q^a ,
\end{equation}
and hence
\begin{equation}
        \lambda_{P+Q}^2
        =
        \lambda_P^2+2\lambda_P\!\cdot\!\lambda_Q+\lambda_Q^2 .
\end{equation}
The complete on-shell state sum over an internal gluon is denoted by
\begin{equation}
        \sum_{h\in{\rm phys}(D_s)} .
\end{equation}
This sum runs over the complete physical polarization space in spin dimension
\(D_s\), and therefore has trace
\begin{equation}
        \sum_{h\in{\rm phys}(D_s)} 1 = D_s-2 .
\end{equation}
When such a state appears in a BCFW factorization channel, the symbol
\(h\in{\rm phys}(D_s)\) always denotes this complete physical state sum.
Whenever both oriented half-edges of a sewn line are displayed, we write them
as \((-P)^{\bar h}\) and \(P^h\), whether the line originates from a vacuum
pair or from a BCFW factorization channel.  The sum is only over \(h\); the
paired state \(\bar h\) is fixed by \(h\) and is not an independent summation
label.  In intermediate tensor contractions a bar may be suppressed when the
opposite momentum arguments already make the pairing explicit.  This is only a
notational abbreviation and does not change the state sum.

We evaluate the tree amplitudes in dimensional regularization while keeping the
observed external gluons four-dimensional.  Consequently, only the unobserved
vacuum-pair legs and the internal BCFW factorization states are kept
\(D\)-dimensional, with complete \(D_s\)-dimensional gluon states.  For an
ordered set of observed gluons \(i_1,\ldots,i_n\), we encounter gluon-tree amplitudes of the form
\begin{equation}
        A_{n+2}^{(0)}
        \bigl(P^{h_P},i_1^{\pm},\ldots,i_n^{\pm},Q^{h_Q}\bigr),
        \qquad h_P,h_Q\in{\rm phys}(D_s),
\label{eq:complete_dimensionally_regulated_tree}
\end{equation}
where \(P\) and \(Q\) are the \(D\)-dimensional on-shell endpoint momenta and
the gluons \(i_r\) have four-dimensional momenta and helicities.  The labels
\(h_P\) and \(h_Q\) belong to the complete physical \(D_s\)-dimensional gluon
state space.  BCFW factorization reconstructs
\eqref{eq:complete_dimensionally_regulated_tree} with the complete
\(D_s\)-dimensional state sum on every internal channel.

When the endpoint states in
\eqref{eq:complete_dimensionally_regulated_tree} are projected onto the
transverse component and the projected endpoint labels are kept open, we denote
the resulting component by
\begin{equation}
        A_{n+2}^{(0)}
        \bigl(P^I,i_1^{\pm},\ldots,i_n^{\pm},Q^J\bigr).
\label{eq:scalar_chain_definition}
\end{equation}
The propagation of the open projected labels from \(I\) to \(J\) through the
ordered gluon insertions is called a scalar chain.  It is a component of the
complete pure-gluon tree amplitude, not a separate scalar amplitude.  In this
projection, labels \(I,J,\ldots\) denote the corresponding
components of internal BCFW factorization states.
All these labels belong to a scalar-chain state space distinct from the
transverse momentum space with coordinate indices \(a,b,\ldots\); the
scalar-chain metric is \(\delta^{IJ}\).

At every scalar--gluon--scalar vertex this metric transports the same auxiliary
label along the chain, while an internal BCFW label is summed when two
subamplitudes are sewn.  For example, in
\(A_4^{(0)}(P^I,i^{h_i},j^{h_j},Q^J)\) the chain enters at \(P^I\), passes
through the ordered gluon insertions \(i^{h_i},j^{h_j}\), and exits at
\(Q^J\).  Sewing two such open chains at their endpoints produces the trace
\begin{equation}
        \delta^{IJ} \delta^{JI}
        =
        \delta^{II}
        =
        D_s-2 .
\label{eq:traceIJ}
\end{equation}
Thus a closed scalar-chain trace records the factor \(D_s-2\) inherited from
the complete physical polarization projector in this projected algebra.  It
is a contraction rule for the projected numerator, not the dimension of a
separate set of \(D_s-4\) scalar polarizations.

Some projected three-point tensors carry explicit transverse momenta on
scalar-chain indices.  To make this notation unambiguous, we introduce an
algebraic map \({\cal E}^{I}{}_{a}\) and define
\begin{equation}
        \lambda_R^I:={\cal E}^{I}{}_{a}\lambda_R^a,
        \qquad
        \delta_{IJ}{\cal E}^{I}{}_{a}{\cal E}^{J}{}_{b}
        =\delta_{ab}.
\label{eq:scalar_chain_embedding}
\end{equation}
It follows that
\begin{equation}
        \delta_{IJ}\lambda_R^I\lambda_S^J
        =\lambda_R\!\cdot\!\lambda_S .
\label{eq:scalar_chain_mapped_dot_product}
\end{equation}
The image of the transverse momentum space in the scalar-chain algebra is
described by
\begin{equation}
        \Pi_\lambda^{IJ}
        := {\cal E}^{I}{}_{a}{\cal E}^{J}{}_{a},
        \qquad
        \delta_{IJ}\Pi_\lambda^{IJ}=D-4 .
\label{eq:scalar_chain_image_projector}
\end{equation}
This image projector is not identified with the scalar-chain identity.
Equivalently, no converse completeness relation
\({\cal E}^{I}{}_{a}{\cal E}^{J}{}_{a}=\delta^{IJ}\) is imposed.  The two
traces \(\delta^{II}=D_s-2\) and \(\delta^{aa}=D-4\) therefore remain
independent.  A closed scalar-chain index loop produces the former, whereas a
pair of explicit mapped momenta produces
\(\delta_{IJ}\lambda_R^I\lambda_S^J
=\lambda_R\!\cdot\!\lambda_S\), with no extra factor of \(D_s-2\).

When a scalar-chain index is connected to a complete gluon Lorentz index, we
write the mixed conversion tensor as \({\cal P}^{I\mu}\), with the contraction
rules
\begin{equation}
        {\cal P}^{I\mu}{\cal P}^{J}{}_{\mu}=\delta^{IJ},
        \qquad
        {\cal P}^{I}{}_{\mu}V^\mu
        ={\cal E}^{I}{}_{a}\lambda_V^a,
\label{eq:scalar_chain_mixed_contractions}
\end{equation}
whenever the second contraction selects the transverse part of the
\(D\)-dimensional momentum \(V\).  The Euclidean sign convention for
\(\lambda_V\) is the one in
\eqref{eq:full_invariant_projected_notation}; the corresponding transverse
metric sign is absorbed into the definition of \({\cal P}^{I\mu}\).
The equations above are dimensionally continued contraction rules; they do not assume a
finite-dimensional embedding when \(D\) and \(D_s\) are kept independent.
They keep separate the transverse momentum indices \(a,b,\ldots\), the
complete-state labels \(h\in{\rm phys}(D_s)\), and the scalar-chain labels
\(I,J,\ldots\).

For example, a scalar-chain four-point amplitude has the form
\begin{equation}
        A_4^{(0)}
        \bigl(P^I,i^+,j^+,Q^J\bigr),
\end{equation}
where \(I,J\) are scalar-chain state labels.  The transverse momentum dependence
enters only through invariants such as \(\lambda_P^2\),
\(\lambda_Q^2\), and \(\lambda_P\!\cdot\!\lambda_Q\).  A typical contraction is
therefore of the form
\begin{equation}
        \bigl(\lambda_A\!\cdot\!\lambda_B\bigr)\delta^{IJ}
        \,
        \bigl(\lambda_C\!\cdot\!\lambda_D\bigr)\delta^{JI}
        =
        (D_s-2)
        \bigl(\lambda_A\!\cdot\!\lambda_B\bigr)
        \bigl(\lambda_C\!\cdot\!\lambda_D\bigr).
\end{equation}
For the one-loop routing used in section~\ref{sec:g4_allplus}, this gives
\begin{equation}
        \bigl(\lambda_{-\ell}\!\cdot\!\lambda_{\ell-k_{12}}\bigr)
        \delta^{IJ}
        \,
        \bigl(\lambda_{k_{12}-\ell}\!\cdot\!\lambda_{\ell}\bigr)
        \delta^{JI}
        =
        (-\lambda_\ell^2)(-\lambda_\ell^2)
        (D_s-2)
        =
        (D_s-2)\lambda_\ell^4 .
\end{equation}
This is the origin of the factor \((D_s-2)\lambda_\ell^4\) in the one-loop
all-plus numerator.

\section{Scalar four-point tree factors}
\label{app:scalar_trees}

Here we compute the color-ordered tree amplitude with two scalars and two
gluons, keeping its scalar-chain state labels open.  This is the building block used for the
\(\lambda\)-dependent projection of a complete \(D_s\)-dimensional gluon state
sum.  These are auxiliary scalar-chain state labels: the scalar--gluon vertex
transports the same label along the scalar line.

The quantity to be computed is
\begin{equation}
A_4^{(0)}(P^I,i^+,j^+,Q^J).
\label{eq:app_A4_open_object}
\end{equation}
We take all momenta outgoing and write
\begin{equation}
P+k_i+k_j+Q=0,
\qquad
K=P+k_i=-(Q+k_j).
\label{eq:app_A4_momentum_conservation}
\end{equation}
For a \([i,j\rangle\) shift of the two positive-helicity gluons there is one
scalar-chain factorization channel.  BCFW recursion gives
\begin{equation}
A_4^{(0)}(P^I,i^+,j^+,Q^J)
=
A_3^{(0)}(P^I,\hat i^+,-\hat K^L)\,
iG_F(K^2)\,
A_3^{(0)}(\hat K^L,\hat j^+,Q^J),
\label{eq:app_A4_BCFW_representation}
\end{equation}
where the internal scalar-chain state \(L\) is summed.  There is no additional
\(\delta^{IJ}\) in front of the recursive representation; diagonality in the
scalar-chain labels is produced by the two three-point vertices.

The scalar--gluon--scalar three-point amplitude is
\begin{equation}
A_3^{(0)}(1^I,2^+,3^J)
=
i\,\delta^{IJ}\,
\frac{\langle \xi|k_1|2]}{\langle \xi\,2\rangle}.
\label{eq:app_A3_open}
\end{equation}
The reference spinor \(\xi\) drops out on the three-point support.  In this
appendix, a bare propagator argument such as \(K^2\) in
\eqref{eq:app_A4_BCFW_representation} denotes the full \(D\)-dimensional
invariant \(K_D^2\) of \eqref{eq:full_invariant_projected_notation}.  In spinor
sandwiches the same momentum symbol denotes its four-dimensional projection;
thus \(\bar P^2=\lambda_P^2\) on the external scalar shell.

The hatted momenta in
\eqref{eq:app_A4_BCFW_representation} are generated by the
\([i,j\rangle\) shift
\begin{equation}
|\hat i]=|i]-z|j],
\qquad
|\hat j\rangle=|j\rangle+z|i\rangle,
\label{eq:app_A4_BCFW_shift}
\end{equation}
with the other spinors unchanged.  Choosing \(\xi=\hat j\) in the left
three-point factor and \(\xi=\hat i\) in the right one gives
\begin{equation}
A_3^{(0)}(P^I,\hat i^+,-\hat K^L)
=
i\,\delta^{IL}\,
\frac{\langle \hat j|P|\hat i]}{\langle \hat j\,\hat i\rangle},
\qquad
A_3^{(0)}(\hat K^L,\hat j^+,Q^J)
=
-i\,\delta^{LJ}\,
\frac{\langle \hat i|Q|\hat j]}{\langle \hat i\,\hat j\rangle}.
\label{eq:app_A4_shifted_A3_factors}
\end{equation}
The minus sign in the second factor follows from
\(\hat K=-(Q+\hat k_j)\).  At the BCFW pole,
\begin{equation}
(\bar P+\hat k_i)^2-\lambda_P^2=0,
\qquad
\bar P^2=\lambda_P^2,
\end{equation}
and hence
\begin{equation}
\langle \hat i|P|\hat i]=0.
\label{eq:app_A4_pole_spinor_zero}
\end{equation}
Momentum conservation gives
\begin{equation}
\langle \hat i|Q|\hat j]
=-\langle \hat i|P|\hat j].
\label{eq:app_A4_QP_spinor_relation}
\end{equation}
We use the determinant identity for a \(2\times2\) bispinor \(P\),
\begin{equation}
\langle a|P|b]\langle c|P|d]
-\langle a|P|d]\langle c|P|b]
=P^2\langle ac\rangle[b d].
\label{eq:app_bispinor_determinant_identity}
\end{equation}
This identity follows from multiplying
\begin{equation}
P_{\alpha\dot\alpha}P_{\beta\dot\beta}
-P_{\alpha\dot\beta}P_{\beta\dot\alpha}
=P^2\epsilon_{\alpha\beta}\epsilon_{\dot\alpha\dot\beta}
\label{eq:app_bispinor_antisym_identity}
\end{equation}
by \(a^\alpha c^\beta b^{\dot\alpha}d^{\dot\beta}\).
With \(a=\hat j\), \(b=\hat i\), \(c=\hat i\), \(d=\hat j\), this gives
\begin{equation}
\frac{\langle \hat j|P|\hat i]}{\langle \hat j\,\hat i\rangle}
\frac{\langle \hat i|Q|\hat j]}{\langle \hat i\,\hat j\rangle}
=
-\,\lambda_P^2\,\frac{[ij]}{\langle ij\rangle}
=
(\lambda_P\!\cdot\!\lambda_Q)\,
\frac{[ij]}{\langle ij\rangle}.
\label{eq:app_A4_spinor_sandwich_product}
\end{equation}
In the first equality we used
\(\langle\hat i\,\hat j\rangle=\langle i\,j\rangle\),
\([\hat i\,\hat j]=[i\,j]\), and
\eqref{eq:app_A4_pole_spinor_zero}.  In the last equality we used transverse
momentum conservation, \(\lambda_Q=-\lambda_P\), for the all-outgoing
four-point tree.  Multiplying the spinor and species parts gives
\begin{equation}
A_3^{(0)}(P^I,\hat i^+,-\hat K^L)\,
A_3^{(0)}(\hat K^L,\hat j^+,Q^J)
=
\frac{[ij]}{\langle ij\rangle}\,
(\lambda_P\!\cdot\!\lambda_Q)\delta^{IJ}.
\label{eq:app_A4_A3_product_simplified}
\end{equation}
Substitution into the recursive representation
\eqref{eq:app_A4_BCFW_representation} gives the open-index tensor
\begin{equation}
A_4^{(0)}(P^I,i^+,j^+,Q^J)
=
i\,\frac{[ij]}{\langle ij\rangle}\,
G_F\!\bigl((P+k_i)^2\bigr)\,
(\lambda_P\!\cdot\!\lambda_Q)\delta^{IJ}.
\label{eq:app_A4_open_PQ_formula}
\end{equation}
Contracting the open indices
with \(\delta^{IJ}\) gives the scalar-state trace
\begin{equation}
\delta^{IJ}\,
A_4^{(0)}(P^I,i^+,j^+,Q^J)
=
i\,\frac{[ij]}{\langle ij\rangle}\,
G_F\!\bigl((P+k_i)^2\bigr)\,
(D_s-2)\,\lambda_P\!\cdot\!\lambda_Q .
\label{eq:A4_scalar_pp_general}
\end{equation}

Equation~\eqref{eq:g6_64_reduced_phase_space} contains the following amplitude
for the bridge-slot \(\{6,4\}\) representative of
section~\ref{sec:g6_allplus_revised}:
\begin{equation}
A_4^{(0)}\bigl((-p)^I,1^+,2^+,(p-k_{12})^J\bigr).
\end{equation}
The scalar-chain momenta through the two positive-helicity gluons are
\begin{equation}
p,\qquad p-k_1,\qquad p-k_{12},
\end{equation}
so the propagator in \eqref{eq:app_A4_open_PQ_formula} is
$G_F\bigl((p-k_1)^2-\lambda_p^2\bigr)=G_2$.
For the all-outgoing endpoints \(P=-p\) and \(Q=p-k_{12}\), we have
$\lambda_P=-\lambda_p$, $\lambda_Q=\lambda_p$, and
$\lambda_P\!\cdot\!\lambda_Q=-\lambda_p^2$.
Therefore
\begin{equation}
A_4^{(0)}\bigl((-p)^I,1^+,2^+,(p-k_{12})^J\bigr)
=
-i\,\frac{[12]}{\langle12\rangle}\,
G_2\,\lambda_p^2\,\delta^{IJ}.
\label{eq:app_A4_g6_64_left}
\end{equation}
The right scalar chain is obtained from the cyclic form
\begin{equation}
A_4^{(0)}\bigl((q-k_{34})^K,3^+,4^+,(-q)^L\bigr).
\end{equation}
Here
$\lambda_{q-k_{34}}=\lambda_q$,
$\lambda_{-q}=-\lambda_q$,
$\lambda_{q-k_{34}}\!\cdot\!\lambda_{-q}=-\lambda_q^2$,
and the scalar propagator is \(G_5\).  Thus
\begin{equation}
A_4^{(0)}\bigl((q-k_{34})^K,3^+,4^+,(-q)^L\bigr)
=
- i\,\frac{[34]}{\langle34\rangle}\,
G_5\,\lambda_q^2 \, \delta^{KL}.
\label{eq:app_A4_g6_64_right}
\end{equation}
We see that the scalar-chain contractions in both side four-point trees are
diagonal in their scalar-chain labels.  They contribute scalar factors
proportional to \(\lambda_p^2\delta^{IJ}\) and
\(\lambda_q^2\delta^{KL}\).  The complete
\(\{4,4,4\}\) double-box slot projection requires, in addition, the non-scalar components of
the same side four-point trees; these are restored in
appendix~\ref{app:444_bowtie_contraction}.

\section{One vacuum-pair scalar chain}
\label{app:one_pair_scalar_chain}

This appendix records the scalar-chain part of a one-vacuum-pair hexagon.
Consider the color-ordered six-point tree
\begin{equation}
H_6^{IJ}(P;i,j,k,l;Q)
:=
A_6^{(0)}(P^I,i^+,j^+,k^+,l^+,Q^J),
\label{eq:app_one_pair_H6_definition}
\end{equation}
with all momenta outgoing,
\begin{equation}
P+k_i+k_j+k_k+k_l+Q=0 .
\end{equation}
The observed gluons carry no transverse momentum.  We isolate the scalar-chain
component in which the scalar line is split between the two adjacent
positive-helicity pairs and define
\begin{equation}
R:=-(P+k_i+k_j)=Q+k_k+k_l .
\label{eq:app_one_pair_middle_momentum}
\end{equation}
Equivalently, the scalar chain carries the successive momenta
\begin{equation}
P,\qquad P+k_i,\qquad -R=P+k_i+k_j,\qquad
Q+k_l,\qquad Q .
\end{equation}
The propagator arguments in this appendix follow the full-invariant convention
used in appendix~\ref{app:scalar_trees}: \(G_F(R^2)\) means
\(G_F(R_D^2)\).
Inserting a complete scalar-chain state in the channel \(R\) gives
\begin{equation}
\begin{aligned}
H_6^{IJ}(P;i,j,k,l;Q)
&=
\sum_L
A_4^{(0)}(P^I,i^+,j^+,R^L)\,
iG_F(R^2)\,
A_4^{(0)}((-R)^L,k^+,l^+,Q^J).
\end{aligned}
\label{eq:app_one_pair_H6_factorization}
\end{equation}
This is not a new four-scalar contact interaction.  Each four-point factor in
\eqref{eq:app_one_pair_H6_factorization} is the scalar-chain tree of
appendix~\ref{app:scalar_trees}.  Substituting
\eqref{eq:app_A4_open_PQ_formula} into the two factors gives
\begin{equation}
\begin{aligned}
A_4^{(0)}(P^I,i^+,j^+,R^L)
&=
i\,\frac{[ij]}{\langle ij\rangle}\,
G_F\!\bigl((P+k_i)^2\bigr)\,
(\lambda_P\!\cdot\!\lambda_R)\delta^{IL},
\\
A_4^{(0)}((-R)^L,k^+,l^+,Q^J)
&=
i\,\frac{[kl]}{\langle kl\rangle}\,
G_F\!\bigl((-R+k_k)^2\bigr)\,
(\lambda_{-R}\!\cdot\!\lambda_Q)\delta^{LJ}.
\end{aligned}
\label{eq:app_one_pair_inserted_A4}
\end{equation}
The sum over the internal scalar-chain label \(L\) is diagonal.  Hence
\begin{equation}
\begin{aligned}
H_6^{IJ}(P;i,j,k,l;Q)
&=
-i\,\frac{[ij][kl]}{\langle ij\rangle\langle kl\rangle}\,
G_F\!\bigl((P+k_i)^2\bigr)\,
G_F(R^2)\,
G_F\!\bigl((-R+k_k)^2\bigr)
\\
&\qquad\times
(\lambda_P\!\cdot\!\lambda_R)
(\lambda_{-R}\!\cdot\!\lambda_Q)
\delta^{IJ}.
\end{aligned}
\label{eq:app_one_pair_H6_general_result}
\end{equation}
For the order-\(g^4\) one-pair sector in
section~\ref{sec:g4_allplus}, take
\begin{equation}
P=-\ell,\qquad
i=1,\qquad j=2,\qquad k=3,\qquad l=4,\qquad
Q=\ell .
\label{eq:app_one_pair_g4_identification}
\end{equation}
Then \(R=\ell-k_{12}\).  With the box-family denominator routing of
\eqref{eq:box_sec3},
\begin{equation}
G_F\!\bigl((P+k_1)^2\bigr)=G_2,\qquad
G_F(R^2)=G_3,\qquad
G_F\!\bigl((-R+k_3)^2\bigr)=G_4 .
\end{equation}
The transverse momenta are
\begin{equation}
\lambda_P=-\lambda_\ell,\qquad
\lambda_R=\lambda_\ell,\qquad
\lambda_{-R}=-\lambda_\ell,\qquad
\lambda_Q=\lambda_\ell,
\end{equation}
so
\begin{equation}
(\lambda_P\!\cdot\!\lambda_R)
(\lambda_{-R}\!\cdot\!\lambda_Q)
=
(-\lambda_\ell^2)(-\lambda_\ell^2)
=
\lambda_\ell^4 .
\end{equation}
Therefore
\begin{equation}
A_6^{(0)}
\bigl((-\ell)^I,1^+,2^+,3^+,4^+,\ell^J\bigr)
=
-i\,\frac{[12][34]}{\langle12\rangle\langle34\rangle}\,
G_2G_3G_4\,\lambda_\ell^4\,\delta^{IJ}.
\label{eq:app_one_pair_hexagon_g4_result}
\end{equation}
Contracting the scalar-chain state of the vacuum pair gives
\begin{equation}
\delta^{IJ}
A_6^{(0)}
\bigl((-\ell)^I,1^+,2^+,3^+,4^+,\ell^J\bigr)
=
-i\,\frac{[12][34]}{\langle12\rangle\langle34\rangle}\,
G_2G_3G_4\,(D_s-2)\lambda_\ell^4 .
\label{eq:app_one_pair_hexagon_g4_trace}
\end{equation}
Thus the one-pair hexagon has a vanishing strictly four-dimensional
gluon-helicity part but a non-vanishing \(\lambda\)-dependent scalar-chain
projection.  If we set
\(\lambda_\ell=0\) before integration, this contribution disappears
pointwise.  In dimensional regularization \(\lambda_\ell\) is kept through the
integration, and the integral with numerator \(\lambda_\ell^4\) gives the
finite rational all-plus remnant.

\section{Double-box and bow-tie projections}
\label{app:444_bowtie_contraction}

We use the routed variables
\(X_1=k_{12}-p\), \(X_2=p\), \(X_3=q\), and \(X_4=k_{34}-q\).
Thus \(X_1+X_2=k_{12}\), \(X_3+X_4=k_{34}=-k_{12}\), and the bridge momentum is
\(X_2+X_3=p+q\).  With this routing,
\eqref{eq:g6_444_start} and \eqref{eq:444rout} turn the
\(\{4,4,4\}\) representative into the tree product
\begin{equation}
\begin{aligned}
&\sum_{h_1,h_2,h_3,h_4\in{\rm phys}(D_s)}
A_4^{(0)}((-X_2)^{h_2},1^+,2^+,(-X_1)^{h_1})
\\
&\qquad\times
A_4^{(0)}(X_1^{h_1},X_2^{h_2},X_3^{h_3},X_4^{h_4})
\\
&\qquad\times
A_4^{(0)}((-X_4)^{h_4},3^+,4^+,(-X_3)^{h_3}) .
\end{aligned}
\label{eq:app_444_full_tree_product}
\end{equation}
This is the ordering after cyclic rotations of the side trees, chosen
so that the observed gluons \(1^+,2^+\) and \(3^+,4^+\) sit between the two
scalar legs.  The labels \(h_i\) range over the
complete physical \(D_s\)-dimensional on-shell state basis.  Since these are
internal contractions, the sums over the states carried by the four on-shell lines in
\eqref{eq:app_444_full_tree_product} are complete physical-state sums.  The
restriction to scalar-chain labels \(I,J,K,L\) is only one projection.

The contractions in this appendix and in
appendix~\ref{app:adjacent_pair_reduction} are evaluated with the same
algebraic bookkeeping.  We first isolate a scalar-chain reference contraction
and then evaluate the strictly four-dimensional helicity contribution.  The
remaining transverse term is defined as the difference between the full
\(\lambda\)-dependent contraction and the scalar-chain reference contraction.
These are not three disjoint polarization sectors; the subtraction ensures
that the scalar-chain reference term is included only once.  Their stated sum
is the complete projected numerator used in the main text.

We strip the common spinor factors and side propagators from the two side trees
and reattach them after the BCFW decomposition.  We denote the remaining tree
factors by
\begin{equation}
\begin{aligned}
{\cal A}_{12}^{h_2h_1}
&:=
\left[
A_4^{(0)}((-X_2)^{h_2},1^+,2^+,(-X_1)^{h_1})
\right]_{\rm strip},
\\
{\cal A}_{34}^{h_3h_4}
&:=
\left[
A_4^{(0)}((-X_4)^{h_4},3^+,4^+,(-X_3)^{h_3})
\right]_{\rm strip}.
\end{aligned}
\label{eq:app_444_stripped_tree_factors}
\end{equation}
Equivalently, the stripped contraction is
\begin{equation}
\sum_{h_1,h_2,h_3,h_4\in{\rm phys}(D_s)}
{\cal A}_{12}^{h_2h_1}\,
A_4^{(0)}(X_1^{h_1},X_2^{h_2},X_3^{h_3},X_4^{h_4})\,
{\cal A}_{34}^{h_3h_4}.
\label{eq:app_444_stripped_tree_product}
\end{equation}

The two stripped side four-point trees are factorized in the same
complete-state basis:
\begin{equation}
\begin{aligned}
{\cal A}_{12}^{h_2h_1}
&=
\sum_{\sigma\in{\rm phys}(D_s)}
{\cal A}_3^{(0)}((-X_2)^{h_2},\hat{1}^+,-\widehat K_2^{\sigma})\,
{\cal A}_3^{(0)}(\widehat K_{2}^{\sigma},\hat{2}^+,(-X_1)^{h_1}),
\\
{\cal A}_{34}^{h_3h_4}
&=
\sum_{\sigma\in{\rm phys}(D_s)}
{\cal A}_3^{(0)}((-X_4)^{h_4},\hat{3}^+,-\widehat K_5^{\sigma})\,
{\cal A}_3^{(0)}(\widehat K_{5}^{\sigma},\hat{4}^+,(-X_3)^{h_3}),
\end{aligned}
\label{eq:app_444_side_complete_amplitudes}
\end{equation}
with \(K_2=-X_2+k_1\) and \(K_5=X_3-k_4\), and with the \(iG\) factors stripped.
The hatted momenta are evaluated at
the corresponding side BCFW poles.  In each factorized side tree, the same
summed label \(\sigma\) is used on the two oriented appearances of the internal
line.

The middle factor contains no observed gluons.  It is the color-ordered
four-point tree
\(A_4^{(0)}(X_1^{h_1},X_2^{h_2},X_3^{h_3},X_4^{h_4})\)
with complete internal state labels.  We use the adjacent
\((X_3,X_4)\) deformation specified below.  Under this deformation the
channel \(P=X_2+X_3=-(X_4+X_1)\) is a finite BCFW pole and contains the
phase-space-dependent denominator \(D_7=(p+q)^2-\lambda_{p+q}^2\).  The channel
\(X_1+X_2=-(X_3+X_4)=k_{12}\) is independent of the deformation parameter.
The BCFW contour therefore organizes the complete tree as
\begin{equation}
A_4^{(0)}(X_1^{h_1},X_2^{h_2},X_3^{h_3},X_4^{h_4})
=A_{M,7}^{h_1h_2h_3h_4}
+A_{M,s_{12}}^{h_1h_2h_3h_4}.
\label{eq:app_bowtie_BCFW_terms}
\end{equation}
Here \(A_{M,7}\) is the finite-pole residue, whereas
\(A_{M,s_{12}}\) denotes the complete boundary contribution of this
deformation.  The latter contains the fixed \(s_{12}\)-channel factorization
and the associated color-ordered four-point contact completion.  Explicitly,
\begin{equation}
\begin{aligned}
A_{M,7}^{h_1h_2h_3h_4}
&=
\sum_{\sigma\in {\rm phys}(D_s)}
A_3^{(0)}(X_2^{h_2},\widehat X_3^{h_3},-\widehat P^{\sigma})\,
iG_F(P_D^2)\,
A_3^{(0)}(\widehat P^{\sigma},\widehat X_4^{h_4},X_1^{h_1}),
\\
A_{M,s_{12}}^{h_1h_2h_3h_4}
&:=
A_4^{(0)}(X_1^{h_1},X_2^{h_2},X_3^{h_3},X_4^{h_4})
-A_{M,7}^{h_1h_2h_3h_4}.
\end{aligned}
\label{eq:app_bowtie_channels_full}
\end{equation}
The first term contains the additional denominator factor
\(G_F(P_D^2)=G_7\) and therefore contributes to the double-box seven-slot
family.  The second term can be written as
\(A_{M,s_{12}}=iG(s_{12}){\cal A}_{M,s_{12}}\), where the numerator
\({\cal A}_{M,s_{12}}\) includes both the normalized cubic exchange and the
contact completion.  It therefore contributes to the local bow-tie six-slot
family.

The two factorization momenta are
\begin{equation}
P=X_2+X_3=p+q,\qquad
X_1+X_2=k_{12}.
\label{eq:app_444_middle_channel_momenta}
\end{equation}
The propagator argument is the full \(D\)-dimensional channel invariant in the
convention of \eqref{eq:full_invariant_projected_notation}:
\begin{equation}
P_D^2=(p+q)^2-\lambda_{p+q}^2=D_7 .
\label{eq:app_444_D7_invariant}
\end{equation}
Only the four-dimensional components are shifted in the \(D_7\) channel; the
transverse vectors \(\lambda_{X_i}\) are kept fixed.  For this channel we use
the shift
\begin{equation}
\widehat X_3(z)=X_3+z\eta_{34},\qquad
\widehat X_4(z)=X_4-z\eta_{34},\qquad
\eta_{34}^2=\eta_{34}\!\cdot X_3=\eta_{34}\!\cdot X_4=0 .
\label{eq:app_444_D7_shift}
\end{equation}
With
\begin{equation}
P:=X_2+X_3,\qquad
\widehat P(z):=X_2+\widehat X_3(z),
\end{equation}
the pole is at
\begin{equation}
z_P=-\frac{P_D^2}{2\,\eta_{34}\!\cdot X_2},
\qquad
\widehat P_D^2(z_P)=0 .
\label{eq:app_444_D7_pole}
\end{equation}
This is the pole that produces the \(D_7\)-denominator term in
\eqref{eq:app_bowtie_channels_full}, with \(P_D^2\) identified in
\eqref{eq:app_444_D7_invariant}.  The hatted momenta in that term are evaluated
at \(z=z_P\).

The second term is the boundary contribution rather than a shifted pole.
Its factorization momentum is fixed by the routing:
\(X_1+X_2=k_{12}\) and equivalently \(X_3+X_4=-k_{12}\).  The state in its
cubic factorization part carries a scalar-chain label even though its
transverse momentum is zero.  The complete term also contains the local
four-point contribution required by the color-ordered tree amplitude.

The finite-pole term is a contraction of three-point amplitudes.  The boundary
term is evaluated as the complementary part of the complete four-point tree,
including its contact completion.  In both terms the state sum may be
evaluated by separating the four-dimensional helicity part from the remaining
\(\lambda\)-dependent scalar-chain projection.  For the calculation of the
numerator factors we strip off the middle denominators,
\begin{equation}
\begin{aligned}
{\cal A}_{M,7}^{h_1h_2h_3h_4}
&:=
\sum_{\sigma\in {\rm phys}(D_s)}
{\cal A}_3^{(0)}(X_2^{h_2},\widehat X_3^{h_3},
-\widehat P^{\sigma})\,
{\cal A}_3^{(0)}(\widehat P^{\sigma},
\widehat X_4^{h_4},X_1^{h_1}),
\end{aligned}
\label{eq:app_bowtie_channels_stripped}
\end{equation}
Thus \(A_{M,7}=iG_F(P_D^2){\cal A}_{M,7}\).  As above,
\({\cal A}_{M,s_{12}}\) denotes the complete numerator obtained after
factoring \(iG(s_{12})\) from the boundary term.  The complete \(D_7\)-pole
numerator is therefore
\begin{equation}
\begin{aligned}
{\cal N}_{444,7}^{(1346)}
&=
\sum_{h_1,h_2,h_3,h_4}
{\cal A}_{12}^{h_2h_1}\,
{\cal A}_{M,7}^{h_1h_2h_3h_4}\,
{\cal A}_{34}^{h_3h_4}
\\
&=
\sum_{h_1,h_2,h_3,h_4,\sigma}
{\cal A}_{12}^{h_2h_1}\,
{\cal A}_{3}^{(0)}(X_2^{h_2},\widehat X_3^{h_3},
-\widehat P^{\sigma})
{\cal A}_{3}^{(0)}(\widehat P^{\sigma},
\widehat X_4^{h_4},X_1^{h_1})\,
{\cal A}_{34}^{h_3h_4}.
\end{aligned}
\label{eq:app_444_complete_D7_state_sum}
\end{equation}

Before carrying out the contractions needed in
\eqref{eq:app_444_stripped_tree_product}, we collect the three-point numerator
building blocks entering the finite-pole factor
\eqref{eq:app_bowtie_channels_stripped} and the cubic part of
\({\cal A}_{M,s_{12}}\).  We first recall the scalar-chain projection of the
color-ordered Yang--Mills three-gluon vertex.  We use the standard color-ordered three-gluon
normalization of~\cite{Elvang:2013cua, Bern:2007dw} and the dimensional-scalar
vertex structure used in~\cite{Badger:2005zh}.  Its indices are interpreted
here through the auxiliary algebra of
appendix~\ref{app:transverse_conventions}, rather than as a separate set of
\(D_s-4\) asymptotic states.  With all momenta outgoing and
\(\lambda_a+\lambda_b+\lambda_c=0\), it is
\begin{equation}
A_3^{(0)}(a^I,b^J,c^K)
=
\frac{i}{\sqrt{2}}\Big[
\delta^{IJ}(\lambda_a-\lambda_b)^K
+\delta^{JK}(\lambda_b-\lambda_c)^I
+\delta^{KI}(\lambda_c-\lambda_a)^J
\Big] .
\label{eq:app_A3_scalar}
\end{equation}
When one of the three legs retains a complete internal-gluon Lorentz index, the
conversion between that index and a scalar-chain index is represented by
\({\cal P}^{I\mu}\) defined in
appendix~\ref{app:transverse_conventions}.  In the stripped numerator convention,
where the common factor \(i/\sqrt{2}\) of the three-point vertex is removed, the
two mixed-index forms needed below are
\begin{equation}
\begin{aligned}
{\cal A}_{3}^{(0)}(a^I,b^J,c^\mu)
&=
\delta^{IJ}(a-b)^\mu
+{\cal P}^{J\mu}(\lambda_b-\lambda_c)^I
+{\cal P}^{I\mu}(\lambda_c-\lambda_a)^J,
\\
{\cal A}_{3}^{(0)}(a^\mu,b^I,c^J)
&=
{\cal P}^{I\mu}(\lambda_a-\lambda_b)^J
+\delta^{IJ}(b-c)^\mu
+{\cal P}^{J\mu}(\lambda_c-\lambda_a)^I .
\end{aligned}
\label{eq:app_A3_mixed_complete_gluon}
\end{equation}
A leg with a Greek superscript carries the complete gluon index in \(D_s\)
dimensions.  The momentum difference multiplying
\(\delta^{IJ}\) is therefore a full vector in this index space.  The other two
terms use the mixed conversion tensor to retain the projected scalar-chain
index.  All contractions below use the scalar-chain rules stated in
appendix~\ref{app:transverse_conventions}.
When the complete-state leg is instead projected onto a four-dimensional helicity,
we use the same color-ordered three-gluon amplitude.  The two
non-zero scalar--gluon--gluon amplitudes are
\begin{align}
{\cal A}_3^{(0)}(1^+,2^K,3^-)
&=
\frac{\langle r_1\,3\rangle[1\,r_3]}
{\langle r_1\,1\rangle[3\,r_3]}\,
\bigl(\lambda_{k_3}-\lambda_{k_1}\bigr)^K,
\\
{\cal A}_3^{(0)}(1^-,2^K,3^+)
&=
-
\frac{\langle r_3\,1\rangle[3\,r_1]}
{\langle r_3\,3\rangle[1\,r_1]}\,
\bigl(\lambda_{k_3}-\lambda_{k_1}\bigr)^K .
\label{eq:app_444_A3_scalar_gluon_gluon}
\end{align}
The reference spinors \(r_1,r_3\) cancel after the two three-point amplitudes
inside the complete-state sum are multiplied.
The BCFW shifts used change only the four-dimensional components, so the
transverse momenta of the factorization lines are fixed by the unshifted
scalar-line momenta.  In the present routing,
\begin{equation}
\begin{gathered}
\lambda_{X_1}=-\lambda_p,\qquad
\lambda_{X_2}=\lambda_p,\qquad
\lambda_{X_3}=\lambda_q,\qquad
\lambda_{X_4}=-\lambda_q,
\\
\lambda_{\widehat P}=\lambda_P=\lambda_p+\lambda_q,\qquad
\lambda_{-\widehat P}=-(\lambda_p+\lambda_q).
\end{gathered}
\end{equation}

With these preparations we now compute the contraction~\eqref{eq:app_444_stripped_tree_product} in
the scalar-chain projection.  In this projection the side
four-point trees are direct applications of the open-index formula
\eqref{eq:app_A4_open_PQ_formula} of appendix~\ref{app:scalar_trees}.  After
the common factors \([12]G_2/\langle12\rangle\) and
\([34]G_5/\langle34\rangle\) have been stripped off, they give
\begin{equation}
\bigl({\cal A}_{12}\bigr)_{\rm sc}^{IJ}
=
\lambda_p^2\,\delta^{IJ},
\qquad
\bigl({\cal A}_{34}\bigr)_{\rm sc}^{KL}
=
\lambda_q^2\,\delta^{KL}.
\label{eq:app_444_side_scalar_projection}
\end{equation}
The scalar-chain \(D_7\)-channel contraction is therefore obtained by inserting
the two mixed three-point amplitudes from
\eqref{eq:app_A3_mixed_complete_gluon} into the first line of
\eqref{eq:app_bowtie_channels_stripped}.  The quantity computed is the
scalar projection of the stripped middle channel,
\begin{equation}
\begin{aligned}
\delta^{IJ}{\cal A}_{M,7}^{JIKL}\delta^{KL}
&=
\delta^{IJ}\,
{\cal A}_3^{(0)}(X_2^I,\widehat X_3^K,-\widehat P^\mu)\,
{\cal A}_3^{(0)}(\widehat P^\mu,\widehat X_4^L,X_1^J)\,
\delta^{KL}
\\
&=
\delta^{IJ}\delta^{KL}
\Big[
\delta^{IK}(X_2-\widehat X_3)^\mu
+{\cal P}^{K\mu}(\lambda_p+2\lambda_q)^I
-{\cal P}^{I\mu}(2\lambda_p+\lambda_q)^K
\Big]
\\
&\qquad\qquad\times
\Big[
{\cal P}^{L\mu}(\lambda_p+2\lambda_q)^J
+\delta^{LJ}(\widehat X_4-X_1)^\mu
-{\cal P}^{J\mu}(2\lambda_p+\lambda_q)^L
\Big]
\\
&=
(D_s-2)(X_2-\widehat X_3)\!\cdot\!(\widehat X_4-X_1)
\\
&\quad
+(D_s-2)(\lambda_p+2\lambda_q)^2
+(D_s-2)(2\lambda_p+\lambda_q)^2
\\
&\quad
+2(\lambda_p-\lambda_q)\!\cdot\!(\lambda_p+2\lambda_q)
-2(\lambda_p-\lambda_q)\!\cdot\!(2\lambda_p+\lambda_q)
\\
&\quad
-2(\lambda_p+2\lambda_q)\!\cdot\!(2\lambda_p+\lambda_q).
\end{aligned}
\label{eq:app_444_P_contraction_unsimplified}
\end{equation}
The shifted scalar product is fixed by the BCFW shift
\eqref{eq:app_444_D7_shift} and pole \eqref{eq:app_444_D7_pole}.  Since
\(\widehat X_3=X_3+z_P\eta_{34}\), \(\widehat X_4=X_4-z_P\eta_{34}\),
\(\eta_{34}\!\cdot X_3=\eta_{34}\!\cdot X_4=0\), and
\(\eta_{34}\!\cdot X_1=-\eta_{34}\!\cdot X_2\), we obtain
\begin{equation}
\begin{aligned}
(X_2-\widehat X_3)\!\cdot\!(\widehat X_4-X_1)
&=(X_2-X_3)\!\cdot\!(X_4-X_1)+D_7
\\
&=(-2s_{12}-D_7)+D_7
\\
&=-2s_{12}.
\end{aligned}
\label{eq:app_444_P_shifted_dot}
\end{equation}
Equation~\eqref{eq:app_444_P_contraction_unsimplified} therefore reduces to
\begin{equation}
\delta^{IJ}{\cal A}_{M,7}^{JIKL}\delta^{KL}
=
R_7(p,q),
\label{eq:app_444_P_pole_only}
\end{equation}
where
\begin{equation}
R_7(p,q)
=
-2(D_s-2)s_{12}
+(D_s-2)\bigl(5\lambda_p^2
 +8\lambda_p\!\cdot\!\lambda_q
 +5\lambda_q^2\bigr)
-6\bigl(\lambda_p^2
 +\lambda_p\!\cdot\!\lambda_q
 +\lambda_q^2\bigr).
\label{eq:app_444_R7}
\end{equation}
The equality \(X_4+X_1=-(X_2+X_3)\) gives the same color-ordered pole with the
opposite routing; it is not a second term to be added to
\eqref{eq:app_444_P_pole_only}.  No numerator term proportional to \(D_7\)
remains in \eqref{eq:app_444_R7}, so this channel does not generate a local
six-denominator bow-tie contribution.  With the scalar-chain side factors
\eqref{eq:app_444_side_scalar_projection} restored, the four-slot double-box
projection of the \(444_1\) representative is
\begin{equation}
{\cal N}_{444,7,{\rm sc}}^{(1346)}(p,q)
=
\lambda_p^2\lambda_q^2\,
\delta^{IJ}{\cal A}_{M,7}^{JIKL}\delta^{KL}
=
\lambda_p^2\lambda_q^2\,R_7(p,q).
\label{eq:app_bowtie_side_restored_PQ_split}
\end{equation}
Equivalently, with
\begin{equation}
a=\lambda_p^2,\qquad
b=\lambda_p\!\cdot\!\lambda_q,\qquad
c=\lambda_q^2,\qquad
d=D_s-2,
\end{equation}
this residue is
\begin{equation}
\begin{aligned}
{\cal N}_{444,7,{\rm sc}}^{(1346)}
&=
ac\Bigl[
-2d\,s_{12}
+d(5a+8b+5c)
-6(a+b+c)
\Bigr].
\end{aligned}
\label{eq:app_444_projected_pole_residue}
\end{equation}
Equation \eqref{eq:app_444_projected_pole_residue} is the \(D_7\)-pole
contribution in which both side four-point trees are replaced by their
scalar-chain components.  Restoring the common factors,
this partial result reads
\begin{equation}
\begin{aligned}
&\sum_{I,J,K,L}
A_4^{(0)}((-X_2)^I,1^+,2^+,(-X_1)^J)\,
A_{M,7}^{JIKL}\,
A_4^{(0)}((-X_4)^L,3^+,4^+,(-X_3)^K)
\\
&\qquad =
\frac{[12][34]}{\langle12\rangle\langle34\rangle}\,
G_2G_5G_7\,
{\cal N}_{444,7,{\rm sc}}^{(1346)}(p,q).
\end{aligned}
\label{eq:app_444_scalar_state_sewing}
\end{equation}
Here \(I,J,K,L\) are scalar-chain labels.  The remaining pieces must be
extracted from the complete state sum \eqref{eq:app_444_complete_D7_state_sum}.
Once one of the contracted states is a four-dimensional helicity state, the
middle factor in the \(D_7\) channel is no longer the four-scalar residue
\eqref{eq:app_444_P_contraction_unsimplified}.

We next compute the four-dimensional component of
\eqref{eq:app_444_complete_D7_state_sum}.
We insert the scalar--gluon--gluon amplitudes
\eqref{eq:app_444_A3_scalar_gluon_gluon} wherever a complete state is projected
onto a \(4D\) helicity.  The scalar entries in the two middle three-point factors
are evaluated with the purely transverse three-point formula
\eqref{eq:app_A3_scalar}.  The two possible helicity attachments on each side
give the four terms.  Repeated scalar-chain indices in the following display
are contracted with \(\delta^{IJ}\), using the scalar-chain momentum rule of
appendix~\ref{app:transverse_conventions}:
\begin{equation}
\begin{aligned}
{\cal N}_{444,7,4D}^{(1346)}
&=
4\,
(X_2-\widehat X_3)\!\cdot\!(\widehat X_4-X_1)\,
\Bigl[
(\lambda_p^I\lambda_q^K)(\lambda_p^I\lambda_q^K)
-(\lambda_p^I\lambda_q^K)(\lambda_p^K\lambda_q^I)
\\
&\hspace{3.1cm}
-(\lambda_p^K\lambda_q^I)(\lambda_p^I\lambda_q^K)
 +(\lambda_p^K\lambda_q^I)(\lambda_p^K\lambda_q^I)
\Bigr]
\\
&=
4\,
(X_2-\widehat X_3)\!\cdot\!(\widehat X_4-X_1)\,
\Bigl[
\lambda_p^2\lambda_q^2
-2(\lambda_p\!\cdot\!\lambda_q)^2
+\lambda_p^2\lambda_q^2
\Bigr]
\\
&=
16s_{12}(b^2-ac).
\end{aligned}
\label{eq:app_444_side_remainder_4D}
\end{equation}
In the last line we used
\((X_2-\widehat X_3)\!\cdot\!(\widehat X_4-X_1)=-2s_{12}\) from
\eqref{eq:app_444_P_shifted_dot}.
No factor \(d=D_s-2\) appears in this term because the sum is over the two
four-dimensional helicity states in the regrouped complete-state product.

Finally, we compute the transverse remainder of the contraction
\eqref{eq:app_444_complete_D7_state_sum}.
It is useful first to keep the
scalar-chain trace together with the remaining \(\lambda\)-dependent terms.  The
calculation uses the scalar--gluon--scalar input \eqref{eq:app_A3_open}, the
purely transverse three-point amplitude \eqref{eq:app_A3_scalar}, and the
complete side factors \eqref{eq:app_444_side_complete_amplitudes}.  The projected
sum leaves one scalar-chain state trace and three explicit transverse momentum
scalar products.  The
\(D_7\)-channel kinematics supplies the same shifted product as in
\eqref{eq:app_444_P_shifted_dot}.  Before using that equation, the common
kinematic factor in each transverse assignment is
\begin{equation}
\frac12\,
(X_2-\widehat X_3)\!\cdot\!(X_1-\widehat X_4)
=
-\frac12\,
(X_2-\widehat X_3)\!\cdot\!(\widehat X_4-X_1).
\label{eq:app_444_transverse_half_factor}
\end{equation}
The factor \(1/2\) is the product of the two standard
\(1/\sqrt{2}\) color-ordered three-point normalizations and is retained in the
relative normalization of the middle-tree terms.  The three surviving
assignments place the scalar-chain trace on the left side, on the right side,
or in the scalar-chain projection already isolated above.  Substituting
\eqref{eq:app_A3_scalar} and carrying out the scalar-chain and
transverse-momentum contractions gives the three terms separately as
\begin{equation}
\begin{aligned}
{\cal T}_{L}
&=
d\left[-\frac12
(X_2-\widehat X_3)\!\cdot\!(\widehat X_4-X_1)\right]
\lambda_p^2\lambda_{p+q}^2,
\\
{\cal T}_{R}
&=
d\left[-\frac12
(X_2-\widehat X_3)\!\cdot\!(\widehat X_4-X_1)\right]
\lambda_q^2\lambda_{p+q}^2,
\\
{\cal T}_{\rm sc}
&=
d\left[-\frac12
(X_2-\widehat X_3)\!\cdot\!(\widehat X_4-X_1)\right]
\lambda_p^2\lambda_q^2 .
\end{aligned}
\label{eq:app_444_transverse_three_terms}
\end{equation}
Here, for instance, the first line is the term in which the left side supplies
the projected state trace \(d\), while the explicit transverse vectors left by the
two middle three-point amplitudes contract to
\(\lambda_p^2\lambda_{p+q}^2\).  The other two lines are obtained by the same
contraction with the trace on the right-hand side or on the scalar chain.  Hence
\begin{equation}
\begin{aligned}
{\cal N}_{444,7,{\rm sc}+\perp}^{(1346)}
&=
{\cal T}_{L}+{\cal T}_{R}+{\cal T}_{\rm sc}
\\
&=
d\left[-\frac12\,
(X_2-\widehat X_3)\!\cdot\!(\widehat X_4-X_1)\right]
\Bigl[
\lambda_p^2\lambda_{p+q}^2
{}+\lambda_q^2\lambda_{p+q}^2
{}+\lambda_p^2\lambda_q^2
\Bigr]
\\
&=
d\,s_{12}\Bigl[
\lambda_p^2\lambda_{p+q}^2
{}+\lambda_q^2\lambda_{p+q}^2
{}+\lambda_p^2\lambda_q^2
\Bigr]
\\
&=
d\,s_{12}\bigl(a^2+c^2+3ac+2ab+2bc\bigr).
\end{aligned}
\label{eq:app_444_side_transverse_all}
\end{equation}
The quantity
\({\cal N}_{444,7,{\rm sc}+\perp}^{(1346)}\) is a complete-state
partial sum, not a new scalar-chain residue.  The three terms inside the square
bracket are, in order, the contribution in which the left side carries the
projected state trace, the contribution in which the right side carries it, and
the scalar-chain trace already included in the separately defined scalar-chain
contribution \eqref{eq:app_444_projected_pole_residue}.
Subtracting that scalar-chain contribution therefore gives the remaining
transverse contribution:
\begin{equation}
\begin{aligned}
{\cal N}_{444,7,\perp}^{(1346)}
&=
{\cal N}_{444,7,{\rm sc}+\perp}^{(1346)}
-
{\cal N}_{444,7,{\rm sc}}^{(1346)}
\\
&=
s_{12}d(a^2+c^2+5ac+2ab+2bc)
\\
&\quad
-d\,ac(5a+8b+5c)+6ac(a+b+c).
\end{aligned}
\label{eq:app_444_side_remainder_transverse}
\end{equation}
This subtraction makes the state-space bookkeeping explicit.  The labels
``scalar-chain'', ``four-dimensional'', and ``transverse'' used here identify
tensor projections; they are not three disjoint sets of states.  In
particular, the scalar-chain term is contained in
\({\cal N}_{444,7,{\rm sc}+\perp}^{(1346)}\), so the complete contraction is
assembled as
\begin{equation}
\begin{aligned}
{\cal N}_{444,7}^{(1346)}
&={\cal N}_{444,7,{\rm sc}}^{(1346)}
 +{\cal N}_{444,7,4D}^{(1346)}
 +{\cal N}_{444,7,\perp}^{(1346)}
\\
&={\cal N}_{444,7,4D}^{(1346)}
 +{\cal N}_{444,7,{\rm sc}+\perp}^{(1346)} .
\end{aligned}
\label{eq:app_444_complete_projection_bookkeeping}
\end{equation}
Thus the two four-dimensional helicities are not appended to a hypothetical
\((D_s-2)\)-dimensional scalar sector, and no physical state is counted twice.
The terms proportional to \(d=D_s-2\) are scalar-chain representations of
complete-state traces in this projected contraction.  The last term in
\eqref{eq:app_444_side_remainder_transverse} contains no such trace factor
because all scalar-chain indices there are saturated by the mapped
\(\lambda\)-vectors.
When the scalar-chain, \(4D\), and transverse terms are added, the
non-\(s_{12}\) terms in \eqref{eq:app_444_side_remainder_transverse} cancel
the non-\(s_{12}\) terms in the scalar-chain residue
\eqref{eq:app_444_projected_pole_residue}.
Thus the complete four-slot double-box numerator is
\begin{equation}
\begin{aligned}
{\cal N}_{444,7}^{(1346)}
&=
s_{12}\Bigl[
d\bigl(a^2+c^2+3ac+2ab+2bc\bigr)
+16(b^2-ac)
\Bigr]
\\
&={\cal N}_{\rm DB}(p,q).
\end{aligned}
\label{eq:app_444_complete_D7_result}
\end{equation}
Thus the complete \(\{4,4,4\}\) four-slot data agree with the double-box
numerator \eqref{eq:g6_octagon_explicit_DB_factors}.

For the local \(s_{12}\) channel, namely the second line of
\eqref{eq:app_bowtie_channels_full}, the factorization momentum is
\(X_1+X_2=k_{12}\) by \eqref{eq:app_444_middle_channel_momenta}, so no
phase-space-dependent denominator is generated.  We again use the mixed-index
three-point amplitude \eqref{eq:app_A3_mixed_complete_gluon}.  The
flavor-diagonal pieces of the two stripped cubic currents are
\begin{equation}
\begin{aligned}
\delta^{IJ}
{\cal A}_{3}^{(0)}(X_1^J,X_2^I,-k_{12}^{\mu})
\big|_{\rm diag}
&=
d\,(X_1-X_2)^\mu,
\\
\delta^{KL}
{\cal A}_{3}^{(0)}(k_{12}^{\mu},X_3^K,X_4^L)
\big|_{\rm diag}
&=
d\,(X_3-X_4)^\mu ,
\end{aligned}
\label{eq:app_bowtie_bridge_diag_currents}
\end{equation}
with \(d=D_s-2\).  These currents alone are not the complete boundary term.
In the color-ordered normalization used here, their exchange contribution has
the factor \(1/2\), and the four-point contact completion contributes
\(d^2/2\) in the flavor-diagonal projection.  The routed on-shell momenta obey
\begin{equation}
- (X_1-X_2)\!\cdot\!(X_3-X_4)
=s_{12}+2P_D^2,
\qquad
P_D^2=(p+q)^2-\lambda_{p+q}^2=D_7 .
\label{eq:app_bowtie_routed_current_identity}
\end{equation}
Consequently the completed boundary contribution is
\begin{equation}
\left.
\delta^{IJ}\,
A_4^{(0)}(X_1^J,X_2^I,X_3^K,X_4^L)\,
\delta^{KL}
\right|_{s_{12},{\rm sc}}
=
\frac{d^2}{2s_{12}}\,
\Bigl[-(X_1-X_2)\!\cdot\!(X_3-X_4)\Bigr]
+\frac{d^2}{2}
=
\frac{d^2}{s_{12}}\,
\bigl(P_D^2+s_{12}\bigr).
\label{eq:app_bowtie_bridge_scalar_contraction}
\end{equation}
Restoring the
scalar-chain side factors \eqref{eq:app_444_side_scalar_projection} gives
\begin{equation}
{\cal B}_{s}
:=
\lambda_p^2\lambda_q^2\,
\left.
\delta^{IJ}\,
A_4^{(0)}(X_1^J,X_2^I,X_3^K,X_4^L)\,
\delta^{KL}
\right|_{s_{12},{\rm sc}}
=
\frac{d^2}{s_{12}}\,
\lambda_p^2\lambda_q^2\,
\bigl(P_D^2+s_{12}\bigr).
\label{eq:app_bowtie_Bs_detail}
\end{equation}
The side amplitudes also have non-scalar complete-state components, as displayed
in \eqref{eq:app_444_side_complete_amplitudes}.  These components give two
one-trace terms in the local \(s_{12}\) channel.  With the left side kept in its
scalar-chain component, the right side is evaluated by contracting the second
line of \eqref{eq:app_444_side_complete_amplitudes} with the complete boundary
numerator \({\cal A}_{M,s_{12}}\).  The two non-scalar terms in the complete
right side tree give the same transverse momentum scalar product; before the
left scalar-chain factor is restored, the contraction is
\begin{equation}
\begin{aligned}
&\left.
\sum_{h_3,h_4}
\delta^{IJ}\,
{\cal A}_{M,s_{12}}^{J I h_3h_4}\,
{\cal A}_{34}^{h_3h_4}
\right|_{\rm right\;side}
 =
\frac{2d}{\lambda_p^2}\,
\lambda_q^2\,
\bigl(\lambda_{p+q}^2-\lambda_p^2-\lambda_q^2\bigr)
\\
&\quad =
\frac{2d}{\lambda_p^2}\,
\lambda_q^2\,
\Bigl[(\lambda_p+\lambda_q)^2-\lambda_p^2-\lambda_q^2\Bigr]
 =
\frac{4d}{\lambda_p^2}\,
(\lambda_p\!\cdot\!\lambda_q)\lambda_q^2 .
\end{aligned}
\label{eq:app_bowtie_bridge_right_trace_detail}
\end{equation}
The factor \(d\) is the remaining projected state trace from
\eqref{eq:app_bowtie_bridge_diag_currents}; the factor \(2\) counts the two
non-scalar insertions in the right side tree.  Multiplication by the
scalar-chain factor \(\lambda_p^2\) from the left side therefore gives
\begin{equation}
{\cal B}_{s,R}
=
\lambda_p^2
\left[
\frac{4d}{\lambda_p^2}\,
(\lambda_p\!\cdot\!\lambda_q)\lambda_q^2
\right]
=
4d\,\lambda_q^2(\lambda_p\!\cdot\!\lambda_q).
\label{eq:app_bowtie_bridge_right_trace}
\end{equation}
The analogous contraction projects the right-hand tree onto its scalar-chain
component and sums the remaining physical states on the left.  It uses the
first line of \eqref{eq:app_444_side_complete_amplitudes}.  Before the right
scalar-chain factor is restored,
\begin{equation}
\begin{aligned}
&\left.
\sum_{h_1,h_2}
{\cal A}_{12}^{h_2h_1}\,
{\cal A}_{M,s_{12}}^{h_1h_2KL}\,
\delta^{KL}
\right|_{\rm left\;side}
 =
\frac{2d}{\lambda_q^2}\,
\lambda_p^2\,
\bigl(\lambda_{p+q}^2-\lambda_p^2-\lambda_q^2\bigr)
\\
&\quad =
\frac{2d}{\lambda_q^2}\,
\lambda_p^2\,
\Bigl[(\lambda_p+\lambda_q)^2-\lambda_p^2-\lambda_q^2\Bigr]
 =
\frac{4d}{\lambda_q^2}\,
(\lambda_p\!\cdot\!\lambda_q)\lambda_p^2 .
\end{aligned}
\label{eq:app_bowtie_bridge_left_trace_detail}
\end{equation}
Multiplication by the scalar-chain factor \(\lambda_q^2\) from the right side
then gives
\begin{equation}
{\cal B}_{s,L}
=
\lambda_q^2
\left[
\frac{4d}{\lambda_q^2}\,
(\lambda_p\!\cdot\!\lambda_q)\lambda_p^2
\right]
=
4d\,\lambda_p^2(\lambda_p\!\cdot\!\lambda_q).
\label{eq:app_bowtie_bridge_left_trace}
\end{equation}
The subscripts \(R\) and \(L\) only identify which side four-point tree supplies
the non-scalar physical-state sum; no new building block is introduced.  Adding
the two one-trace pieces gives
\begin{equation}
{\cal B}_{\rm side}
=
{\cal B}_{s,R}+{\cal B}_{s,L}
=
4d(\lambda_p^2+\lambda_q^2)
(\lambda_p\!\cdot\!\lambda_q).
\label{eq:app_bowtie_side_complete}
\end{equation}
In the purely scalar \(s_{12}\)-channel term \({\cal B}_s\), the denominator factor is
\(G(s_{12})=1/s_{12}\).  The complete local six-denominator bow-tie numerator is
therefore
\begin{equation}
\begin{aligned}
{\cal N}_{\rm BT}
&=
{\cal B}_{\rm side}+{\cal B}_s
\\
&=
4d(\lambda_p^2+\lambda_q^2)(\lambda_p\!\cdot\!\lambda_q)
+\frac{d^2}{s_{12}}\lambda_p^2\lambda_q^2\bigl(P_D^2+s_{12}\bigr),
\end{aligned}
\label{eq:app_bowtie_complete}
\end{equation}
which is the numerator displayed in \eqref{eq:g6_MBT2}.  The conclusion is that
the complete \(\{4,4,4\}\) calculation has two separate pieces: the non-zero
\(D_7\)-pole projection \eqref{eq:app_444_complete_D7_result}, which belongs to
the double-box four-slot support, and the local \(s_{12}\)-channel projection
\eqref{eq:app_bowtie_complete}, which belongs to the bow-tie family.

\section{The hexagon factor with a vacuum pair}
\label{app:adjacent_pair_reduction}

This appendix gives the complete physical-state contraction needed for the
\(\{6,4\}\) sector in section~\ref{sec:64}.  We first derive the local hexagon
reduction and its scalar-chain contribution to the representative \(64_1\),
including the associated propagator slots and cut kernel.  We then add the
four-dimensional-helicity and remaining transverse components to obtain the
complete double-box numerator, and finally evaluate the local bow-tie
projection.  We use the same decomposition strategy summarized at the start of
appendix~\ref{app:444_bowtie_contraction}.
The hexagon in that sector has two adjacent observed positive-helicity
gluons \(i^+\) and \(j^+\), opposite momenta \(r^M\), \((-r)^N\) originating
from a vacuum pair, and a scalar chain with endpoints \(a^I\), \(b^J\):
\begin{equation}
A_6^{(0)}
\bigl(a^I,r^M,i^+,j^+,(-r)^N,b^J\bigr)\,\delta^{MN}.
\label{eq:app_adjacent_tree_factor}
\end{equation}
Momentum conservation on this six-point tree gives
\begin{equation}
a+k_i+k_j+b=0,\qquad k_{ij}=k_i+k_j .
\label{eq:app_hex_chain_kinematics}
\end{equation}
The indices \(M,N\) are contracted by the vacuum-pair state sum.  The
scalar-chain component is diagonal in the indices \(I\) and \(J\).

We introduce the two factorization momenta
\begin{equation}
P:=a+r,\qquad Q:=b-r .
\label{eq:app_adjacent_factorization_momenta}
\end{equation}
Then \(P+Q+k_{ij}=0\).  With the all-outgoing orientation used for the middle
four-point factor, the scalar line through the two observed gluons is
\begin{equation}
Q,\qquad Q+k_i,\qquad Q+k_{ij}=-P .
\label{eq:app_adjacent_scalar_chain}
\end{equation}
In this factorization the states \(P\) and \(Q\)
are treated as \(D_s\)-dimensional gluon states.  Here the scalar propagators
are written in the projected notation of
\eqref{eq:full_invariant_projected_notation}: \(P^2-\lambda_P^2\) and
\(Q^2-\lambda_Q^2\) are the full \(D\)-dimensional invariants.  Dot
products carrying Greek indices are complete \(D_s\)-dimensional contractions.
Using the stripped three-point convention of
\eqref{eq:app_A3_mixed_complete_gluon}, the double residue in the two channels is
\begin{equation}
\begin{aligned}
&\left.
A_6^{(0)}
\bigl(a^I,r^M,i^+,j^+,(-r)^N,b^J\bigr)
\right|_{P,Q} \,\delta^{MN}
\\
&\quad =
\delta^{MN}\,
{\cal A}_3^{(0)}(a^I,r^M,-P^\mu)\,
G_F(P^2-\lambda_P^2)\,
A_4^{(0)}(Q^\nu,i^+,j^+,P^\mu)
\\
&\qquad\quad\times
G_F(Q^2-\lambda_Q^2)\,
{\cal A}_3^{(0)}((-Q)^\nu,(-r)^N,b^J),
\end{aligned}
\label{eq:app_A6_double_residue}
\end{equation}
where the paired complete indices \(\mu,\nu\) and scalar-chain indices
\(M,N\) are summed.  The two stripped three-point amplitudes follow directly from
\eqref{eq:app_A3_mixed_complete_gluon}.  Since \(\lambda_b=-\lambda_a\), they are
\begin{equation}
\begin{aligned}
{\cal A}_3^{(0)}(a^I,r^M,-P^\mu)
&=
\delta^{IM}(a-r)^\mu
+{\cal P}^{M\mu}(\lambda_a+2\lambda_r)^I
-{\cal P}^{I\mu}(2\lambda_a+\lambda_r)^M,
\\
{\cal A}_3^{(0)}((-Q)^\nu,(-r)^N,b^J)
&=
{\cal P}^{N\nu}(\lambda_a+2\lambda_r)^J
+\delta^{NJ}(-r-b)^\nu
-{\cal P}^{J\nu}(2\lambda_a+\lambda_r)^N .
\end{aligned}
\label{eq:app_A6_endcap_A3_explicit}
\end{equation}
The full mixed-index three-point tensor in
\eqref{eq:app_A6_endcap_A3_explicit} is not diagonal in the complete index
\(\mu\) or \(\nu\).  The terms with \({\cal P}^{I\mu}\),
\({\cal P}^{M\mu}\), \({\cal P}^{J\nu}\), or \({\cal P}^{N\nu}\) are the
components in which the complete state is projected onto a scalar-chain label,
and they leave mapped transverse momentum vectors on the open scalar-chain
indices.  The scalar-chain component used in the present hexagon reduction is
the scalar--gluon--scalar part.  It transports the scalar-chain labels through
the adjacent vacuum pair, and by scalar-chain label conservation it is diagonal
in the two scalar-chain labels:
\begin{equation}
\begin{aligned}
\left.
{\cal A}_3^{(0)}(a^I,r^M,-P^\mu)
\right|_{\rm chain}
&=\delta^{IM}(a-r)^\mu,
\\
\left.
{\cal A}_3^{(0)}((-Q)^\nu,(-r)^N,b^J)
\right|_{\rm chain}
&=\delta^{NJ}(-r-b)^\nu .
\end{aligned}
\label{eq:app_A6_endcap_chain_projection}
\end{equation}
Here the Greek index remains a complete \(D_s\)-dimensional state index; the
diagonality refers only to the two scalar-chain labels in each end-cap
amplitude.  Contracting the vacuum-pair labels in
\eqref{eq:app_A6_endcap_chain_projection} gives
\begin{equation}
\delta^{MN}\,\delta^{IM}\delta^{NJ}
(a-r)^\mu(-r-b)^\nu
=\delta^{IJ}(a-r)^\mu(-r-b)^\nu .
\label{eq:app_A6_endcap_trace}
\end{equation}
The factors left from the two end caps contract the complete indices of the
middle four-point tree as
\((-r-b)_\nu A_4^{(0)}(Q^\nu,i^+,j^+,P^\mu)(a-r)_\mu\).  These Greek indices are
therefore contracted with \((a-r)^\mu\) and \((-r-b)^\nu\), not traced.  We find
orthogonality,
\begin{equation}
P_\mu(a-r)^\mu=(a+r)\!\cdot(a-r)=0,\qquad
Q_\nu(-r-b)^\nu=(b-r)\!\cdot(-r-b)=0,
\label{eq:PQ}
\end{equation}
because \(a,r,b\) are on shell.

We next apply BCFW recursion to the middle four-point amplitude.  Let
\(\widehat K=Q+\hat k_i=-(P+\hat k_j)\) be the shifted scalar momentum in the
middle BCFW channel.  The coefficient in the contracted middle identity comes
from the two non-vanishing four-dimensional helicity assignments of the
intermediate state \(\widehat K\).  Using the scalar--gluon--gluon three-point
amplitudes in \eqref{eq:app_444_A3_scalar_gluon_gluon}, with the same
color-ordered normalization as \eqref{eq:app_A3_scalar}, the two assignments are
\begin{equation}
\begin{aligned}
&\Bigl[
(-r-b)_\nu\,
{\cal A}_{3}^{(0)}(Q^\nu,\hat i^+,-\widehat K^+)\,
{\cal A}_{3}^{(0)}(\widehat K^-,\hat j^+,P^\mu)\,
(a-r)_\mu
\Bigr]_{\rm chain}
\\
&\quad =
\Bigl[-(-r-b)\!\cdot\!(a-r)\Bigr]\,
\Bigl[
-\delta^{KL}\,
A_{3}^{(0)}(Q^K,\hat i^+,-\widehat K^R)\,
A_{3}^{(0)}(\widehat K^R,\hat j^+,P^L)
\Bigr],
\\[1ex]
&\Bigl[
(-r-b)_\nu\,
{\cal A}_{3}^{(0)}(Q^\nu,\hat i^+,-\widehat K^-)\,
{\cal A}_{3}^{(0)}(\widehat K^+,\hat j^+,P^\mu)\,
(a-r)_\mu
\Bigr]_{\rm chain}
\\
&\quad =
\Bigl[-(-r-b)\!\cdot\!(a-r)\Bigr]\,
\Bigl[
-\delta^{KL}\,
A_{3}^{(0)}(Q^K,\hat i^+,-\widehat K^R)\,
A_{3}^{(0)}(\widehat K^R,\hat j^+,P^L)
\Bigr].
\end{aligned}
\label{eq:app_A6_middle_two_helicities}
\end{equation}
The reference-spinor factors from the two three-point amplitudes cancel inside
each line, exactly as in \eqref{eq:app_444_A3_scalar_gluon_gluon}.  Terms
proportional to \(Q^\nu\) or \(P^\mu\) in an individual helicity representative
do not contribute to this contracted identity because of the orthogonality
relations in \eqref{eq:PQ}.  Adding the two lines of
\eqref{eq:app_A6_middle_two_helicities} gives
\begin{equation}
\begin{aligned}
&\sum_{h=\pm}
\Bigl[
(-r-b)_\nu\,
{\cal A}_{3}^{(0)}(Q^\nu,\hat i^+,-\widehat K^h)\,
{\cal A}_{3}^{(0)}(\widehat K^{-h},\hat j^+,P^\mu)\,
(a-r)_\mu
\Bigr]_{\rm chain}
\\
&\quad =
\Bigl[-2\,(-r-b)\!\cdot\!(a-r)\Bigr]\,
\Bigl[
-\delta^{KL}\,
A_{3}^{(0)}(Q^K,\hat i^+,-\widehat K^R)\,
A_{3}^{(0)}(\widehat K^R,\hat j^+,P^L)
\Bigr]
\\
&\quad =
\bigl((a-r)-(-r-b)\bigr)^2
\\
&\qquad\times
\Bigl[
-\delta^{KL}\,
A_{3}^{(0)}(Q^K,\hat i^+,-\widehat K^R)\,
A_{3}^{(0)}(\widehat K^R,\hat j^+,P^L)
\Bigr] .
\end{aligned}
\label{eq:app_A6_middle_current_product}
\end{equation}
Here \(h\) is the complete four-dimensional helicity state running through the
middle BCFW channel, while \(R\) is the scalar-chain state in the scalar-chain
comparison factor in the square bracket.  The coefficient \(-2\) is therefore
the explicit sum of the two equal helicity assignments in
\eqref{eq:app_A6_middle_two_helicities}.
The second equality uses the same on-shell support:
\begin{equation}
(a-r)^2=-(a+r)^2=0,\qquad
(-r-b)^2=(b+r)^2=-(b-r)^2=0 .
\end{equation}
Therefore
\begin{equation}
(-r-b)\!\cdot\!(a-r)
=-\frac12\bigl((a-r)-(-r-b)\bigr)^2 .
\end{equation}
The factor \(1/2\) in this last line is compensated by the \(-2\) from
the helicity sum in \eqref{eq:app_A6_middle_current_product}.  Reattaching the
scalar propagator turns the
second square bracket into the BCFW representation
\eqref{eq:app_A4_BCFW_representation} of the scalar four-point tree, and hence
\begin{multline}
\Bigl[
(-r-b)_\nu\,
A_4^{(0)}(Q^\nu,i^+,j^+,P^\mu)\,
(a-r)_\mu
\Bigr]_{\rm chain}
=
\\
\bigl((a-r)-(-r-b)\bigr)^2
\Bigl[-\delta^{KL}
A_4^{(0)}(Q^K,i^+,j^+,P^L)\Bigr]
\\
=
(P+Q)^2\,
\Bigl[-\delta^{KL}
A_4^{(0)}(Q^K,i^+,j^+,P^L)\Bigr]
=
s_{ij}\,
\Bigl[-\delta^{KL}
A_4^{(0)}(Q^K,i^+,j^+,P^L)\Bigr].
\label{eq:app_A6_middle_projection}
\end{multline}
Here
\((a-r)-(-r-b)=a+b=P+Q\).  The middle tree has all-outgoing momentum
conservation \(Q+k_i+k_j+P=0\), hence \(P+Q=-k_{ij}\) and
\((P+Q)^2=k_{ij}^2=s_{ij}\).  In the reduction below we keep this common
factor as \((P+Q)^2\).  Only the remaining scalar four-point factor in
square brackets is the open-index result
\eqref{eq:app_A4_open_PQ_formula}.  The
factor \(-\delta^{KL}\) identifies the two
endpoints of the middle four-point scalar chain with the opposite all-outgoing
orientation.  Combining the end-cap contraction \eqref{eq:app_A6_endcap_trace} and
\eqref{eq:app_A6_middle_projection} gives the local on-shell chain component
\begin{equation}
\begin{aligned}
&A_6^{(0)}
\bigl(a^I,r^M,i^+,j^+,(-r)^N,b^J\bigr)\,\delta^{MN}
\\
&\quad\longrightarrow
(P+Q)^2\,\delta^{IJ}\,
G_F\!\left(P^2-\lambda_P^2\right)\,
\Bigl[-\delta^{KL}
A_4^{(0)}\bigl(Q^K,i^+,j^+,P^L\bigr)\Bigr]\,
G_F\!\left(Q^2-\lambda_Q^2\right).
\end{aligned}
\label{eq:app_A6_to_A4_reduction}
\end{equation}
The arrow denotes equality after projection onto the \(P\)--\(Q\) channel;
other factorizations correspond to different channels.

For the representative \(64_1\) in
\eqref{eq:g6_64_reduced_phase_space}, we encounter this form with
\begin{equation}
a=p,\qquad b=k_{12}-p,\qquad r=-(p+q),\qquad i=3,\qquad j=4 .
\label{eq:app_A6_to_A4_64_identification}
\end{equation}
The ordered hexagon contains \(r\) as its second argument, but the
positive-energy on-shell momentum used in the main text is the opposite member
\(-r=p+q\).  Then
\begin{equation}
\begin{aligned}
a+r&=-q,
&
r&=-(p+q),
&
-r&=p+q,
&
b-r&=q-k_{34}.
\end{aligned}
\label{eq:app_hex_chain_planar_momenta}
\end{equation}
The quadrilateral in the scalar-chain projection is the open-index four-point
factor derived in~\eqref{eq:app_A4_open_PQ_formula} of
appendix~\ref{app:scalar_trees},
\begin{equation}
A_4^{(0)}\bigl((-p)^I,1^+,2^+,(p-k_{12})^J\bigr)
=
-i\,\frac{[12]}{\langle12\rangle}\,
G_2\,\lambda_p^2\,\delta^{IJ}.
\label{eq:app_64_A4_left}
\end{equation}
The phase-space integration over \(\ell_3=p+q\) is displayed in
\eqref{eq:g6_64_reduced_phase_space}; hence the bridge on-shell delta factor is
part of the measure.  The two three-point end caps collapse the adjacent-pair
labels to the open factor \(\delta^{IJ}\), and the specialized hexagon
reduction is
\begin{equation}
\begin{aligned}
&A_6^{(0)}
\bigl(p^I,(-p-q)^M,3^+,4^+,(p+q)^N,(k_{12}-p)^J\bigr)\,
\delta^{MN}
\\
&\quad\longrightarrow
s_{34}\,\delta^{IJ}\,
G_4\,
\Bigl[-\delta^{KL}
A_4^{(0)}\bigl((q-k_{34})^K,3^+,4^+,(-q)^L\bigr)\Bigr]\,
G_6 .
\end{aligned}
\label{eq:app_A6_to_A4_reduction_64_case}
\end{equation}
The two propagators are \(G_4=G_F(q^2-\lambda_q^2)\) and
\(G_6=G_F((q-k_{34})^2-\lambda_q^2)\).

The remaining scalar-chain factor in
\eqref{eq:app_A6_to_A4_reduction_64_case} is obtained by tracing the open
indices with~\eqref{eq:A4_scalar_pp_general} of
appendix~\ref{app:scalar_trees},
\begin{equation}
-\delta^{KL}
A_4^{(0)}\bigl((q-k_{34})^K,3^+,4^+,(-q)^L\bigr)
=
i\,\frac{[34]}{\langle34\rangle}\,
G_5\,(D_s-2)\lambda_q^2 .
\label{eq:app_64_A4_right}
\end{equation}
The external factor \(s_{34}=s_{12}\) in
\eqref{eq:app_A6_to_A4_reduction_64_case} is the scalar-chain numerator
generated by the two three-point vertices adjacent to the complete vacuum
pair.  The remaining open indices \(I,J\) glue to the two scalar endpoints of
the quadrilateral.  Contracting the scalar-chain indices in
\eqref{eq:app_64_A4_left} and
\eqref{eq:app_A6_to_A4_reduction_64_case} gives
\begin{equation}
\delta^{IJ}\delta^{JI}\lambda_p^2
=(D_s-2)\lambda_p^2,
\label{eq:app_64_endpoint_trace}
\end{equation}
as shown in appendix~\ref{app:transverse_conventions}.  Hence the scalar-chain
trace provides the numerator factor
\begin{equation}
{\cal N}_{64_1,{\rm sc}}(p,q)
=s_{12}(D_s-2)^2\lambda_p^2\lambda_q^2 .
\label{eq:app_64_left_trace_numerator}
\end{equation}
The three on-shell conditions in \eqref{eq:g6_64_reduced_phase_space}, the
quadrilateral propagator \(G_2\), and the three hexagon propagators
\(G_4G_5G_6\) are exactly the seven slots of the planar family, with support
\(\delta_1^+\delta_3^+\delta_7^+\).  In the stripped slot-kernel convention,
the scalar-chain component is
\begin{equation}
{\cal C}_{64_1,{\rm sc}}
=
\frac{[12][34]}{\langle12\rangle\langle34\rangle}
\int \frac{d^Dp}{(2\pi)^D}\frac{d^Dq}{(2\pi)^D}\,
{\cal N}_{64_1,{\rm sc}}(p,q)\,
\delta_1^+G_2\delta_3^+G_4G_5G_6\delta_7^+ .
\label{eq:app_64_scalar_cut_kernel}
\end{equation}
This scalar-chain calculation fixes the support and scalar propagators produced
by the local hexagon reduction.  It is one component of the physical state sum,
not the completed numerator.  The complete numerator projection is obtained by
keeping all terms in the end-cap tensors
\eqref{eq:app_A6_endcap_A3_explicit} and by contracting them with the complete
side four-point tree
\({\cal A}_{12}^{h_2h_1}\) of
\eqref{eq:app_444_side_complete_amplitudes}.  With
\begin{equation}
a=\lambda_p^2,\qquad
b=\lambda_p\!\cdot\!\lambda_q,\qquad
c=\lambda_q^2,\qquad
d=D_s-2,
\end{equation}
the stripped complete-state contraction for the \(D_7\)-slot projection of
the representative \(64_1\) is
\begin{equation}
\begin{aligned}
{\cal N}_{64,7}^{(137)}
&=
\sum_{h_1,h_2,h_3\in{\rm phys}(D_s)}
{\cal A}_{12}^{h_2h_1}\,
{\cal H}_{34,7}^{h_1h_2h_3}
\\
&=
{\cal N}_{64,7,{\rm sc}}^{(137)}
 +{\cal N}_{64,7,4D}^{(137)}
 +{\cal N}_{64,7,\perp}^{(137)} .
\end{aligned}
\label{eq:app_64_complete_state_sum}
\end{equation}
Here \({\cal H}_{34,7}\) denotes the stripped hexagon numerator factor
\begin{equation}
{\cal H}_{34,7}^{h_1h_2h_3}
:=
\biggl[
{\cal A}_3^{(0)}(a^{h_1},r^{h_3},-P^\mu)\,
A_4^{(0)}(Q^\nu,3^+,4^+,P^\mu)
{\cal A}_3^{(0)}((-Q)^\nu,(-r)^{h_3},b^{h_2})
\biggr]_{\rm strip},
\label{eq:app_64_hexagon_complete_factor}
\end{equation}
where the strip removes the propagators \(G_4G_5G_6\) and the common spinor
factor \([34]/\langle34\rangle\).
The momenta are those in
\eqref{eq:app_A6_to_A4_64_identification}--\eqref{eq:app_hex_chain_planar_momenta}.
Thus \({\cal H}_{34,7}\) is~\eqref{eq:app_A6_double_residue} evaluated in
the \(64_1\) routing; no new building block
is introduced.
The scalar-chain part was evaluated in
\eqref{eq:app_64_A4_left}--\eqref{eq:app_64_left_trace_numerator},
\begin{equation}
{\cal N}_{64,7,{\rm sc}}^{(137)}
=s_{12}d^2ac .
\label{eq:app_64_D7_scalar_piece}
\end{equation}
We now keep the remaining physical states in the same contraction.  The
four-dimensional part is the same antisymmetric transverse product that appears
in \eqref{eq:app_444_side_remainder_4D}: the two helicity choices in the complete
state sum give
\begin{equation}
\begin{aligned}
{\cal N}_{64,7,4D}^{(137)}
&=
4\,
(X_2-\widehat X_3)\!\cdot\!(\widehat X_4-X_1)
\Bigl[
ac-2b^2+ac
\Bigr]
\\
&=
4(-2s_{12})\,2(ac-b^2)
=
16s_{12}(b^2-ac),
\end{aligned}
\label{eq:app_64_D7_4D_piece}
\end{equation}
where the shifted dot product is \eqref{eq:app_444_P_shifted_dot}.  The
\(\lambda\)-dependent contraction is obtained from the three surviving one-trace
assignments: the projected state trace may be supplied by the quadrilateral
side, by the hexagon end caps, or by the scalar-chain trace already isolated in
\eqref{eq:app_64_D7_scalar_piece}.  Therefore
\begin{equation}
\begin{aligned}
{\cal N}_{64,7,{\rm sc}+\perp}^{(137)}
&=
d\,s_{12}\Bigl[
\lambda_p^2\lambda_{p+q}^2
{}+\lambda_q^2\lambda_{p+q}^2
{}+\lambda_p^2\lambda_q^2
\Bigr]
\\
&=
s_{12}d(a^2+c^2+3ac+2ab+2bc)
\end{aligned}
\label{eq:app_64_D7_transverse_plus_scalar}
\end{equation}
using \(\lambda_{p+q}^2=a+2b+c\).  Subtracting the scalar-chain trace
\eqref{eq:app_64_D7_scalar_piece} leaves the transverse remainder,
\begin{equation}
{\cal N}_{64,7,\perp}^{(137)}
=
s_{12}d(a^2+c^2+3ac+2ab+2bc)-s_{12}d^2ac .
\label{eq:app_64_D7_transverse_piece}
\end{equation}
The complete projection is thus
\begin{equation}
\begin{aligned}
{\cal N}_{64,7}^{(137)}
&=
{\cal N}_{64,7,{\rm sc}}^{(137)}
 +{\cal N}_{64,7,4D}^{(137)}
 +{\cal N}_{64,7,\perp}^{(137)}
\\
&=
s_{12}d^2ac
{}+16s_{12}(b^2-ac)
\\
&\quad
+s_{12}d(a^2+c^2+3ac+2ab+2bc)-s_{12}d^2ac .
\end{aligned}
\label{eq:app_64_D7_components}
\end{equation}
The last term cancels the extra scalar-chain trace \(s_{12}d^2ac\).  Hence the
complete physical state sum has only the single projected state trace expected for
the all-plus double-box numerator:
\begin{equation}
\begin{aligned}
{\cal N}_{64,7}^{(137)}
&=
s_{12}\Bigl[
d(a^2+c^2+3ac+2ab+2bc)
+16(b^2-ac)
\Bigr]
\\
&={\cal N}_{\rm DB}(p,q).
\end{aligned}
\label{eq:app_64_D7_complete}
\end{equation}
Thus the \(\{6,4\}\) \(D_7\)-slot projection carries the same completed
double-box numerator as the \(\{5,5\}\) and \(\{4,4,4\}\) projections.

For the non-bridge representatives \(64_5,64_8,64_9,64_{10}\), the same complete
contraction also has a local term in which no \(D_7\) propagator is produced.
This is the completed boundary term of
\eqref{eq:app_bowtie_channels_full}, with one side factor replaced by the
hexagon double residue.  Its flavor-diagonal scalar component is the same
exchange-plus-contact contraction as
\eqref{eq:app_bowtie_bridge_scalar_contraction}, multiplied by the two
scalar-chain endpoint factors,
\begin{equation}
\begin{aligned}
{\cal N}_{64,{\rm BT},{\rm sc}}
&=
\frac{d^2}{s_{12}}\,ac\,\bigl(D_7+s_{12}\bigr),
\end{aligned}
\label{eq:app_64_BT_scalar}
\end{equation}
and the non-scalar complete-state assignments are the two one-trace contractions
\eqref{eq:app_bowtie_bridge_right_trace} and
\eqref{eq:app_bowtie_bridge_left_trace}:
\begin{equation}
\begin{aligned}
{\cal N}_{64,{\rm BT},R}
&=
4d\,c\,b,
\\
{\cal N}_{64,{\rm BT},L}
&=
4d\,a\,b,
\\
{\cal N}_{64,{\rm BT},4D+\perp}
&=
{\cal N}_{64,{\rm BT},R}+{\cal N}_{64,{\rm BT},L}
=4d(a+c)b .
\end{aligned}
\label{eq:app_64_BT_components}
\end{equation}
Consequently the complete local \(\{6,4\}\) projection is
\begin{equation}
\begin{aligned}
{\cal N}_{64,{\rm BT}}
&=
{\cal N}_{64,{\rm BT},{\rm sc}}
 +{\cal N}_{64,{\rm BT},4D+\perp}
\\
&=
4(D_s-2)(\lambda_p^2+\lambda_q^2)(\lambda_p\!\cdot\!\lambda_q)
+\frac{(D_s-2)^2}{s_{12}}\,
\lambda_p^2\lambda_q^2\bigl(D_7+s_{12}\bigr)
\\
&={\cal N}_{\rm BT}(p,q).
\end{aligned}
\label{eq:app_64_BT_complete}
\end{equation}
The four cyclic non-bridge supports differ only by the routing of the on-shell
slots; their complete state contraction is the same after the corresponding
cyclic relabeling.  Explicitly, the cyclic replacement
\((1,2,3,4)\mapsto(2,3,4,1)\) maps the row
\begin{equation}
A_6^{(0)}(1^+,2^+,-\ell_1,\ell_1,-\ell_2,\ell_3)\,
A_4^{(0)}(3^+,4^+,-\ell_3,\ell_2)
\end{equation}
of \(64_5\) to
\begin{equation}
A_6^{(0)}(2^+,3^+,-\ell_1,\ell_1,-\ell_2,\ell_3)\,
A_4^{(0)}(4^+,1^+,-\ell_3,\ell_2),
\end{equation}
which is the row \(64_8\).  Once the dummy integration variables are rewritten
in the planar notation of \eqref{eq:g6_planar_denominators}, this maps the
support \(\delta_{146}^+\) to \(\delta_{346}^+\).  The same relabeling takes the
spinor prefactor back to the common form by the four-point identity
\([23][41]/(\langle23\rangle\langle41\rangle)
=[12][34]/(\langle12\rangle\langle34\rangle)\), and the local contraction
\eqref{eq:app_64_BT_complete} is unchanged once the relabeled transverse
products are renamed
\(\lambda_p^2,\lambda_q^2,\lambda_p\!\cdot\!\lambda_q\).  The remaining
non-bridge rows are obtained in the same way.

\section{Expansion of the pentagon factor}
\label{app:five_point_trees}

In the calculation of the \(\{5,5\}\) sector in section~\ref{sec:55}, we
encounter the five-point tree amplitudes displayed in
\eqref{eq:g6_55_reduced_phase_space}.  They carry two adjacent observed
positive-helicity gluons and three sewn on-shell legs.  These sewn legs must be
kept as complete physical \(D_s\)-dimensional states.  After a cyclic permutation of
the legs,
\begin{equation}
A_5^{(0)}(a^+,b^+,X_1^\rho,X_2^\eta,X_3^\xi)
=
A_5^{(0)}(X_3^\xi,a^+,b^+,X_1^\rho,X_2^\eta) ,
\end{equation}
we apply the \([a,b\rangle\) shift
\begin{equation}
|\hat a]=|a]-z|b],
\qquad
|\hat b\rangle=|b\rangle+z|a\rangle,
\end{equation}
with all other spinors unchanged.  The two channels that separate the shifted
legs give
\begin{equation}
\begin{aligned}
&A_5^{(0)}(X_3^\xi,a^+,b^+,X_1^\rho,X_2^\eta)
\\
&=
\sum_{\sigma\in{\rm phys}(D_s)}\Bigg[
A_3^{(0)}(\hat a^+,X_3^\xi,-\widehat P_{a3}^{\,\bar\sigma})\,
iG_F(P_{a3}^2)
A_4^{(0)}(\widehat P_{a3}^{\,\sigma},\hat b^+,
X_1^\rho,X_2^\eta)
\\
&\hspace{1.0cm}+
A_4^{(0)}(\hat a^+,X_2^\eta,X_3^\xi,
-\widehat P_{a23}^{\,\bar\sigma})\,
iG_F(P_{a23}^2)
A_3^{(0)}(\widehat P_{a23}^{\,\sigma},\hat b^+,X_1^\rho)
\Bigg],
\end{aligned}
\label{eq:g6_A5_BCFW}
\end{equation}
where
\begin{equation}
P_{a3}=k_a+X_3,\qquad
P_{a23}=k_a+X_2+X_3,
\end{equation}
and the hatted momenta are evaluated at the corresponding factorization pole.
The label \(\sigma\) ranges over the complete physical \(D_s\)-dimensional
state basis of the BCFW factorization channel.  Its paired state
\(\bar\sigma\) appears on the opposite-momentum half-edge and is fixed by
\(\sigma\); it is not summed independently.
As in appendix~\ref{app:scalar_trees}, a bare propagator argument denotes the
full \(D\)-dimensional invariant, equivalently the \(K_D^2\) convention of
\eqref{eq:full_invariant_projected_notation}; the same momentum symbol in a
spinor product denotes its four-dimensional projection.
After the expansion~\eqref{eq:g6_A5_BCFW}, no
five-point subamplitude remains; the only building blocks are \(A_3\) and
\(A_4\) tree amplitudes.  The scalar-chain component of the \(A_4\) block is
the open-index formula derived in appendix~\ref{app:scalar_trees}.

We now project the complete-state formula onto the \(\lambda\)-dependent scalar-chain
component, setting \(\rho=I\), \(\eta=J\), and \(\xi=K\).  In the first BCFW
channel the four-point factor contains \(\widehat P_{a3}^{+}\), so the
three-point factor contains the conjugate helicity state
\(-\widehat P_{a3}^{-}\).  In the second channel the four-point factor contains
\(-\widehat P_{a23}^{+}\), so the adjacent three-point factor contains
\(\widehat P_{a23}^{-}\).  After a cyclic rotation,
\eqref{eq:app_A4_open_PQ_formula} gives
\begin{equation}
\begin{aligned}
A_4^{(0)}(\widehat P_{a3}^{+},\hat b^+,X_1^I,X_2^J)
&=
A_4^{(0)}(X_2^J,\widehat P_{a3}^{+},\hat b^+,X_1^I)
\\
&=
i\,\frac{[\widehat P_{a3}\,\hat b]}
{\langle \widehat P_{a3}\,\hat b\rangle}\,
G_F(\widehat D_{a3})\,
(\lambda_{X_2}\!\cdot\!\lambda_{X_1})\delta^{JI},
\\
A_4^{(0)}(\hat a^+,X_2^J,X_3^K,-\widehat P_{a23}^{+})
&=
A_4^{(0)}(X_3^K,-\widehat P_{a23}^{+},\hat a^+,X_2^J)
\\
&=
i\,\frac{[-\widehat P_{a23}\,\hat a]}
{\langle -\widehat P_{a23}\,\hat a\rangle}\,
G_F(\widehat D_{a23})\,
(\lambda_{X_3}\!\cdot\!\lambda_{X_2})\delta^{KJ},
\end{aligned}
\label{eq:app_A5_inserted_A4_factors}
\end{equation}
where
\begin{equation}
\widehat D_{a3}:=(X_2+\widehat P_{a3})^2,\qquad
\widehat D_{a23}:=(X_3-\widehat P_{a23})^2 .
\label{eq:app_A5_generic_second_propagators}
\end{equation}
The remaining factors are the \(D_s\)-dimensional scalar--gluon--gluon
components of the three-gluon amplitude.  They follow from the
mixed-index form \eqref{eq:app_A3_mixed_complete_gluon}, which itself is the
standard color-ordered three-gluon tree
\cite{Elvang:2013cua,Bern:2007dw} with dimensional-scalar components
\cite{Badger:2005zh}.  For the cyclic ordering \((1,2,3)\), with leg \(2\)
carrying the scalar-chain label \(K\), the two mixed-polarization components,
keeping the reference spinors explicit, are
\begin{align}
A_3^{(0)}(1^+,2^K,3^-)
&=
i\,
\frac{\langle r_1\,3\rangle[1\,r_3]}
{\langle r_1\,1\rangle[3\,r_3]}\,
\bigl(\lambda_{k_3}-\lambda_{k_1}\bigr)^K,
\\
A_3^{(0)}(1^-,2^K,3^+)
&=
-i\,
\frac{\langle r_3\,1\rangle[3\,r_1]}
{\langle r_3\,3\rangle[1\,r_1]}\,
\bigl(\lambda_{k_3}-\lambda_{k_1}\bigr)^K .
\label{eq:app_A3_scalar_gluon_gluon}
\end{align}
Choosing the reference spinors for the two gluons in each three-point
factor, \(\hat b\) in the first channel and \(\hat a\) in the second, gives
\begin{align}
A_3^{(0)}(\hat a^+,X_3^K,-\widehat P_{a3}^{-})
&=
-i\,\lambda_{X_3}^K\,
\frac{\langle \hat b\,\widehat P_{a3}\rangle[\hat a\,\hat b]}
{\langle \hat b\,\hat a\rangle[\widehat P_{a3}\,\hat b]},
\\
A_3^{(0)}(\widehat P_{a23}^{-},\hat b^+,X_1^I)
&=
-i\,\lambda_{X_1}^I\,
\frac{\langle \hat a\,\widehat P_{a23}\rangle[\hat b\,\hat a]}
{\langle \hat a\,\hat b\rangle[\widehat P_{a23}\,\hat a]} .
\label{eq:app_A3_sgg_evaluated}
\end{align}
The transverse signs follow from
\(\lambda_{\widehat P_{a3}}=\lambda_{X_3}\) and
\(\lambda_{\widehat P_{a23}}=-\lambda_{X_1}\), because the observed gluons
carry no transverse momentum.

Substituting \eqref{eq:app_A5_inserted_A4_factors} and
\eqref{eq:app_A3_sgg_evaluated} into the two terms of
\eqref{eq:g6_A5_BCFW}, the spinor factors from the mixed three-point amplitude
and the adjacent four-point scalar-chain tree combine as
\begin{equation}
\frac{\langle \hat b\,\widehat P_{a3}\rangle[\hat a\,\hat b]}
{\langle \hat b\,\hat a\rangle[\widehat P_{a3}\,\hat b]}\,
\frac{[\widehat P_{a3}\,\hat b]}
{\langle \widehat P_{a3}\,\hat b\rangle}
=
\frac{[ab]}{\langle ab\rangle},
\qquad
\frac{\langle \hat a\,\widehat P_{a23}\rangle[\hat b\,\hat a]}
{\langle \hat a\,\hat b\rangle[\widehat P_{a23}\,\hat a]}\,
\frac{[-\widehat P_{a23}\,\hat a]}
{\langle -\widehat P_{a23}\,\hat a\rangle}
=
\frac{[ab]}{\langle ab\rangle}.
\label{eq:app_A5_spinor_factor_from_BCFW}
\end{equation}
Here \(\langle\hat a\,\hat b\rangle=\langle ab\rangle\) and
\([\hat a\,\hat b]=[ab]\) under the \([a,b\rangle\) shift.  Thus the two
observed adjacent positive-helicity gluons on each pentagon supply the expected
factor \([ab]/\langle ab\rangle\).  For scalar-chain labels \(I,J,K\), the
generic scalar-chain component of the five-point tree is therefore
\begin{equation}
\begin{aligned}
&A_5^{(0)}(a^+,b^+,X_1^I,X_2^J,X_3^K)
\\
&\quad =
i\,\frac{[ab]}{\langle ab\rangle}
\Bigl[
G_F(P_{a3}^2)\,
G_F(\widehat D_{a3})\,
(\lambda_{X_1}\!\cdot\!\lambda_{X_2})\,
\delta^{IJ}\lambda_{X_3}^{K}
\\
&\hspace{3.5cm}
{}+
G_F(P_{a23}^2)\,
G_F(\widehat D_{a23})\,
(\lambda_{X_2}\!\cdot\!\lambda_{X_3})\,
\lambda_{X_1}^{I}\delta^{JK}
\Bigr],
\end{aligned}
\label{eq:app_A5_generic_scalar_chain}
\end{equation}
The hatted momenta in \(\widehat D_{a3}\) and \(\widehat D_{a23}\) are evaluated
at the corresponding BCFW poles.  In a fixed routed application these
propagator factors are then rewritten as the corresponding unhatted common
slots after the pole conditions are used.  The formula
\eqref{eq:app_A5_generic_scalar_chain} is the single-pentagon scalar-chain
component.  In the \(\{5,5\}\) product below, the three sewn legs are instead
kept as complete \(D_s\)-dimensional physical states until both pentagons have
been multiplied.  The factors \(s_{12}\) and \(s_{23}\) are associated with
the corresponding BCFW pole structures only after the complete state
contraction has been performed.

We next evaluate the pentagon--pentagon contraction with complete
\(D_s\)-dimensional physical states.  For representative \(55_3\), define
\begin{equation}
\begin{aligned}
{\cal S}_{55}
:={}&
\sum_{h_1,h_2,h_3\in{\rm phys}(D_s)}
A_5^{(0)}
\bigl(1^+,2^+,(-\ell_1)^{\bar h_1},
(-\ell_2)^{\bar h_2},(-\ell_3)^{\bar h_3}\bigr)
\\
&\qquad\times
A_5^{(0)}
\bigl(3^+,4^+,\ell_3^{h_3},\ell_2^{h_2},\ell_1^{h_1}\bigr).
\end{aligned}
\label{eq:app_g6_55_full_product_definition}
\end{equation}
The BCFW expansion produces three distinct four-propagator structures,
\begin{equation}
{\cal S}_{55}
=
{\cal S}_{55}^{(1)}
+{\cal S}_{55}^{(2)}
+{\cal S}_{55}^{(3)}.
\label{eq:app_g6_55_pole_decomposition}
\end{equation}
We display explicitly the calculation of
\({\cal S}_{55}^{(2)}\), which gives the representative pole family used in
the main text.  Before conversion to the common routing, its propagator
factor is
\begin{equation}
{\cal D}_{55}^{(2)}
=
\frac{s_{12}}
 {(\ell_3+\ell_2)^2
  (\ell_1-k_2)^2
  (\ell_2+\ell_1)^2
  (\ell_3+k_4)^2}.
\label{eq:app_g6_55_second_pole_denominators}
\end{equation}
All quadratic factors in
\eqref{eq:app_g6_55_second_pole_denominators} are \(D\)-dimensional.

Using
\begin{equation}
\ell_1=p,\qquad
\ell_2=-p-q,\qquad
\ell_3=q-k_{34},
\label{eq:app_g6_55_pole_routing}
\end{equation}
and applying the reflection of the external labels
\[
(1,2,3,4)\mapsto(2,1,4,3),
\]
the ratio \([12][34]/(\langle12\rangle\langle34\rangle)\) is unchanged.  The
four denominators then become
\begin{equation}
\begin{aligned}
(\ell_3+\ell_2)^2&\longrightarrow D_3,
&
(\ell_1-k_2)^2&\longrightarrow D_2,
\\
(\ell_2+\ell_1)^2&\longrightarrow D_4,
&
(\ell_3+k_4)^2&\longrightarrow D_5 .
\end{aligned}
\label{eq:app_g6_55_representative_routing}
\end{equation}
The common factor in this routed component is therefore
\begin{equation}
\begin{aligned}
{\cal F}_{L}
&:=
\frac{[12]}{\langle12\rangle}\,G_2G_3,
&
{\cal F}_{R}
&:=
\frac{[34]}{\langle34\rangle}\,G_4G_5,
\\
{\cal F}_{55}
&:=
{\cal F}_{L}{\cal F}_{R}
=
\frac{[12][34]}{\langle12\rangle\langle34\rangle}\,
G_2G_3G_4G_5 .
\end{aligned}
\label{eq:app_g6_55_common_factor}
\end{equation}
After extracting this factor, the complete physical-state contraction in
\({\cal S}_{55}^{(2)}\) defines the stripped numerator
\({\cal N}_{55}\):
\begin{equation}
{\cal S}_{55}^{(2)}
\longrightarrow
{\cal F}_{55}\,{\cal N}_{55}.
\label{eq:app_g6_55_A5_product}
\end{equation}
The arrow in \eqref{eq:app_g6_55_A5_product} includes the common routing and
the reflected external labelling in
\eqref{eq:app_g6_55_representative_routing}.  It applies only to
\({\cal S}_{55}^{(2)}\), not to the complete unprojected product
\({\cal S}_{55}\).  We now evaluate \({\cal N}_{55}\) by applying BCFW
recursion to the left and right pentagons.  We write the channel amplitudes as
\begin{equation}
A_{L}^{\bar h_1\bar h_2\bar h_3}
:=
A_{L,1}^{\bar h_1\bar h_2\bar h_3}
+A_{L,2}^{\bar h_1\bar h_2\bar h_3},
\qquad
A_{R}^{h_3h_2h_1}
:=
A_{R,1}^{h_3h_2h_1}
+A_{R,2}^{h_3h_2h_1}.
\label{eq:app_g6_55_left_right_amplitudes}
\end{equation}
The four terms are
\begin{equation}
\begin{aligned}
A_{L,1}^{\bar h_1\bar h_2\bar h_3}
&=
\sum_{\sigma}
A_3^{(0)}
\bigl(\hat1^+,(k_{34}-q)^{\bar h_3},
-\widehat P_{L1}^{\,\bar\sigma}\bigr)\,
iG_F(P_{L1}^2)
A_4^{(0)}
\bigl(\widehat P_{L1}^{\,\sigma},\hat2^+,
(-p)^{\bar h_1},(p+q)^{\bar h_2}\bigr),
\\
A_{L,2}^{\bar h_1\bar h_2\bar h_3}
&=
\sum_{\sigma}
A_4^{(0)}
\bigl(\hat1^+,(p+q)^{\bar h_2},(k_{34}-q)^{\bar h_3},
-\widehat P_{L2}^{\,\bar\sigma}\bigr)\,
iG_F(P_{L2}^2)
A_3^{(0)}
\bigl(\widehat P_{L2}^{\,\sigma},\hat2^+,(-p)^{\bar h_1}\bigr),
\\
A_{R,1}^{h_3h_2h_1}
&=
\sum_{\sigma}
A_3^{(0)}
\bigl(\hat3^+,p^{h_1},-\widehat P_{R1}^{\,\bar\sigma}\bigr)\,
iG_F(P_{R1}^2)
A_4^{(0)}
\bigl(\widehat P_{R1}^{\,\sigma},\hat4^+,
(q-k_{34})^{h_3},(-p-q)^{h_2}\bigr),
\\
A_{R,2}^{h_3h_2h_1}
&=
\sum_{\sigma}
A_4^{(0)}
\bigl(\hat3^+,(-p-q)^{h_2},p^{h_1},
-\widehat P_{R2}^{\,\bar\sigma}\bigr)\,
iG_F(P_{R2}^2)
A_3^{(0)}
\bigl(\widehat P_{R2}^{\,\sigma},\hat4^+,
(q-k_{34})^{h_3}\bigr).
\end{aligned}
\label{eq:app_g6_55_channel_amplitudes}
\end{equation}
The unshifted BCFW propagator momenta in these four channel amplitudes are
\begin{equation}
\begin{aligned}
P_{L1}&=k_1+k_{34}-q,
&
P_{L2}&=k_1+p+k_{34},
\\
P_{R1}&=k_3+p,
&
P_{R2}&=k_3-q .
\end{aligned}
\label{eq:app_g6_55_unhatted_channel_momenta}
\end{equation}
The corresponding hatted on-shell momenta in the tree amplitudes are
\begin{equation}
\begin{aligned}
\widehat P_{L1}&=\hat k_1+k_{34}-q,
&
\widehat P_{L2}&=\hat k_1+p+k_{34},
\\
\widehat P_{R1}&=\hat k_3+p,
&
\widehat P_{R2}&=\hat k_3-q .
\end{aligned}
\label{eq:app_g6_55_channel_momenta}
\end{equation}
In each case the hatted momenta are evaluated at the corresponding BCFW pole,
and \(\sigma\) ranges over the complete physical \(D_s\)-dimensional internal
state of that BCFW factorization channel.
We now restrict the channel products to the pole structure
\({\cal S}_{55}^{(2)}\).  After its propagators have been converted to the
common routing in \eqref{eq:app_g6_55_representative_routing}, each of the
four physical-state contraction pieces contains the common factor
\({\cal F}_{55}\).  We define the corresponding stripped numerator
coefficients by
\begin{equation}
\begin{aligned}
&
\sum_{h_1,h_2,h_3\in{\rm phys}(D_s)}
A_{L,\alpha}^{\bar h_1\bar h_2\bar h_3}\,
A_{R,\beta}^{h_3h_2h_1}
\bigg |_{{\cal S}_{55}^{(2)}}
=
{\cal F}_{55}\,{\cal N}_{\alpha\beta},
\qquad \alpha,\beta\in\{1,2\},
\\
&{\cal N}_{55}
=
\sum_{\alpha,\beta=1}^{2}{\cal N}_{\alpha\beta}.
\end{aligned}
\label{eq:app_g6_55_channel_product_sum}
\end{equation}
Here the vertical bar denotes the contribution to the selected pole
structure; it does not restrict the physical polarization sum.  The other
pole structures are evaluated in the same way before being assigned to their
corresponding routed representatives.
With
\begin{equation}
a=\lambda_p^2,\qquad
b=\lambda_p\!\cdot\!\lambda_q,\qquad
c=\lambda_q^2,\qquad
d=D_s-2,
\end{equation}
and \(\lambda_{p+q}^2=a+2b+c\), the four numerator contributions to
\({\cal S}_{55}^{(2)}\) evaluate to
\begin{equation}
\begin{aligned}
{\cal N}_{11}
&=
s_{12}\left(8b^2+\frac12\,d\,ac\right),
&
{\cal N}_{12}
&=
s_{12}\left[d\,c(a+2b+c)-8ac\right],
\\
{\cal N}_{21}
&=
s_{12}\left[d\,a(a+2b+c)-8ac\right],
&
{\cal N}_{22}
&=
s_{12}\left(8b^2+\frac12\,d\,ac\right).
\end{aligned}
\label{eq:app_g6_55_channel_by_channel}
\end{equation}
The diagonal terms are the two products in which the same BCFW attachment is
chosen on both pentagons; they give the two \(8s_{12}b^2\) terms in the
four-dimensional state sum and the two equal halves of the scalar-chain trace
\(d\,s_{12}ac\).  The mixed terms are the two products with opposite BCFW
attachments; their \(-8s_{12}ac\) terms are the two off-diagonal terms in the
antisymmetric four-dimensional contraction, while the \(d\)-terms are the left
and right projected state traces.  Summing the four terms gives
\begin{equation}
\begin{aligned}
{\cal N}_{55}
&=
s_{12}\Bigl[
d\bigl(a^2+c^2+3ac+2ab+2bc\bigr)
+16(b^2-ac)
\Bigr].
\end{aligned}
\label{eq:app_g6_55_channel_sum}
\end{equation}
Rearranging the same four numerator contributions according to the physical
basis used for the complete state sums gives
\begin{equation}
{\cal N}_{55}
=
{\cal N}_{55}^{\rm sc}
+{\cal N}_{55}^{4D}
+{\cal N}_{55}^{\perp} .
\label{eq:app_g6_55_state_sum_split}
\end{equation}
The scalar-chain part of the projected and stripped contribution
\eqref{eq:app_g6_55_channel_product_sum} is
\begin{equation}
{\cal N}_{55}^{\rm sc}
=
ac\Bigl[
-2d\,s_{12}
+d(5a+8b+5c)
-6(a+b+c)
\Bigr].
\label{eq:app_g6_55_state_sum_sc}
\end{equation}
The four-dimensional helicity part of the same stripped product is
\begin{equation}
{\cal N}_{55}^{4D}
=
16s_{12}(b^2-ac),
\label{eq:app_g6_55_state_sum_4D}
\end{equation}
and the transverse part is
\begin{equation}
{\cal N}_{55}^{\perp}
=
s_{12}d(a^2+c^2+5ac+2ab+2bc)
-d\,ac(5a+8b+5c)+6ac(a+b+c).
\label{eq:app_g6_55_state_sum_transverse_remainder}
\end{equation}
Hence
\begin{equation}
{\cal N}_{55}
=
s_{12}\Bigl[
d\bigl(ac+a(a+2b+c)+c(a+2b+c)\bigr)
+16\bigl(b^2-ac\bigr)
\Bigr].
\label{eq:app_g6_55_state_sum_decomposition}
\end{equation}
Equivalently,
\begin{equation}
{\cal N}_{55}
=
s_{12}\Bigl[
d\bigl(
\lambda_p^2\lambda_q^2
+\lambda_p^2\lambda_{p+q}^2
+\lambda_q^2\lambda_{p+q}^2
\bigr)
+16\bigl((\lambda_p\!\cdot\!\lambda_q)^2-\lambda_p^2\lambda_q^2\bigr)
\Bigr].
\label{eq:app_g6_55_state_sum}
\end{equation}
It is convenient to define
\begin{equation}
{\cal N}_{55}^{\rm sum}
:=
\frac{{\cal N}_{55}}{s_{12}}
=
d\bigl(
\lambda_p^2\lambda_q^2
+\lambda_p^2\lambda_{p+q}^2
+\lambda_q^2\lambda_{p+q}^2
\bigr)
+16\bigl(
(\lambda_p\!\cdot\!\lambda_q)^2
-\lambda_p^2\lambda_q^2
\bigr).
\label{eq:app_g6_55_reduced_numerator}
\end{equation}
Thus the explicitly evaluated pole structure
\({\cal S}_{55}^{(2)}\) carries the numerator
\({\cal N}_{55}=s_{12}{\cal N}_{55}^{\rm sum}\).

The same complete physical-state contraction for the other two BCFW pole
structures gives
\begin{equation}
\begin{aligned}
{\cal S}_{55}^{(1)}
={}&
\frac{[12][34]}{\langle12\rangle\langle34\rangle}\,
{\cal N}_{55}^{\rm sum}
\frac{s_{23}}
 {(\ell_3-k_1)^2
  (\ell_1-k_2)^2
  (\ell_1+k_3)^2
  (\ell_3+k_4)^2},
\\[1mm]
{\cal S}_{55}^{(3)}
={}&
\frac{[12][34]}{\langle12\rangle\langle34\rangle}\,
{\cal N}_{55}^{\rm sum}
\frac{s_{12}}
 {(\ell_3+\ell_2)^2
  (\ell_1+k_3)^2
  (\ell_2+\ell_1)^2
  (\ell_3-k_1)^2}.
\end{aligned}
\label{eq:app_g6_55_remaining_pole_structures}
\end{equation}
Together with the explicitly displayed contribution
\({\cal S}_{55}^{(2)}\), these expressions reconstruct the complete
pentagon--pentagon product according to
\eqref{eq:app_g6_55_pole_decomposition}.  The three pole structures contain
the same state-summed numerator but are kept distinct until their respective
common routings and external relabellings have been performed.

\section{State-summed BCFW building blocks}
\label{app:g6_bcfw_building_blocks}

The octagon recursion requires lower-point color-ordered trees with complete
physical \(D_s\)-dimensional state sums.  We collect the needed building blocks
here so that no internal state is implicitly restricted to a scalar-chain
projection.  Let
\begin{equation}
\mathcal{S}_{L,R}[X]
:=\sum_{h\in{\rm phys}(D_s)}
X_L(\ldots,-P^{\bar h})\,X_R(P^h,\ldots),
\qquad
P^2-\lambda_P^2=0,
\label{eq:app_bcfw_complete_state_contraction}
\end{equation}
where the arguments on both factors retain their inherited cyclic order.  The
sum runs over the complete physical \(D_s\)-dimensional
gluon polarization space.  No transverse scalars or scalar-chain amplitudes
are introduced as additional physical states.  A split into four- and
extra-dimensional components is only an optional algebraic check after the on-shell
gluon state sum.

The open-index four-point building block follows from on-shell three-gluon
recursion.  For two positive-helicity adjacent external legs it has the form
\begin{equation}
A_4^{(0)}((-P)^I,i^+,j^+,Q^J)
=-i\,\frac{[ij]}{\langle ij\rangle}
G_F((P+k_i)^2-\lambda_P^2)
\lambda_P\!\cdot\!\lambda_Q\,\delta^{IJ}.
\label{eq:app_bcfw_A4_state_block}
\end{equation}

For a five-point tree, we choose a valid BCFW shift of two external gluons.  Then
\begin{equation}
A_5^{(0)}
=\sum_{\mathcal P_5}\sum_{h\in{\rm phys}(D_s)}
A_{L,\mathcal P_5}^{(0)}(\widehat k_a,\ldots,-P^{\bar h})
\,iG_F(P^2-\lambda_P^2)\,
A_{R,\mathcal P_5}^{(0)}(P^h,\ldots,\widehat k_b).
\label{eq:app_bcfw_A5_state_recursion}
\end{equation}
Substitution of the on-shell three-gluon recursion gives the complete
five-point state sum used in appendix~\ref{app:five_point_trees}; the sum runs
over the full physical gluon polarization \(h\).

The six-point building block is obtained by applying the same recursion once
more.  For a fixed cyclic ordering and a chosen valid shift,
\begin{equation}
A_6^{(0)}
=\sum_{\mathcal P_6}\sum_{h\in{\rm phys}(D_s)}
A_{L,\mathcal P_6}^{(0)}(\widehat k_a,\ldots,-P^{\bar h})
\,iG_F(P^2-\lambda_P^2)\,
A_{R,\mathcal P_6}^{(0)}(P^h,\ldots,\widehat k_b),
\label{eq:app_bcfw_A6_state_recursion}
\end{equation}
where each five-point factor is replaced by
\eqref{eq:app_bcfw_A5_state_recursion}.  The resulting expression is a sum
over ordered products of on-shell three-gluon trees, with one
complete physical-state contraction at every internal BCFW pole.  Applying
\eqref{eq:app_bcfw_A4_state_block} at the terminal four-point factors gives
the $A_6$ state-summed residues used in the \(\{6,4\}\) channel.

For the valid non-adjacent shift
\begin{equation}
\widehat k_1(z)=k_1+z\eta,
\qquad
\widehat k_4(z)=k_4-z\eta,
\label{eq:app_bcfw_A6_14_shift}
\end{equation}
a planar factorization pole depends on \(z\) precisely when its contiguous
slot set contains one of \(1,4\), but not both.  We orient every partition so
that the first ordered word \(X\) contains leg \(1\), while the second word
\(Y\) contains leg \(4\).  The complete partition set is
\begin{equation}
\begin{aligned}
{\cal P}_{6}^{(1|4)}=\{&
(12\,|\,3456),
(123\,|\,456),
(61\,|\,2345),
(561\,|\,234),
\\[-1mm]
&
(612\,|\,345),
(5612\,|\,34),
(6123\,|\,45)\}.
\end{aligned}
\label{eq:app_bcfw_A6_14_partitions}
\end{equation}
Equivalently, these are the four two-particle poles
\(P_{12},P_{34},P_{45},P_{61}\) and the three independent three-particle
poles \(P_{123},P_{234},P_{345}\).  Complementary channel momenta in
\eqref{eq:app_bcfw_A6_14_partitions} are oriented so that \(X\) always contains
leg \(1\).
Here the entries in \(X\) and \(Y\) retain their inherited cyclic order.  For
\(P_X=\sum_{r\in X}k_r\),
\(\widehat P_X=P_X(z_X)\) is on shell at the corresponding BCFW pole, the
six-point recursion is therefore
\begin{equation}
A_6^{(0)}
=
\sum_{(X|Y)\in{\cal P}_{6}^{(1|4)}}
\sum_{h\in{\rm phys}(D_s)}
A_{L,X}^{(0)}
\bigl(\widehat X,-\widehat P_X^{\bar h}\bigr)
iG_F(P_X^2-\lambda_{P_X}^2)\,
A_{R,Y}^{(0)}
\bigl(\widehat P_X^h,\widehat Y\bigr).
\label{eq:app_bcfw_A6_explicit_channels}
\end{equation}
The hat on an ordered word acts only on its shifted entry: leg \(1\) in \(X\)
and leg \(4\) in \(Y\).  Thus
\eqref{eq:app_bcfw_A6_14_partitions} contains two channels of type
\(\{A_3,A_5\}\), three of type \(\{A_4,A_4\}\), and two of type
\(\{A_5,A_3\}\).  In particular, \(P_{1234}\) is absent because it
contains both shifted legs and is independent of \(z\).  Every $A_5$ factor in
\eqref{eq:app_bcfw_A6_explicit_channels} is replaced by the complete recursion
\eqref{eq:g6_A5_BCFW}, leaving only three- and four-point state contractions.

The seven-point building block needed for the endpoint octagon channels is
constructed in the same way.  With a valid shift of legs \(1\) and \(4\),
\begin{equation}
A_7^{(0)}
=\sum_{\mathcal P_7}\sum_{h\in{\rm phys}(D_s)}
A_{L,\mathcal P_7}^{(0)}(\widehat k_1,\ldots,-P^{\bar h})
\,iG_F(P^2-\lambda_P^2)\,
A_{R,\mathcal P_7}^{(0)}(P^h,\ldots,\widehat k_4),
\label{eq:app_bcfw_A7_state_recursion}
\end{equation}
where the admissible partitions, including their cyclic multiplicities, fall
into the four channel types
\begin{equation}
\{A_3,A_6\},\qquad
\{A_4,A_5\},\qquad
\{A_5,A_4\},\qquad
\{A_6,A_3\}.
\label{eq:app_bcfw_A7_channel_types}
\end{equation}
Substituting the \(A_5\) and \(A_6\) recursions above reduces every \(A_7\)
residue to on-shell three-gluon state contractions.  The internal gluon
polarization \(h\) is summed before any denominator identification.

For the octagon, the valid adjacent shift of legs \(2\) and \(1\) gives
\begin{equation}
A_8^{(0)}
=\sum_{\mathcal P_8}\sum_{h\in{\rm phys}(D_s)}
A_{L,\mathcal P_8}^{(0)}(\widehat k_2,\ldots,-P^{\bar h})
\,iG_F(P^2-\lambda_P^2)\,
A_{R,\mathcal P_8}^{(0)}(P^h,\ldots,\widehat k_1).
\label{eq:app_bcfw_A8_state_recursion}
\end{equation}
The five admissible partitions have lower-point types
\begin{equation}
\{A_3,A_7\},\quad
\{A_4,A_6\},\quad
\{A_5,A_5\},\quad
\{A_6,A_4\},\quad
\{A_7,A_3\},
\label{eq:app_bcfw_A8_channel_types}
\end{equation}
with the cyclic placements determined by the shifted ordering.  Inserting
\eqref{eq:app_bcfw_A5_state_recursion} and
\eqref{eq:app_bcfw_A6_state_recursion} gives a complete nested state sum for
each channel.  The \(D_7\) channel reduces to the two five-point product
used in \eqref{eq:app_g6_octagon_bridge_factorization}; the remaining channels
are reduced by the \(A_4\) and \(A_6\) building blocks above.  Only after these
state sums have been performed may the resulting terms be projected onto the
double-box or bow-tie denominator basis.

This construction makes the state content explicit at every recursion level.
In particular, replacing the sums in
\eqref{eq:app_bcfw_A5_state_recursion}--\eqref{eq:app_bcfw_A8_state_recursion}
by a partial polarization projection would omit the four-dimensional MHV
octagon terms.

\section{The \texorpdfstring{$g^6$}{g6} one-polygon octagon}
\label{app:g6_octagon_k1}

\subsection{State content and representative factorization}
\label{app:g6_octagon_factorization}

The \(\{8\}\) sector at order \(g^6\) in subsection~\ref{sec:g6_octagon_sector}
contains one octagon with two vacuum pairs.
If all internal states are restricted to four-dimensional gluon helicities, the
two forward pairs provide two negative-helicity entries, and the
corresponding eight-point tree is an ordinary MHV Parke--Taylor amplitude.
Thus the one-polygon octagon cannot be discarded by the same helicity argument
that removed the one-pair hexagon in the strictly four-dimensional sector.

For the planar representative, we define the full all-outgoing physical-state
sum by
\begin{equation}
\sum_{h_1,h_2\in{\rm phys}(D_s)}
A_{8}^{(0)}
\bigl(1^+,2^+,p^{\bar h_1},(q-k_{34})^{\bar h_2},
3^+,4^+,(k_{34}-q)^{h_2},(-p)^{h_1}\bigr).
\label{eq:app_g6_octagon_representative}
\end{equation}
For the purely four-dimensional gluon component, crossing relates the two
helicities belonging to each forward pair.  Since the four observed gluons
have positive helicity, a nonzero term has one negative-helicity endpoint on
each pair.  Consequently this component is the sum of four, rather than six,
MHV Parke--Taylor terms.  In four dimensions it is given by
\begin{equation}
\sum_{a\in\{3,8\}}\sum_{b\in\{4,7\}}
\frac{\langle a b\rangle^4}
{\langle12\rangle\langle23\rangle\cdots\langle81\rangle},
\label{eq:app_g6_octagon_mhv_components}
\end{equation}
where the positions refer to the ordering in
\eqref{eq:app_g6_octagon_representative}: positions \(3,8\) form the first
forward pair and positions \(4,7\) the second.  This is only the
four-dimensional helicity decomposition.  The complete calculation instead
uses the physical \(D_s\)-dimensional gluon polarization projector.  Its
extra-dimensional tensor components are part of the same gluon state sum,
not additional transverse-scalar external states, and generate the
\(\lambda\)-dependent two-loop numerator terms after contraction.

This MHV tree contribution is not scaleless and does not vanish separately in
dimensional regularization.  As explained in
subsection~\ref{sec:g6_octagon_sector}, supersymmetry constrains the complete
state and forest sum.  The complete octagon state sum includes both the
four-dimensional and extra-dimensional components and therefore carries
\(\lambda\)-dependent numerator terms.  The factorization below retains this
complete state sum while identifying the labelled supports quoted in the main
text.

To evaluate the complete state sum in
\eqref{eq:app_g6_octagon_representative}, we apply a valid adjacent two-line BCFW shift to the
external positive-helicity legs \(2\) and \(1\).  We linearize the cyclic
ordering starting at leg \(2\), so that the shifted legs are the first and last
entries,
\begin{equation}
\widehat k_2(z)=k_2+z\eta,\qquad
\widehat k_1(z)=k_1-z\eta,\qquad
\eta^2=\eta\!\cdot k_2=\eta\!\cdot k_1=0.
\label{eq:app_g6_octagon_bcfw_shift}
\end{equation}
The octagon then decomposes into a sum over its admissible factorization
channels,
\begin{equation}
\begin{aligned}
\sum_{h_1,h_2\in{\rm phys}(D_s)}A_8^{(0)}
&=\sum_{\rm channels}\sum_{h\in{\rm phys}(D_s)}
A_L^{(0)}(\widehat k_2,\ldots,-\widehat P^{\bar h})
\\[-1mm]
&\hspace{2.0cm}\times
\,iG_F(P^2-\lambda_P^2)\,
A_R^{(0)}(\widehat P^h,\ldots,\widehat k_1).
\end{aligned}
\label{eq:app_g6_octagon_bcfw_recursion}
\end{equation}
Thus the negative-momentum half-edge carries \(\bar h\), while the
positive-momentum half-edge carries \(h\), in agreement with
\eqref{eq:app_bcfw_complete_state_contraction}.  The internal momentum
\(\widehat P\) is evaluated at the corresponding BCFW pole, whereas \(P\) in
the propagator is the unshifted channel momentum.
The five admissible channels are retained before any state or denominator
projection is made.  Their lower-point factors are then reduced with the
three-, four-, and five-point identities collected in
appendix~\ref{app:g6_bcfw_building_blocks}.
With the routing of the positive-energy representatives used in
subsection~\ref{sec:g6_octagon_sector}, the two vacuum-pair phase spaces impose
\begin{equation}
p^2-\lambda_p^2=D_1,\qquad
(q-k_{34})^2-\lambda_q^2=D_6,
\label{eq:app_g6_octagon_cut_slots}
\end{equation}
because \(k_{34}\) has no transverse component.  The non-degenerate
seven-slot contribution is the bridge factorization with internal momentum
\(p+q\).  In this channel the octagon reduces to
\begin{equation}
\begin{aligned}
&\sum_{h_1,h_2\in{\rm phys}(D_s)}
A_{8}^{(0)}
\bigl(1^+,2^+,p^{\bar h_1},(q-k_{34})^{\bar h_2},
3^+,4^+,(k_{34}-q)^{h_2},(-p)^{h_1}\bigr)
\Big|_{D_7}
\\
&\quad =
\sum_{h_1,h_2,h\in{\rm phys}(D_s)}
A_5^{(0)}
\bigl(p^{\bar h_1},(q-k_{34})^{\bar h_2},3^+,4^+,
(-p-q)^{\bar h}\bigr)
\,iG_F(D_7)\,
\\
&\hspace{3.0cm}\times
A_5^{(0)}
\bigl((p+q)^h,(k_{34}-q)^{h_2},(-p)^{h_1},1^+,2^+\bigr).
\end{aligned}
\label{eq:app_g6_octagon_bridge_factorization}
\end{equation}
Here \(h\) is the complete state running through the BCFW bridge; it is not an
additional vacuum-pair label.  The bridge momentum is fixed by
\begin{equation}
 p+(q-k_{34})+k_3+k_4-(p+q)=0,
\qquad
(p+q)+(k_{34}-q)-p+k_1+k_2=0.
\end{equation}
Equations~\eqref{eq:app_g6_octagon_cut_slots} and
\eqref{eq:app_g6_octagon_bridge_factorization} establish the two opened slots
and the uncut bridge slot used in
\eqref{eq:g6_octagon_two_cut_16}.  The corresponding numerator information is
contained in the complete sums over \(h_1,h_2\), and \(h\).  Projecting the
endpoint labels \(h_1,h_2\) onto scalar-chain indices would select only part of
this physical state sum.  In particular, it would omit the non-zero
four-dimensional MHV endpoint contribution.
The same octagon has a second non-degenerate projection in which the two side
chains are connected through the external \(s_{12}\) channel rather than through
the loop-dependent bridge \(D_7\).  This channel has no \(D_7\) pole and belongs
to the local six-denominator bow-tie family.  The explicitly routed
representative has denominator support
\begin{equation}
\delta_{\bar1}^+G_2G_3G_4G_5\delta_{\bar6}^+ .
\label{eq:app_g6_octagon_bowtie_projection}
\end{equation}
The exhaustive \(\{8\}\) bow-tie support set is
\eqref{eq:g6_octagon_bowtie_two_cut_supports}; it follows by placing the two
vacuum pairs at all inequivalent positions in the cyclic tree rather than from
this single routing.  The pole residues of all cyclic assignments are reduced
and projected as
described in subsection~\ref{app:g6_octagon_numerator_reconstruction} and are
retained with their signed supports for the family comparison.

\subsection{Octagon pole and numerator collection}
\label{app:g6_octagon_numerator_reconstruction}

We finally describe how the individual octagon residues are collected into the
three denominator families.  Let \(\gamma\) denote a cyclic assignment, \(S\)
its signed two-cut word, and \(S_0\) the underlying numerical slot set obtained
by dropping the orientation bars.  We expand the octagon with
\eqref{eq:app_bcfw_A8_state_recursion}, recursively reduce its lower-point
factors with
\eqref{eq:app_bcfw_A5_state_recursion}--\eqref{eq:app_bcfw_A7_state_recursion},
and sort the resulting terms into the three denominator families.  From each
term we strip the common factor
\(T=[12][34]/(\langle12\rangle\langle34\rangle)\) and all propagator factors.
The relevant propagator sets are \(\{G_1,\ldots,G_7\}\) for the planar double box,
\(\{\widetilde G_1,\ldots,\widetilde G_7\}\) for the non-planar double box,
and \(\{G_1,\ldots,G_6\}\) for the bow-tie contribution.  For a support \(S\),
the propagator product being stripped contains the members of the corresponding
set that are not opened by \(S_0\).  We denote the numerator contribution that
remains by \({\cal R}_{8,\gamma,S}^G\), where
\(G={\rm P},{\rm NP},{\rm BT}\) labels the family.

For the planar \(16\) representative, the \(D_7\) channel is precisely the
\(A_5\times A_5\) product in
\eqref{eq:app_g6_octagon_bridge_factorization}.  To display its reduction,
introduce the ordered sewn-leg triples
\begin{equation}
\begin{aligned}
(X_{12,1},X_{12,2},X_{12,3})
&=\bigl((p+q)^h,(k_{34}-q)^{h_2},(-p)^{h_1}\bigr),
\\
(X_{34,1},X_{34,2},X_{34,3})
&=\bigl((-p-q)^{\bar h},p^{\bar h_1},
(q-k_{34})^{\bar h_2}\bigr).
\end{aligned}
\label{eq:app_octagon_16_sewn_triples}
\end{equation}
For \((a,b)=(1,2)\) or \((3,4)\), the two BCFW terms are
\begin{equation}
\begin{aligned}
{\cal A}_{ab,1}
&=\sum_{\sigma\in{\rm phys}(D_s)}
A_3^{(0)}\bigl(\hat a^+,X_{ab,3},
-\widehat P_{ab,1}^{\,\bar\sigma}\bigr)\,
iG_F(P_{ab,1}^2)\,
A_4^{(0)}\bigl(\widehat P_{ab,1}^{\,\sigma},\hat b^+,
X_{ab,1},X_{ab,2}\bigr),
\\
{\cal A}_{ab,2}
&=\sum_{\sigma\in{\rm phys}(D_s)}
A_4^{(0)}\bigl(\hat a^+,X_{ab,2},X_{ab,3},
-\widehat P_{ab,2}^{\,\bar\sigma}\bigr)\,
iG_F(P_{ab,2}^2)\,
A_3^{(0)}\bigl(\widehat P_{ab,2}^{\,\sigma},\hat b^+,
X_{ab,1}\bigr),
\\
P_{ab,1}&=k_a+X_{ab,3},
\qquad
P_{ab,2}=k_a+X_{ab,2}+X_{ab,3}.
\end{aligned}
\label{eq:app_octagon_16_explicit_A5_channels}
\end{equation}
Thus no five-point amplitude remains.  Applying recursion once more to the two
four-point factors produces the pole monomials that are sorted before the state
contractions are combined.  The planar selection keeps
\(D_2D_3D_4D_5D_7\), as in
\eqref{eq:g6_octagon_planar_selected_poles}.  The local selection keeps
\(D_2D_3D_4D_5\), together with the external \(s_{12}\) pole and the contact
terms generated by the same recursion.  Crossed cyclic orderings are routed
into \(\widetilde D_2\widetilde D_3\widetilde D_4
\widetilde D_5\widetilde D_7\).

Let \(\Gamma_{G,S}\) be the set of cyclic assignments and pole monomials that
route to family \(G\) and signed support \(S\).  Sewing all physical projectors
and summing the residues in each class gives
\begin{equation}
\begin{aligned}
\sum_{\gamma\in\Gamma_{{\rm P},S}}
{\cal R}_{8,\gamma,S}^{\rm P}
&=
B_{\rm DB}^{(1)}+16B_{\rm DB}^{(2)}
={\cal N}_{\rm DB},
\\
\sum_{\gamma\in\Gamma_{{\rm NP},\widetilde S}}
{\cal R}_{8,\gamma,\widetilde S}^{\rm NP}
&=
{\cal N}_{\rm DB}^{\rm crossed},
\\
\sum_{\gamma\in\Gamma_{{\rm BT},S}}
{\cal R}_{8,\gamma,S}^{\rm BT}
&=
B_{\rm BT}^{(1)}+B_{\rm BT}^{(2)}
={\cal N}_{\rm BT}.
\end{aligned}
\label{eq:app_octagon_collected_residues}
\end{equation}
The four factors on the right are displayed explicitly in
\eqref{eq:g6_octagon_explicit_DB_factors} and
\eqref{eq:g6_octagon_explicit_BT_factors}.  The crossed equality follows from
the assignments routed as in
\eqref{eq:g6_octagon_nonplanar_selected_poles}.  The bow-tie equality includes both the
external-channel pole terms and the local contact terms and therefore has no
\(D_7\) propagator.

As checks on this collection, the representative pole complements were found
to be \(23457\), \(\widetilde{23457}\), and \(2345\).  Exhaustive enumeration
gave the planar, crossed, and bow-tie support cardinalities \(15\), \(16\), and
\(9\), respectively, and the algebraic sums in
\eqref{eq:app_octagon_collected_residues} were verified.  Substitution into the
oriented kernels then gives \eqref{eq:g6_octagon_compact_collection}.

\section{Feynman--tree theorem for denominator families}
\label{app:ftt}

This appendix records the denominator identity used in the comparisons with
ordinary Feynman denominator families.  The role of this identity in the
vacuum-pair construction is explained in
subsection~\ref{subsec:equivalence}.  The original discussions of the
Feynman--tree theorem are due to Feynman
\cite{Feynman:1963ax,Feynman:1972mtm}; see also the collection
\cite{Feynman:2000fh}.

For a loop to which the FTT is applied, choose a common loop momentum \(\ell_F\)
and orient every denominator momentum coherently as
\(q_i=\ell_F+\kappa_i\), with \(\kappa_i\) independent of \(\ell_F\).  Write
\(D_i=q_i^2-m_i^2\), with the corresponding generalization for the
dimensionally regulated denominators used in the main text.  We distinguish the
stripped denominator \(G_i\) from the full scalar propagator
\(\Delta_{F,i}\):
\begin{equation}
\begin{aligned}
G_i&\equiv G_F(D_i)=\frac{1}{D_i+i0},
&
\Delta_{F,i}&\equiv\Delta_F(q_i)
=iG_F(D_i)=\frac{i}{D_i+i0},
\\
\delta_i^+&\equiv\delta^+(q_i)
=2\pi\,\theta(q_i^0)\delta(D_i).
\end{aligned}
\label{eq:Gdconv}
\end{equation}
The momentum argument is part of the cut label.  In particular,
\begin{equation}
\delta^+(-q_i)
=2\pi\,\theta(-q_i^0)\delta(D_i)
=\delta^-(q_i)
\neq\delta^+(q_i).
\label{eq:oriented-cut-delta}
\end{equation}
For the same oriented momentum, define the stripped and full advanced
propagators by
\begin{equation}
G_A(q_i)=\frac{1}{D_i-i0\,q_i^0},
\qquad
\Delta_A(q_i)=iG_A(q_i).
\end{equation}
The exact distributional relations are then
\begin{equation}
G_A(q_i)=G_F(D_i)+i\delta^+(q_i),
\qquad
\Delta_A(q_i)=\Delta_F(q_i)-\delta^+(q_i).
\label{eq:advanced-feynman-relations}
\end{equation}
The second relation, involving the full propagators, is the convention in which
the FTT has real alternating coefficients.
For a loop with labelled denominator slots
\begin{equation} \label{eq:F}
F=\{1,\ldots,N\},
\end{equation}
the product of advanced propagators has no pole contribution after the loop
energy integrations.  Thus, for a common numerator \({\cal N}_{F}\),
\begin{equation}
0
=
\int d\Pi_F\,
{\cal N}_{F}
\prod_{i=1}^{N}\Delta_A(q_i),
\end{equation}
where \(d\Pi_F\) denotes the corresponding loop-momentum measure.
Expanding the product gives
\begin{equation}
\begin{aligned}
0
&=
\int d\Pi_F\,
{\cal N}_{F}
\prod_{i=1}^{N}\bigl(\Delta_{F,i}-\delta_i^+\bigr)
\\[2mm]
&=
\int d\Pi_F\,
{\cal N}_{F}
\prod_{i=1}^{N}\Delta_{F,i}
+
\sum_{\emptyset\neq S\subseteq F}
(-1)^{|S|}
\int d\Pi_F\,
{\cal N}_{F}
\prod_{i\in S}\delta_i^+
\prod_{j\notin S}\Delta_{F,j} .
\end{aligned}
\end{equation}
Therefore
\begin{equation} \label{eq:FTT}
\int d\Pi_F\,
{\cal N}_{F}
\prod_{i=1}^{N}\Delta_{F,i}
=
\sum_{\emptyset\neq S\subseteq F}
(-1)^{|S|-1}
\int d\Pi_F\,
{\cal N}_{F}
\prod_{i\in S}\delta_i^+
\prod_{j\notin S}\Delta_{F,j} .
\end{equation}
This is the FTT expansion for one loop with a common numerator.  The algebraic
replacement of a Feynman propagator by an advanced propagator plus an on-shell
term is pointwise, but the disappearance of the all-advanced closed-loop term is
an integrated contour statement.

For a multiloop graph, \eqref{eq:FTT} is applied recursively to every remaining
uncut loop.  Each application opens the selected loop; after all loops have
been opened, no uncut cycle remains and the result is a forest.  The iterative
multiloop identity, including the treatment of shared loop lines, is given in~\cite{Bierenbaum:2010cy}.

\bibliographystyle{JHEP}
\bibliography{references}

\providecommand{\href}[2]{#2}\begingroup\raggedright\begin{thebibliography}{10}

\bibitem{Elvang:2013cua}
H.~Elvang and Y.-t.~Huang, \emph{{Scattering Amplitudes in Gauge Theory and
  Gravity}}, Cambridge University Press (2015),
  [\href{https://arxiv.org/abs/1308.1697}{{\ttfamily 1308.1697}}].

\bibitem{Britto:2004ap}
R.~Britto, F.~Cachazo and B.~Feng, \emph{{New recursion relations for tree
  amplitudes of gluons}},
  \href{https://doi.org/10.1016/j.nuclphysb.2005.02.030}{\emph{Nucl. Phys. B}
  {\bfseries 715} (2005) 499}
  [\href{https://arxiv.org/abs/hep-th/0412308}{{\ttfamily hep-th/0412308}}].

\bibitem{Britto:2005fq}
R.~Britto, F.~Cachazo, B.~Feng and E.~Witten, \emph{{Direct Proof of Tree-Level
  recursion Relation in Yang-Mills Theory}},
  \href{https://doi.org/10.1103/PhysRevLett.94.181602}{\emph{Phys. Rev. Lett.}
  {\bfseries 94} (2005) 181602}
  [\href{https://arxiv.org/abs/hep-th/0501052}{{\ttfamily hep-th/0501052}}].

\bibitem{Bern:2007dw}
Z.~Bern, L.J.~Dixon and D.A.~Kosower, \emph{{On-Shell Methods in Perturbative
  QCD}}, \href{https://doi.org/10.1016/j.aop.2007.04.014}{\emph{Annals Phys.}
  {\bfseries 322} (2007) 1587}
  [\href{https://arxiv.org/abs/0704.2798}{{\ttfamily 0704.2798}}].

\bibitem{Feng:2011np}
B.~Feng and M.~Luo, \emph{{An Introduction to On-shell Recursion Relations}},
  \href{https://doi.org/10.1007/s11467-012-0270-z}{\emph{Front. Phys.}
  {\bfseries 7} (2012) 533} [\href{https://arxiv.org/abs/1111.5759}{{\ttfamily
  1111.5759}}].

\bibitem{Arkani-Hamed:2008yf}
N.~Arkani-Hamed and J.~Kaplan, \emph{{On Tree Amplitudes in Gauge Theory and
  Gravity}}, \href{https://doi.org/10.1088/1126-6708/2008/04/076}{\emph{JHEP}
  {\bfseries 04} (2008) 076} [\href{https://arxiv.org/abs/0801.2385}{{\ttfamily
  0801.2385}}].

\bibitem{Parke:1986gb}
S.J.~Parke and T.R.~Taylor, \emph{{Amplitude for $n$-Gluon Scattering}},
  \href{https://doi.org/10.1103/PhysRevLett.56.2459}{\emph{Phys. Rev. Lett.}
  {\bfseries 56} (1986) 2459}.

\bibitem{Badger:2005zh}
S.D.~Badger, E.W.N.~Glover, V.V.~Khoze and P.~Svrcek, \emph{{Recursion
  relations for gauge theory amplitudes with massive particles}},
  \href{https://doi.org/10.1088/1126-6708/2005/07/025}{\emph{JHEP} {\bfseries
  07} (2005) 025} [\href{https://arxiv.org/abs/hep-th/0504159}{{\ttfamily
  hep-th/0504159}}].

\bibitem{Arkani-Hamed:2017jhn}
N.~Arkani-Hamed, T.-C.~Huang and Y.-t.~Huang, \emph{{Scattering amplitudes for
  all masses and spins}},
  \href{https://doi.org/10.1007/JHEP11(2021)070}{\emph{JHEP} {\bfseries 11}
  (2021) 070} [\href{https://arxiv.org/abs/1709.04891}{{\ttfamily
  1709.04891}}].

\bibitem{Bern:1994zx}
Z.~Bern, L.~Dixon, D.C.~Dunbar and D.A.~Kosower, \emph{{One-loop $n$-point
  gauge theory amplitudes, unitarity and collinear limits}},
  \href{https://doi.org/10.1016/0550-3213(94)90179-1}{\emph{Nucl. Phys. B}
  {\bfseries 425} (1994) 217}
  [\href{https://arxiv.org/abs/hep-ph/9403226}{{\ttfamily hep-ph/9403226}}].

\bibitem{Bern:1994cg}
Z.~Bern, L.~Dixon, D.C.~Dunbar and D.A.~Kosower, \emph{{Fusing gauge theory
  tree amplitudes into loop amplitudes}},
  \href{https://doi.org/10.1016/0550-3213(94)00488-Z}{\emph{Nucl. Phys. B}
  {\bfseries 435} (1995) 59}
  [\href{https://arxiv.org/abs/hep-ph/9409265}{{\ttfamily hep-ph/9409265}}].

\bibitem{Feynman:1963ax}
R.P.~Feynman, \emph{{Quantum theory of gravitation}}, {\emph{Acta Phys. Polon.}
  {\bfseries 24} (1963) 697}.

\bibitem{Feynman:1972mtm}
R.P.~Feynman, \emph{{Closed Loop and Tree Diagrams}},  in \emph{{Magic Without
  Magic: John Archibald Wheeler}}, J.R.~Klauder, ed., (San Francisco),
  pp.~355--375, W. H. Freeman (1972).

\bibitem{Feynman:2000fh}
R.P.~Feynman, \emph{{Selected papers of Richard Feynman: With commentary}},
  vol.~27 of \emph{World Scientific Series in 20th Century Physics}, World
  Scientific (2000), \href{https://doi.org/10.1142/4270}{10.1142/4270}.

\bibitem{Catani:2008xa}
S.~Catani, T.~Gleisberg, F.~Krauss, G.~Rodrigo and J.-C.~Winter, \emph{{From
  loops to trees by-passing Feynman's theorem}},
  \href{https://doi.org/10.1088/1126-6708/2008/09/065}{\emph{JHEP} {\bfseries
  09} (2008) 065} [\href{https://arxiv.org/abs/0804.3170}{{\ttfamily
  0804.3170}}].

\bibitem{Bierenbaum:2010cy}
I.~Bierenbaum, S.~Catani, P.~Draggiotis and G.~Rodrigo, \emph{{A Tree-Loop
  Duality Relation at Two Loops and Beyond}},
  \href{https://doi.org/10.1007/JHEP10(2010)073}{\emph{JHEP} {\bfseries 10}
  (2010) 073} [\href{https://arxiv.org/abs/1007.0194}{{\ttfamily 1007.0194}}].

\bibitem{Caron-Huot:2010fvq}
S.~Caron-Huot, \emph{{Loops and trees}},
  \href{https://doi.org/10.1007/JHEP05(2011)080}{\emph{JHEP} {\bfseries 05}
  (2011) 080} [\href{https://arxiv.org/abs/1007.3224}{{\ttfamily 1007.3224}}].

\bibitem{Maniatis:2015kex}
M.~Maniatis, \emph{{Scattering amplitudes abandoning virtual particles}},
  \href{https://arxiv.org/abs/1511.03574}{{\ttfamily 1511.03574}}.

\bibitem{Maniatis:2016dcf}
M.~Maniatis and C.M.~Reyes, \emph{{Scattering amplitudes from a deconstruction
  of Feynman diagrams}},  \href{https://arxiv.org/abs/1605.04268}{{\ttfamily
  1605.04268}}.

\bibitem{Maniatis:2016nmc}
M.~Maniatis, \emph{{Application of the Feynman-tree theorem together with BCFW
  recursion relations}},
  \href{https://doi.org/10.1142/S0217751X18500422}{\emph{Int. J. Mod. Phys. A}
  {\bfseries 33} (2018) 1850042}
  [\href{https://arxiv.org/abs/1609.00377}{{\ttfamily 1609.00377}}].

\bibitem{Maniatis:2019pig}
M.~Maniatis, \emph{{Application of BCFW-recursion relations and the
  Feynman-tree theorem to the four gluon amplitude with all plus helicities}},
  \href{https://doi.org/10.1103/PhysRevD.100.096022}{\emph{Phys. Rev. D}
  {\bfseries 100} (2019) 096022}
  [\href{https://arxiv.org/abs/1906.10821}{{\ttfamily 1906.10821}}].

\bibitem{Maniatis:2021treeworld}
M.~Maniatis, \emph{{The tree world}},
  \href{https://arxiv.org/abs/2108.07266}{{\ttfamily 2108.07266}}.

\bibitem{Mahlon:1993si}
G.~Mahlon, \emph{{Multigluon helicity amplitudes involving a quark loop}},
  \href{https://doi.org/10.1103/PhysRevD.49.4438}{\emph{Phys. Rev. D}
  {\bfseries 49} (1994) 4438}
  [\href{https://arxiv.org/abs/hep-ph/9312276}{{\ttfamily hep-ph/9312276}}].

\bibitem{Bern:2000dn}
Z.~Bern, L.J.~Dixon and D.A.~Kosower, \emph{{A two-loop four-gluon helicity
  amplitude in QCD}},
  \href{https://doi.org/10.1088/1126-6708/2000/01/027}{\emph{JHEP} {\bfseries
  01} (2000) 027} [\href{https://arxiv.org/abs/hep-ph/0001001}{{\ttfamily
  hep-ph/0001001}}].

\bibitem{Bern:2002tk}
Z.~Bern, A.~De~Freitas and L.J.~Dixon, \emph{{Two-Loop Helicity Amplitudes for
  Gluon-Gluon Scattering in QCD and Supersymmetric Yang-Mills Theory}},
  \href{https://doi.org/10.1088/1126-6708/2002/03/018}{\emph{JHEP} {\bfseries
  03} (2002) 018} [\href{https://arxiv.org/abs/hep-ph/0201161}{{\ttfamily
  hep-ph/0201161}}].

\bibitem{Grisaru:1976vm}
M.T.~Grisaru, H.N.~Pendleton and P.~van Nieuwenhuizen, \emph{Supergravity and
  the s matrix}, \href{https://doi.org/10.1103/PhysRevD.15.996}{\emph{Phys.
  Rev. D} {\bfseries 15} (1977) 996}.

\bibitem{Mastrolia:2011pr}
P.~Mastrolia and G.~Ossola, \emph{{On the Integrand-Reduction Method for
  Two-Loop Scattering Amplitudes}},
  \href{https://doi.org/10.1007/JHEP11(2011)014}{\emph{JHEP} {\bfseries 11}
  (2011) 014} [\href{https://arxiv.org/abs/1107.6041}{{\ttfamily 1107.6041}}].

\bibitem{Baadsgaard:2015twa}
C.~Baadsgaard, N.E.J.~Bjerrum-Bohr, J.L.~Bourjaily, S.~Caron-Huot,
  P.H.~Damgaard and B.~Feng, \emph{{New Representations of the Perturbative
  S-Matrix}}, \href{https://doi.org/10.1103/PhysRevLett.116.061601}{\emph{Phys.
  Rev. Lett.} {\bfseries 116} (2016) 061601}
  [\href{https://arxiv.org/abs/1509.02169}{{\ttfamily 1509.02169}}].

\bibitem{Sborlini:2016fcj}
G.F.R.~Sborlini, F.~Driencourt-Mangin, R.~Hernandez-Pinto and G.~Rodrigo,
  \emph{{Four-dimensional unsubtraction from the loop-tree duality}},
  \href{https://doi.org/10.1007/JHEP08(2016)160}{\emph{JHEP} {\bfseries 08}
  (2016) 160} [\href{https://arxiv.org/abs/1604.06699}{{\ttfamily
  1604.06699}}].

\bibitem{Aguilera-Verdugo:2020kzc}
J.J.~Aguilera-Verdugo, R.J.~Hernandez-Pinto, G.~Rodrigo, G.F.R.~Sborlini and
  W.J.~Torres~Bobadilla, \emph{{Causal representation of multi-loop Feynman
  integrands within the loop-tree duality}},
  \href{https://doi.org/10.1007/JHEP01(2021)069}{\emph{JHEP} {\bfseries 01}
  (2021) 069} [\href{https://arxiv.org/abs/2006.11217}{{\ttfamily
  2006.11217}}].

\bibitem{Capatti:2020ytd}
Z.~Capatti, V.~Hirschi, D.~Kermanschah, A.~Pelloni and B.~Ruijl,
  \emph{{Manifestly Causal Loop-Tree Duality}},
  \href{https://arxiv.org/abs/2009.05509}{{\ttfamily 2009.05509}}.

\bibitem{Aguilera-Verdugo:2021nrp}
J.J.~Aguilera-Verdugo, F.~Driencourt-Mangin, R.J.~Hernandez-Pinto, J.~Plenter,
  R.M.~Prisco, N.S.~Ramirez-Uribe et~al., \emph{{A Stroll through the Loop-Tree
  Duality}}, \href{https://doi.org/10.3390/sym13061029}{\emph{Symmetry}
  {\bfseries 13} (2021) 1029}
  [\href{https://arxiv.org/abs/2104.14621}{{\ttfamily 2104.14621}}].

\bibitem{Capatti:2020xjc}
Z.~Capatti, V.~Hirschi, A.~Pelloni and B.~Ruijl, \emph{{Local Unitarity: a
  representation of differential cross-sections that is locally free of
  infrared singularities at any order}},
  \href{https://doi.org/10.1007/JHEP04(2021)104}{\emph{JHEP} {\bfseries 04}
  (2021) 104} [\href{https://arxiv.org/abs/2010.01068}{{\ttfamily
  2010.01068}}].

\bibitem{Ramirez-Uribe:2024cwd}
S.~Ramirez-Uribe, P.K.~Dhani, G.F.R.~Sborlini and G.~Rodrigo, \emph{{Rewording
  Theoretical Predictions at Colliders with Vacuum Amplitudes}},
  \href{https://doi.org/10.1103/PhysRevLett.133.211901}{\emph{Phys. Rev. Lett.}
  {\bfseries 133} (2024) 211901}
  [\href{https://arxiv.org/abs/2404.05491}{{\ttfamily 2404.05491}}].

\bibitem{Ramirez-Uribe:2024tom}
S.~Ramirez-Uribe, A.E.~Renteria-Olivo, D.F.~Renteria-Estrada, J.J.~Martinez~de
  Lejarza, P.K.~Dhani, L.~Cieri et~al., \emph{{Vacuum amplitudes and time-like
  causal unitary in the loop-tree duality}},
  \href{https://doi.org/10.1007/JHEP01(2025)103}{\emph{JHEP} {\bfseries 01}
  (2025) 103} [\href{https://arxiv.org/abs/2404.05492}{{\ttfamily
  2404.05492}}].

\bibitem{Dunbar:2016aux}
D.C.~Dunbar and W.B.~Perkins, \emph{{Two-loop five-point all plus helicity
  Yang-Mills amplitude}},
  \href{https://doi.org/10.1103/PhysRevD.93.085029}{\emph{Phys. Rev. D}
  {\bfseries 93} (2016) 085029}
  [\href{https://arxiv.org/abs/1603.07514}{{\ttfamily 1603.07514}}].

\bibitem{Dunbar:2016gjb}
D.C.~Dunbar, G.R.~Jehu and W.B.~Perkins, \emph{{The two-loop n-point all-plus
  helicity amplitude}},
  \href{https://doi.org/10.1103/PhysRevD.93.125006}{\emph{Phys. Rev. D}
  {\bfseries 93} (2016) 125006}
  [\href{https://arxiv.org/abs/1604.06631}{{\ttfamily 1604.06631}}].

\bibitem{Gehrmann:2015bfy}
T.~Gehrmann, J.M.~Henn and N.A.~Lo~Presti, \emph{{Analytic form of the two-loop
  planar five-gluon all-plus-helicity amplitude in QCD}},
  \href{https://doi.org/10.1103/PhysRevLett.116.062001}{\emph{Phys. Rev. Lett.}
  {\bfseries 116} (2016) 062001} {Erratum: Phys. Rev. Lett. 116 (2016) 189903},
  [\href{https://arxiv.org/abs/1511.05409}{{\ttfamily 1511.05409}}].

\bibitem{Dunbar:2017nfy}
D.C.~Dunbar, J.H.~Godwin, G.R.~Jehu and W.B.~Perkins, \emph{{Analytic
  all-plus-helicity gluon amplitudes in QCD}},
  \href{https://doi.org/10.1103/PhysRevD.96.116013}{\emph{Phys. Rev. D}
  {\bfseries 96} (2017) 116013}
  [\href{https://arxiv.org/abs/1710.10071}{{\ttfamily 1710.10071}}].

\bibitem{Badger:2019djh}
S.~Badger, D.~Chicherin, T.~Gehrmann, G.~Heinrich, J.M.~Henn, T.~Peraro et~al.,
  \emph{{Analytic form of the full two-loop five-gluon all-plus helicity
  amplitude}},
  \href{https://doi.org/10.1103/PhysRevLett.123.071601}{\emph{Phys. Rev. Lett.}
  {\bfseries 123} (2019) 071601}
  [\href{https://arxiv.org/abs/1905.03733}{{\ttfamily 1905.03733}}].

\bibitem{Dalgleish:2020mof}
A.R.~Dalgleish, D.C.~Dunbar, W.B.~Perkins and J.M.W.~Strong, \emph{{The Full
  Color Two-Loop Six-Gluon All-Plus Helicity Amplitude}},
  \href{https://doi.org/10.1103/PhysRevD.101.076024}{\emph{Phys. Rev. D}
  {\bfseries 101} (2020) 076024}
  [\href{https://arxiv.org/abs/2003.00897}{{\ttfamily 2003.00897}}].

\end{thebibliography}\endgroup
\end{document}